\documentstyle[aps,psfig]{revtex}

\newcommand{\braa}[1]{\mbox{$\langle #1 |$}}
\newcommand{\keta}[1]{\mbox{$| #1 \rangle$}}
\newcommand{\bra}[2]{\mbox{$\langle #1 #2 |$}}
\newcommand{\ket}[2]{\mbox{$| #1 #2 \rangle$}}
\newcommand{\Vs}[4]{\mbox{$ \bra{#1}{#2} V^{s} \ket{#3}{#4}$}}

\newcommand{\be}{\begin{equation}}
\newcommand{\ee}{\end{equation}}
\newcommand{\ba}{\begin{array}}
\newcommand{\ea}{\end{array}}
\newcommand{\disp}{\displaystyle}

\begin{document}

\title{A Gapless Theory of Bose-Einstein Condensation in Dilute Gases at Finite Temperature}
\author{S.\ A.\ Morgan}
\address{Clarendon Laboratory, 
Department of Physics, University of Oxford, Parks Road, 
\mbox{Oxford OX1 3PU, United Kingdom.}}
\vspace{6mm}
\maketitle

\begin{abstract}
In this paper we develop a gapless theory of BEC which can be applied to both trapped and homogeneous gases at zero and finite temperature. The starting point for the theory is the second quantized, many-body Hamiltonian for a system of structureless bosons with pairwise interactions. A number conserving approach is used to rewrite this Hamiltonian in a form which is approximately quadratic with higher order cubic and quartic terms. The quadratic part of the Hamiltonian can be diagonalized exactly by transforming to a quasiparticle basis, while requiring that the condensate satisfy the Gross-Pitaevskii equation. The non-quadratic terms are assumed to have a small effect and are dealt with using first and second order perturbation theory. The conventional treatment of these terms, based on factorization approximations, is shown to be inconsistent.

Infra-red divergences can appear in individual terms of the perturbation expansion, but we show analytically that the total contribution beyond quadratic order is finite. The resulting excitation spectrum is gapless and the energy shifts are small for a dilute gas away from the critical region, justifying the use of perturbation theory. Ultra-violet divergences can appear if a contact potential is used to describe particle interactions.  We show that the use of this potential as an approximation to the two-body T-matrix leads naturally to a high-energy renormalization.

The theory developed in this paper is therefore well-defined at both low and high energy and provides a systematic description of Bose-Einstein condensation in dilute gases. It can therefore be used to calculate the energies and decay rates of the excitations of the system at temperatures approaching the phase transition.
\end{abstract}


\section{Introduction}

Dilute Bose-condensed gases provide a rare example of an interacting, many-body system for which a quantitative, microscopic analysis is possible at finite temperature. For this reason such systems have long been the subject of theoretical study \cite{Bogoliubov} and much research was done in the 1950's and 60's with a view to understanding the properties of liquid Helium. However, the strong interactions which are present in that system mask the purely quantum statistical effects of condensation, and it was not until 1995 with the first production of BEC in trapped gases that a quantitative comparison of theory and experiment became possible \cite{BECRb}. A large number of experiments on dilute gas BECs now exist and a wide range of properties have been measured with high precision. These systems therefore provide a very important model for testing the application of quantum field theoretical techniques to coherent, many-body systems at finite temperature.

A wide variety of different (but related) approaches to the theory of BEC were developed for the homogeneous case and have now been extended to trapped gases. 
The first quantitative analysis was given by Bogoliubov in 1947 \cite{Bogoliubov} and was based on approximating the Hamiltonian of the system by one which is quadratic and hence can be diagonalized exactly. The extension of this method to higher order calculations and trapped gases has become known as the Hartree-Fock-Bogoliubov (HFB) theory \cite{Griffin}. Bogoliubov obtained his quadratic Hamiltonian by using a description of BEC in terms of spontaneous symmetry breaking. It has recently been shown, however, that this is not necessary and the same results can be obtained using a number conserving approach \cite{Girardeau_Arnowitt,Girardeau,Gardiner,Castin_Dum}. We will follow such an approach in this paper.
 
In 1958, Beliaev introduced a theory of BEC at $T=0$ based on the Green's functions of quantum field theory \cite{Beliaev}. This approach was further developed by Popov and Fadeev \cite{Popov_Fadeev,Popov} who applied it to calculations at finite temperature, and recently Fedichev and Shlyapnikov \cite{Fedichev_Shlyapnikov} have extended the calculations to higher order and to trapped gases. A similar approach (but restricted to the homogeneous limit) has also been described by Shi and Griffin \cite{Shi}.

Beliaev's approach was also pursued by Hugenholtz and Pines \cite{Hugenholtz_Pines}, who derived the result now known as the Hugenholtz-Pines theorem. This theorem shows that the energy spectrum of a Bose gas is gapless, which means that the energy of an excitation tends to zero as its momentum tends to zero.\footnote{An exception is the charged Bose gas, for which the Hugenholtz-Pines theorem is satisfied but nonetheless there is a gap in the excitation spectrum at low energy because the Fourier transform of the Coulomb potential is proportional to $1/k^{2}$ as $k \rightarrow 0$.} This is an exact result which puts a strong constraint on any theoretical description of BEC, and can be seen as a general consequence of spontaneous symmetry breaking, as shown by Goldstone \cite{Goldstone}.  

An alternative approach to the theory of the dilute Bose gas was introduced by Lee, Huang and Yang \cite{Huang_Yang}-\cite{Lee_Yang} and was based on the use of the pseudopotential as a means of describing low-energy collisions. Higher order calculations using this method were given by Wu \cite{Wu}, who calculated the ground state energy, and by Mohling and Sirlin \cite{Mohling_Sirlin} and Mohling and Morita \cite{Mohling_Morita}, who calculated the excitation spectrum at zero and finite temperature. For some reason these last two papers are not commonly cited in the literature, even though they give an explicit demonstration of the fact that consistent calculations beyond the quadratic Hamiltonian of Bogoliubov produce finite, gapless results. The theory described in this paper is the logical extension of the work of Mohling, Sirlin and Morita to the case of an inhomogeneous gas.

In 1959, a number conserving variational approach which included the effect of all pair correlations in a dilute Bose gas was introduced by Girardeau and Arnowitt \cite{Girardeau_Arnowitt}. However, the direct application of the variational principle violates the Hugenholtz-Pines theorem by predicting a gap in the low-energy excitation spectrum (in fact this approach reproduces the HFB theory, as discussed in Sec.~\ref{HFB}). It was later shown by Takano \cite{Takano}, that the inclusion of cubic terms in the Hamiltonian removes this gap, a result which is consistent with the work of Mohling and Sirlin \cite{Mohling_Sirlin} and this paper. More recently, the variational approach has been used by Bijlsma and Stoof \cite{Bijlsma_Stoof} who considered the properties of homogeneous Bose gases in two and three dimensions. By combining the variational method with an effective Hamiltonian these authors were able to obtain a gapless theory. This approach is essentially equivalent to a gapless extension of the HFB theory which we will discuss in Sec.~\ref{C7}.

The low-temperature properties of trapped gases BECs are well-described by the Gross-Pitaevskii equation (GPE) \cite{GPE}, and properties of the condensate such as its size, shape and energy can be determined from the static solutions to this equation \cite{Dalfovo,Edwards1}. The excitation energies can be found by linearization around these static solutions \cite{Edwards2}, and the results of these calculations are in good agreement with experiment at low temperature \cite{ExRb,ExNa}. An extension of the linear response approach which includes the effect of the noncondensate has recently been developed by Rusch and Burnett \cite{Rusch}.

Current numerical calculations of the properties of BEC at finite temperature are usually based on the Popov approximation to the HFB theory \cite{Griffin}. The results of these calculations are in excellent agreement with experiment for temperatures below about $60 \%$ of the critical temperature \cite{Hutchinson_Popov,Dodd_Popov}. There is a significant discrepancy at higher temperatures, however, and this has motivated the development of theories which go beyond HFB-Popov. In addition there are a number of theoretical difficulties associated with the HFB approach, such as the presence of divergences and the prediction of a gap in the excitation spectrum at low-energy (see below). These difficulties are avoided in the Popov approximation where the term responsible for these effects is neglected. This is a rather ad hoc procedure, however, and there is a need for a systematic approach which goes beyond the HFB theory in a consistent manner. In this paper we will develop such an approach and show that it leads to a theory which is gapless and free of divergences. It is hoped that the numerical implementation of this theory to trapped gases will yield improved agreement with experiment at high temperatures.

\subsection{Problems with HFB}

It is well-known that calculations beyond the quadratic Hamiltonian of Bogoliubov using the HFB theory encounter difficulties associated with the presence of infra-red and ultra-violet divergences and the appearance of a gap in the excitation spectrum, in violation of the Hugenholtz-Pines theorem. Ultra-violet divergences arise if atomic interactions are approximated by a contact potential via
\be
\ba{c@{\hspace{1cm}}c}
{\disp V({\bf r}) \rightarrow U_{0} \delta({\bf r}), } & {\disp U_{0} = \frac{4\pi\hbar^{2}a}{m}, }
\ea
\label{contactpotential}
\ee
where $a$ is the \emph{s}-wave scattering length and $m$ is the atomic mass. The contact potential is the lowest order approximation to the pseudopotential of Huang and Yang \cite{Huang_Yang}, and the problem of ultra-violet divergences can be avoided by using a better approximation,\footnote{Specifically the correct form of the pseudopotential is $V({\bf r}) = U_0 \, \delta({\bf r})(\partial/\partial {\bf r}){\bf r}
+ \mbox{O}[(ka)^{2}]$, where ${\bf r}={\bf r_{1}}-{\bf r_{2}}$ is the separation of the two interacting atoms and $k$ is their relative momentum.} as in Ref.~\cite{Mohling_Sirlin}. This has the disadvantage, however, that it produces a Hamiltonian which is not Hermitian. It is also not sufficiently general for our purposes as it does not allow the theory to be used for the study of two-dimensional or charged gases. We will therefore develop the theory using $V({\bf r})$ and the results we will obtain can be applied to a wide range of physical systems. For the special case of neutral atoms in three dimensions, we will show how the contact potential can be introduced in a manner which does not lead to ultra-violet divergences.

Infra-red divergences also appear in the HFB theory. An example is provided by the perturbative shifts to the quasiparticle energies which scale with momentum $k$ as $\epsilon_{k} \sim 1/k$ as $k \rightarrow 0$ (see Sec.~\ref{C6}). One way of dealing with these difficulties is to use a description of the system in terms of density and phase variables, as shown by Popov \cite{Popov_book}. We will show instead that the infra-red divergences are spurious, in the sense that a consistent treatment of higher order terms leads to finite predictions for physical quantities, and also to a gapless excitation spectrum.

\subsection{Outline of paper}

The development of the theory starts in Sec.~\ref{C2} from the second quantized, many-body Hamiltonian for a system of identical, structureless bosons. Interactions between the atoms are assumed to correspond to binary collisions and are described by a general interatomic potential $V({\bf r})$. When a condensate is present, the Hamiltonian can be rewritten in the form of Eqs.~(\ref{H})-(\ref{H4}) and is quadratic to leading order with higher order terms which are cubic and quartic in noncondensate operators. This BEC form of the Hamiltonian is usually obtained using spontaneous symmetry breaking, but instead we will use arguments based on the conservation of particle number following the approach of Refs.~\cite{Girardeau_Arnowitt,Girardeau,Gardiner,Castin_Dum}.

The BEC Hamiltonian allows a systematic calculation of the properties of dilute gas BECs, because any quadratic Hamiltonian can be diagonalized exactly, while the non-quadratic terms should be small and can be dealt with perturbatively. The exact diagonalization of the quadratic part is the subject of Sec.~\ref{C3}, and is achieved using a transformation to a {\em quasiparticle} basis, with the condensate wave function determined by the time-independent GPE. The quasiparticle energies and transformation coefficients are calculated from the Bogoliubov-de Gennes (BdG) equations.

The cubic and quartic terms in the Hamiltonian produce shifts and widths of the quasiparticle energies and become progressively more important as the temperature increases and a significant noncondensate fraction appears. These terms are treated using first and second order perturbation theory in a quasiparticle basis. Sec.~\ref{C4} is concerned with the explicit calculation of the perturbative expressions in terms of the quasiparticle energies and transformation coefficients. We also show that the conventional treatment of the non-quadratic Hamiltonian in terms of factorization approximations is inconsistent. 

In Sec.~\ref{C5} we discuss the physical interpretation of the higher order terms obtained from perturbation theory. We show that they modify the description of particle scattering and upgrade the bare interatomic potential $V({\bf r})$ to a T-matrix. This allows us to rewrite all interaction matrix elements in terms of the two-body T-matrix which describes particle collisions in a vacuum. It is this T-matrix, not $V({\bf r})$, which can be approximated by a contact potential for low-energy scattering. The difference between the two-body T-matrix and $V({\bf r})$ naturally provides the renormalization which is required to remove ultra-violet divergences in the theory.

The problem of infra-red divergences in the theory of BEC is a feature of the homogeneous limit since in a trapped gas there is a natural low-energy cut-off determined by the size of the trap. In Sec.~\ref{C6} we therefore specialize to this limit and present explicit results for the energies and lifetimes of the quasiparticle excitations. We show analytically that, although infra-red divergences do occur in individual terms of the perturbative expressions, the total contribution beyond quadratic order is finite. In addition, the excitation spectrum is gapless as required by the Hugenholtz-Pines theorem. These results are valid at both zero and finite temperature and prove that the theory developed in this paper provides a consistent description of BEC beyond the approximation of a quadratic Hamiltonian.

In Sec.~\ref{validity} we discuss the expected range of validity of the perturbative treatment of non-quadratic terms. We show that this approach is justified for a dilute gas, but that in the homogeneous limit it must fail at the critical point. For a trapped gas, however, it is possible that the theory will remain valid even in the region of the phase transition. We conclude in Sec.~\ref{C7} with a description of a gapless extension to the usual HFB theory \cite{Proukakis_gapless} which can be used as an approximation to the full theory of this paper. This approach has recently been shown to give improved agreement with experiment at temperatures approaching the critical point \cite{Hutchinson_gapless}.

\section{The BEC Hamiltonian} \label{C2}

The many-body Hamiltonian for a system of structureless bosons with pairwise interactions can be written in the usual second quantized formalism as
\begin{equation}
\hat{H} = \sum_{ij} H^{sp}_{ij}\hat{a}^{\dagger}_{i}\hat{a}_{j} + \frac{1}{2}\sum_{ijkm}\Vs{i}{j}{k}{m}\hat{a}^{\dagger}_{i}\hat{a}^{\dagger}_{j}\hat{a}_{k}\hat{a}_{m}.
\label{Hamiltonian}
\end{equation}
The operators $\hat{a}^{\dagger}_{i}$ and $\hat{a}_{i}$ respectively create or annihilate a particle from the basis state with wave function $\zeta_{i}({\bf r})$ and obey the usual Bose commutation relations. The matrix elements $H^{sp}_{ij}$ are defined by
\begin{equation}
H^{sp}_{ij} = \int \! d^{3}{\bf r} \, \zeta^{*}_{i}({\bf r}) \hat{H}^{sp} \zeta_{j}({\bf r}),
\end{equation}
where the single-particle Hamiltonian $\hat{H}^{sp}$ is
\begin{equation}
\hat{H}^{sp} = - \frac{\hbar^2}{2m}\nabla^{2} + V_{\mbox{\scriptsize Trap}}({\bf r}),
\label{H_sp}
\end{equation}
and $V_{\mbox{\scriptsize Trap}}({\bf r})$ is the magnetic potential that confines the atoms. For current experimental configurations this can be taken to be harmonic, although the formalism can also be used to describe a homogeneous gas if the trap potential is set to zero. We will denote the eigenvalues of $\hat{H}^{sp}$ by $\epsilon^{sp}_{i}$ and the corresponding eigenvectors (normalized to $1$) by $\zeta^{sp}_{i}({\bf r})$. In general, these functions do not correspond to the basis functions used in Eq.~(\ref{Hamiltonian}) [i.e. $\zeta_{i}({\bf r}) \neq \zeta^{sp}_{i}({\bf r})$] and consequently $H^{sp}_{ij}$ is not diagonal (see below).

The interaction matrix elements in Eq.~(\ref{Hamiltonian}) are symmetrized and are defined by
\be
\Vs{i}{j}{k}{m} =  \frac{1}{2} \Big [ \bra{i}{j}V\ket{k}{m} + \bra{j}{i}V\ket{k}{m} \Big ],
\ee
where
\be
\ba{c@{\hspace{1cm}}c}
{\disp \bra{i}{j}V\ket{k}{m} = \int \! d^{3}{\bf r_{1}} \, d^{3}{\bf r_{2}} \, \left \{ \zeta^{*}_{i}({\bf r_{1}})\zeta^{*}_{j}({\bf r_{2}})V({\bf r})\zeta_{k}({\bf r_{2}})\zeta_{m}({\bf r_{1}}) \right \}, }&{\disp
({\bf r} = {\bf r_{1}} - {\bf r_{2}}), }
\ea 
\ee
and $V({\bf r})$ is the bare interatomic potential.

The Hamiltonian of Eq.~(\ref{Hamiltonian}) is written in the basis provided by the orthonormal functions $\{ \zeta_{i}(\bf r) \}$. One of these, $\zeta_{0}({\bf r})$, describes the condensate and we will show in Sec.~\ref{C3} that it is a solution of the time-independent GPE \cite{GPE}. The remaining functions $\zeta_{i}({\bf r})$ (i $\neq 0$) form a complete set orthogonal to the condensate. The choice of these functions is a matter of convenience, and in Sec.~\ref{C3} we will transform to the quasiparticle amplitudes $u_{i}({\bf r})$ and $v_{i}({\bf r})$ which are determined by the Bogoliubov-de Gennes (BdG) equations. Two different choices for the basis functions will lead to the same quasiparticle amplitudes, but via a different transformation. We note, however, that for a trapped gas, the functions $\zeta_{i}({\bf r})$ can not be chosen to be eigenstates of the single-particle Hamiltonian because they must be orthogonal to the condensate and this is not an eigenstate of $\hat{H}^{sp}$ in a trap.\footnote{Possible choices for these functions are given in Eqs.~(\ref{trap_basis}) and (\ref{trap_basis2}).} This issue does not arise in the homogeneous limit where the functions $\zeta_{i}({\bf r})$ can be taken to be the usual plane waves (see Sec.~\ref{C6}).

When a condensate is present the Hamiltonian of Eq.~(\ref{Hamiltonian}) can be rewritten in a form which allows a systematic approximation scheme to be developed. This can be achieved using a number conserving approach, following the arguments of Girardeau and Arnowitt \cite{Girardeau_Arnowitt,Girardeau}, Gardiner \cite{Gardiner} and Castin and Dum \cite{Castin_Dum}. The details of these arguments are given in Ref.~\cite{thesis} and we will merely quote the results here.

Number conservation is achieved by defining the pair operators $\hat{\alpha}_{i}=\hat{\beta}^{\dagger}_{0}\hat{a}_{i}$, where $\hat{\beta}_{0} = (\hat{N}_{0}+1)^{-1/2}\hat{a}_{0}$ and $\hat{N}_{0}$ is the number operator for the condensate. The $\{\hat{\alpha}_{i} \}$ clearly conserve particle number, and they also satisfy Bose commutation relations exactly provided only that the subspace of total condensate depletion is ignored. For all practical purposes they can therefore be treated as ordinary Bose operators and in what follows we will simply denote them by $\hat{a}_{i}$ rather than $\hat{\alpha}_{i}$.

It is straightforward to show that the Hamiltonian can be rewritten in terms of the pair operators in the form \cite{thesis}
\begin{equation}
\hat{H} = H_{0} + \hat{H}_{1} + \hat{H}_{2} + \hat{H}_{3} + \hat{H}_{4},
\label{H}
\end{equation}
where
\begin{eqnarray}
H_0 &=& N_{0} \Big [ H^{sp}_{00} + \frac{1}{2}N_{0}\Vs{0}{0}{0}{0} \Big ], \label{H0} \\[5mm]
\hat{H}_{1} &=& \sqrt{N_{0}} \sum_{i \neq 0} \Big [ H^{sp}_{i0} + N_{0}\Vs{i}{0}{0}{0} \Big ] \hat{a}^{\dagger}_{i} + h.c. \, , \label{H1} \\[5mm]
\hat{H}_{2} &=& \sum_{ij \neq 0} \bigg [ H^{sp}_{ij} - \lambda \delta_{ij} + 2N_{0}\Vs{0}{i}{j}{0} \bigg ] \hat{a}^{\dagger}_{i}\hat{a}_{j} \nonumber \\
&& + \sum_{ij \neq 0} \bigg [ \frac{N_{0}}{2}\Vs{i}{j}{0}{0}\hat{a}^{\dagger}_{i}\hat{a}^{\dagger}_{j} + h.c. \bigg ] + \lambda \langle \hat{N}_{ex} \rangle, \label{H2} \\[5mm]
\hat{H}_{3} &=& \sum_{ijk \neq 0} \bigg [  \sqrt{N_{0}}\Vs{i}{j}{k}{0}\hat{a}^{\dagger}_{i}\hat{a}^{\dagger}_{j}\hat{a}_{k} +  h.c. \bigg ], \label{H3} \\[5mm]
\hat{H}_{4} &=& \sum_{ijkm \neq 0} \frac{1}{2} \Vs{i}{j}{k}{m} \hat{a}^{\dagger}_{i}\hat{a}^{\dagger}_{j}\hat{a}_{k}\hat{a}_{m}.
\label{H4}
\end{eqnarray}
Here the average $\langle \ldots \rangle$ stands for a quantum expectation value in a pure energy eigenstate (which for the purposes of this paper means a quasiparticle number state), while $\hat{N}_{ex}$ is the number operator for the noncondensate, $\hat{N}_{ex} = \sum_{i \neq 0} \hat{a}^{\dagger}_{i}\hat{a}_{i}$. The parameter $N_{0}$ is the mean condensate population and is defined by $N_{0} = N - \langle \hat{N}_{ex} \rangle$. In Eq.~(\ref{H2}) we have also \emph{defined} the parameter $\lambda$ by
\begin{equation}
\lambda \equiv H^{sp}_{00} + N_{0}\Vs{0}{0}{0}{0}.
\label{condensate_eigenvalue}
\end{equation}
We will show in Sec.~\ref{C3} that this quantity is in fact the condensate eigenvalue as calculated from the GPE. The number conserving approach automatically leads to this parameter entering the quadratic Hamiltonian, and we note that it is distinct from the related parameter $\mu$ (the chemical potential) which appears in $\hat{H}_{2}$ in a broken symmetry approach (see below).

The number conserving Hamiltonian given above has essentially the same form as that obtained using spontaneous symmetry breaking. In this approach, the condensate annihilation operator $\hat{a}_{0}$ is replaced with the number $\sqrt{N_{0}}$, on the grounds that the commutation relations of $\hat{a}_{0}$ can be neglected when $N_{0}$ is macroscopic \cite{Bogoliubov}. Since this procedure does not conserve particle number, the grand canonical Hamiltonian $\hat{H}-\mu \hat{N}$ is diagonalized rather than simply $\hat{H}$, the chemical potential $\mu$ being chosen (as a function of temperature) so that the mean number of particles is constant. 

The principle differences between the number conserving and broken symmetry approaches are in the interpretation of the operators, the replacement of the condensate eigenvalue $\lambda$ with the chemical potential $\mu$ in the quadratic Hamiltonian and the fact that the condensate population $N_{0}$ is not an independent parameter in a number conserving approach, but must be determined self-consistently by $N_{0} = N - \langle \hat{N}_{ex} \rangle$. In addition, the BEC Hamiltonian of Eqs.~(\ref{H})-(\ref{H4} describes the energy of the system rather than the `free energy' $\hat{H}-\mu \hat{N}$ which is obtained in spontaneous symmetry breaking. The relation between the two approaches is discussed in more detail in Ref.~\cite{thesis}.

The Hamiltonian of Eqs.~(\ref{H})-(\ref{H4}) is the basis of a systematic treatment of a Bose condensed system. The presence of a condensate means that $N_{0}$ is a large number and we can therefore group terms according to the powers of the condensate population they contain. The dominant terms should correspond to the quadratic Hamiltonian $\hat{H}_{\mbox{\scriptsize Q}} = H_{0} + \hat{H}_{1} + \hat{H}_{2}$ which can be diagonalized exactly, while the non-quadratic terms should have a small effect and can be dealt with perturbatively. The results of this procedure are discussed in the following sections, while a discussion of the validity of perturbation theory is deferred to Sec.~\ref{validity} when the theory has been developed further. It will turn out, however, that at $\mbox{T}=0$ the use of perturbation theory is justified if $n a^{3} \ll 1$, where $n = N/\Omega$ is the particle number density. This is the usual dilute gas criterion and is well-satisfied in current experiments for which $n a^{3} \sim 10^{-5}$. At finite temperature, this condition is replaced by the requirement $(k_{b}T/n_{0}U_{0}).(n_{0} a^{3})^{1/2} \ll 1$, where $n_{0} = N_{0}/\Omega$ is the condensate density. This is the same result as that obtained in Ref.~\cite{Fedichev_Shlyapnikov}.

\section{The Quadratic Hamiltonian} \label{C3}

In this section we discuss the exact diagonalization of the quadratic Hamiltonian.
The first stage is described in Sec.~\ref{T0_condensate} and involves the minimization of the energy functional $H_{0}$. This requires the condensate wave function to  be a solution of the GPE, which in turn means that the linear Hamiltonian $\hat{H}_{1}$ is identically zero. The diagonalization of $\hat{H}_{2}$ is described in Sec.~\ref{Diagonalize_H2}, and is achieved by transforming to a quasiparticle basis. The quasiparticle energies and wave functions are determined by the Bogoliubov-de Gennes equations. These equations are rewritten in the position representation in Sec.~\ref{position}, where we also discuss the issue of orthogonality of the excitations to the condensate. Although the development of the theory in this section applies to pure states, we conclude in Sec.~\ref{Thermal} with a discussion of how thermal averages can be obtained.

\subsection{The GPE} \label{T0_condensate}

The first stage in the diagonalization of the quadratic Hamiltonian is the minimization of $H_{0}$, which is a functional of the condensate wave function $\zeta_{0}({\bf r})$. The eigenstate solutions for the condensate can therefore be found by functional differentiation of this Hamiltonian with respect to $\zeta^{*}_{0}({\bf r})$. This leads immediately to the time-independent GPE which determines the condensate wave function \cite{GPE}. In basis space notation this can be written as the set of equations (for all $k$)
\begin{equation}
H^{sp}_{k0} + N_{0}\Vs{k}{0}{0}{0} = \lambda \delta_{k0}.
\label{zeroT_GPE}
\end{equation}
The GPE is usually written in the position representation using the contact potential approximation of Eq.~(\ref{contactpotential}). In this case Eq.~(\ref{zeroT_GPE}) takes the more familiar form
\begin{equation}
 - \frac{\hbar^2}{2m}\nabla^{2}\zeta_{0}({\bf r}) + V_{\mbox{\scriptsize Trap}}({\bf r})\zeta_{0}({\bf r}) + N_{0}U_{0}|\zeta_{0}({\bf r})|^{2}\zeta_{0}({\bf r}) = \lambda \, \zeta_{0}({\bf r}).
 \label{GPE_position}
\end{equation}

The quantity $\lambda$ in Eqs.~(\ref{zeroT_GPE}) and (\ref{GPE_position}) is introduced as the Lagrange multiplier for the constraint on the normalization ($\int \! d^{3}{\bf r} \, |\zeta_{0}({\bf r})|^{2} = 1$) when we perform the functional differentiation. Comparison with Eq.~(\ref{condensate_eigenvalue}) shows that it is the same parameter as we defined earlier and which appears naturally in the quadratic Hamiltonian of Eq.~(\ref{H2}) in a number conserving approach. Equation~(\ref{zeroT_GPE}) clearly shows that $\lambda$ corresponds to the energy of a particle in the condensate.

The basis wave functions $\zeta_{i}({\bf r})$ which describe the noncondensate must be chosen to be orthogonal to $\zeta_{0}({\bf r})$. A convenient choice for these functions for a trapped gas is therefore provided by the solutions of the linear Schr\"{o}dinger equation with an effective potential produced by the trap and the condensate density \cite{Hutchinson_Popov}
\begin{equation}
 - \frac{\hbar^2}{2m}\nabla^{2}\zeta_{i}({\bf r}) + \bigg [ V_{\mbox{\scriptsize Trap}}({\bf r}) + N_{0}U_{0}|\zeta_{0}({\bf r})|^{2} \bigg ] \zeta_{i}({\bf r}) = \epsilon^{\scriptscriptstyle B}_{i} \, \zeta_{i}({\bf r}),
 \label{trap_basis}
\end{equation}
where $\epsilon^{\scriptscriptstyle B}_{i}$ denotes the energy of the basis state. Comparison with Eq.~(\ref{GPE_position}) shows that the solutions of this equation are indeed orthogonal to the condensate.

\subsection{Diagonalization of $\hat{H}_{2}$} \label{Diagonalize_H2}

If the GPE of Eq.~(\ref{zeroT_GPE}) is satisfied, then the linear Hamiltonian $\hat{H}_{1}$ of Eq.~(\ref{H1}) vanishes. This is what we would expect since the condensate gives a functional minimum of $H_{0}$ and linear variations vanish at extrema. The next order contributions to the energy of the system therefore come from the quadratic Hamiltonian $\hat{H}_{2}$ of Eq.~(\ref{H2}). This can be written as
\begin{equation}
\hat{H}_{2} =  \sum_{ij \neq 0} \left [ {\cal L}_{ij} \hat{a}^{\dagger}_{i}\hat{a}_{j} + \frac{1}{2}{\cal M}_{ij}\hat{a}^{\dagger}_{i}\hat{a}^{\dagger}_{j} + \frac{1}{2}{\cal M}^{*}_{ij}\hat{a}_{i}\hat{a}_{j} \right ] + \lambda \langle \hat{N}_{ex} \rangle,
\label{H2_LM}
\end{equation}
where we have defined the quantities
\begin{equation}
\begin{array}{c@{\hspace{10mm}}c}
{\displaystyle {\cal L}_{ij} \equiv H^{sp}_{ij} - \lambda \delta_{ij} + 2N_{0}\Vs{0}{i}{j}{0}, } &
{\displaystyle {\cal M}_{ij} \equiv N_{0}\Vs{i}{j}{0}{0}. }
\end{array}
\label{LM_zeroT}
\end{equation}
We note that the contribution from $\lambda$ in ${\cal L}_{ij}$ ensures that the excitation (quasiparticle) energies are measured relative to the condensate, while the expression $\lambda \langle \hat{N}_{ex} \rangle$ in $\hat{H}_{2}$ is simply a number which shifts the zero of energy so that the total Hamiltonian gives the absolute energy of the noncondensate.

The interaction terms in Eq.~(\ref{LM_zeroT}) represent the most important scattering processes which occur in a condensed gas. The term $2N_{0}\Vs{0}{i}{j}{0}$ in ${\cal L}_{ij}$ describes direct (Hartree) and exchange (Fock) collisions of excited atoms off the condensate, while ${\cal M}_{ij}$ describes the effect of pair formation out of the condensate. The diagonalization of the quadratic Hamiltonian corresponds to summing over these direct, exchange and pair excitation processes to all orders in the interaction potential $V({\bf r})$ and is therefore equivalent to the Random Phase Approximation \cite{Nozieres}.

$\hat{H}_{2}$ can be diagonalized by transforming to a quasiparticle basis in which new operators $\hat{\beta}_{i}$ are defined by
\begin{equation}
\hat{\beta}_{i} = \sum_{j \neq 0} U^{*}_{ij}\hat{a}_{j} - V^{*}_{ij}\hat{a}^{\dagger}_{j}.
\label{qp}
\end{equation}
We note that the quasiparticle transformation is defined in the subspace orthogonal to the condensate. If we have $m$ basis states spanning this space, then the coefficients $U_{ij}$ and $V_{ij}$ form two $m \times m$ matrices $U$ and $V$. The transformation of Eq.~(\ref{qp}) is required to be canonical, which means that it preserves the commutation relations and leads to bosonic quasiparticles. This is the case if the matrices $U$ and $V$ satisfy the following orthogonality and symmetry conditions
\begin{equation}
\begin{array}{c@{\hspace{1cm}}c}
UU^{\dagger}-VV^{\dagger} = 1, & UV^{T}-VU^{T} = 0.
\end{array}
\label{orthosym1}
\end{equation}
These conditions imply that the inverse quasiparticle transformation has the form
\begin{equation}
\hat{a}_{i} = \sum_{j \neq 0} U_{ji}\hat{\beta}_{j} + V^{*}_{ji}\hat{\beta}^{\dagger}_{j}.
\label{qp_inverse}
\end{equation}

The necessary and sufficient conditions for the quasiparticle transformation to diagonalize $\hat{H}_{2}$ are given by the Bogoliubov-de Gennes (BdG) equations. These can be written as the matrix eigenvalue problem
\begin{equation}
\left (
\begin{array}{cc}
{\cal L} & {\cal M} \\
-{\cal M}^{*} & -{\cal L}^{*}
\end{array}
\right )
\left (
\begin{array}{c}
\vec{u}_{p} \\
\vec{v}_{p}
\end{array}
\right )
= \epsilon_{p}
\left (
\begin{array}{c}
\vec{u}_{p} \\
\vec{v}_{p}
\end{array}
\right ),
\label{BdGmatrix}
\end{equation}
where ${\cal L}$ and ${\cal M}$ are the matrices with elements ${\cal L}_{ij}$ and ${\cal M}_{ij}$ respectively and we have defined the vectors $\vec{u}_{p}$ and $\vec{v}_{p}$ by $\vec{u}_{p} \equiv 
( 
\begin{array}{cccc}
 U_{p1}, & U_{p2}, & \ldots & U_{pm}
\end{array}
)^{T} $ and $
\vec{v}_{p} \equiv 
( 
\begin{array}{cccc}
V_{p1}, & V_{p2}, & \ldots & V_{pm}
\end{array}
)^{T}
$.
For obvious reasons, we will refer to the matrices ${\cal L}$ and ${\cal M}$ as the diagonal and off-diagonal elements of the BdG equations respectively. If Eq.~(\ref{BdGmatrix}) is satisfied, then the Hamiltonian takes the form
\begin{equation}
\hat{H}_{2} = \sum_{i \neq 0} \left [ \epsilon_{i} \hat{\beta}^{\dagger}_{i}\hat{\beta}_{i} + \frac{1}{2}(\epsilon_{i} - {\cal L}_{ii}) \right ] + \lambda \langle \hat{N}_{ex} \rangle,
\end{equation}
while the quasiparticle energies $\epsilon_{p}$ are given by
\begin{equation}
\epsilon_{p} =
\left (
\begin{array}{cc}
\vec{u}^{*}_{p} & -\vec{v}^{*}_{p}
\end{array}
\right )
\left (
\begin{array}{cc}
{\cal L} & {\cal M} \\
-{\cal M}^{*} & -{\cal L}^{*}
\end{array}
\right )
\left (
\begin{array}{c}
\vec{u}_{p} \\
\vec{v}_{p}
\end{array}
\right ).
\label{qp_energy_matrix}
\end{equation}

The detailed properties of the BdG equations are discussed in Ref.~\cite{Blaizot_Ripka}. We note here, however, that the solutions come in pairs with positive and negative energies, because if there is one solution with $\epsilon_{p} > 0$ then there is also another with $\epsilon_{p'}  = - \epsilon_{p}$, $\vec{u}_{p'} = \vec{v}^{*}_{p}$ and $\vec{v}_{p'} = \vec{u}^{*}_{p}$. The problem of a `missing eigenvector' may therefore appear, because if $\epsilon_{p} = 0$ only a single eigenvector is obtained rather than two independent ones. The solution with zero energy is proportional to the condensate wave function, however, so this issue does not arise in our formalism in which the BdG equations are written in the space orthogonal to the condensate. The solutions of Eq.~(\ref{BdGmatrix}) span this space and there is no solution with zero energy and no difficulty with a missing eigenvector. A discussion of the orthogonality of the excitations to the condensate is given in Sec.~\ref{position}.

Despite the fact that there is no solution of the BdG equations with exactly zero energy, we will nonetheless prove that the excitation spectrum is gapless by showing in the homogeneous limit that the energy of a quasiparticle tends to zero as its momentum tends to zero (see Sec.~\ref{C6}). Thus, although there is no solution to Eq.~(\ref{BdGmatrix}) with $\epsilon_{p} = 0$, there may be solutions with an arbitrarily small energy.

The BdG equations can be solved straightforwardly in the high-energy limit ($\epsilon_{p} \rightarrow \infty$) where the condensate has a small effect. In this limit the basis wave functions can be taken to be eigenstates of $\hat{H}^{sp}$ [i.e. $\zeta_{i}({\bf r}) = \zeta^{sp}_{i}({\bf r})$], while the quasiparticle transformation reduces to $U_{ij} \rightarrow \delta_{ij}$ and $V_{ij} \rightarrow 0$. More precisely, the leading order contributions to $U_{ij}$, $V_{ij}$ and $\epsilon_{p}$ are\footnote{At high energy we can replace $\epsilon_{i} + \epsilon_{j}$ with $\epsilon^{sp}_{i} + \epsilon^{sp}_{j}$ in the energy denominator of $V_{ij}$, but we have written the result in this form for the benefit of later comparison.} 
\begin{equation}
\begin{array}{l@{\hspace{1cm}}l}
{\displaystyle U_{ij} \rightarrow \delta_{ij}, } & {\displaystyle V_{ij} \rightarrow \frac{{\cal M}^{*}_{ij}}{-(\epsilon_{i}+ \epsilon_{j})} \, , }
\end{array}
\label{UV_high}\vspace{2mm}
\end{equation}
\begin{equation}
\epsilon_{p} \rightarrow {\cal L}_{pp} = \epsilon^{sp}_{p} - \lambda + 2N_{0}\Vs{0}{p}{p}{0}.
\label{qpenergy_HF}
\end{equation}
The result of Eq.~(\ref{qpenergy_HF}) gives the quasiparticle energy in the Hartree-Fock approximation.

For the further development of the theory, it is convenient to define the quantities
\begin{equation}
\begin{array}{c@{\hspace{1cm}}c}
{\displaystyle \rho_{ij} \equiv \langle \hat{a}^{\dagger}_{j}\hat{a}_{i} \rangle, } &
{\displaystyle \kappa_{ij} \equiv \langle \hat{a}_{j}\hat{a}_{i} \rangle, }
\end{array}
\label{rhokappa}
\end{equation}
with the properties $\rho_{ij} = \rho^{*}_{ji}$ and $\kappa_{ij} = \kappa_{ji}$. We will refer to $\rho_{ij}$ as the {\em one-body density matrix} and to $\kappa_{ij}$ as the {\em anomalous average}. As before $\langle \ldots \rangle$ denotes an ordinary quantum expectation value in a quasiparticle number state. The population of the noncondensate is related to $\rho_{ij}$ by
\begin{equation}
\langle \hat{N}_{ex} \rangle = \sum_{i \neq 0} \rho_{ii}.
\label{noncondensate_pop}
\end{equation}

Using Eq.~(\ref{qp_inverse}), $\rho_{ij}$ and $\kappa_{ij}$ can be written in terms of the quasiparticle populations $n_{p}$ and transformation coefficients $U_{ij}$ and $V_{ij}$ as
\begin{eqnarray}
\rho_{ij} &=& \sum_{p \neq 0} [ U_{pi}U^{*}_{pj} + V^{*}_{pi}V_{pj} ]n_{p} + V^{*}_{pi}V_{pj}, \label{rho_qp} \\[2mm]
\kappa_{ij} &=& \sum_{p \neq 0} [ U_{pi}V^{*}_{pj} + U_{pj}V^{*}_{pi} ]n_{p} + U_{pj}V^{*}_{pi}.
\label{kappa_qp}
\end{eqnarray}
We will also be interested in the change in $\rho_{ij}$ and $\kappa_{ij}$ when a quasiparticle is added to some particular mode $p$ (i.e. $n_{p} \rightarrow n_{p}+1$), which we will denote by $\Delta \rho_{ij}(p)$ and $\Delta \kappa_{ij}(p)$. These quantities are given by the expressions in square brackets in Eqs.~(\ref{rho_qp}) and (\ref{kappa_qp})
\begin{equation}
\begin{array}{c@{\hspace{1cm}}c}
{\displaystyle \Delta \rho_{ij}(p) = U_{pi}U^{*}_{pj} + V^{*}_{pi}V_{pj}, } &
{\displaystyle \Delta \kappa_{ij}(p) = U_{pi}V^{*}_{pj} + U_{pj}V^{*}_{pi}. }
\end{array}
\label{deltarhokappa_qp}
\end{equation}
Eqs.~(\ref{noncondensate_pop}) and (\ref{rho_qp}) show that the zero-temperature depletion of the condensate depends only on the matrix $V$ and hence arises from the off-diagonal part of the BdG equations which describes pair excitation from the condensate.

Since $\hat{H}_{2}$ is diagonal in a quasiparticle basis, we can find its contribution to the total energy of the system by taking its expectation value in a quasiparticle number state. Using Eqs.~(\ref{H2_LM}), (\ref{LM_zeroT}) and (\ref{rhokappa}) this gives
\begin{equation}
E_{2}\{n_{i}\} = \langle \hat{H}_{2} \rangle = \sum_{ij \neq 0} \Bigg \{ \Big [ H^{sp}_{ij} + 2N_{0}\Vs{0}{i}{j}{0} \Big ] \rho_{ji} 
+ \Big [ \frac{N_{0}}{2}\Vs{i}{j}{0}{0}\kappa^{*}_{ji} +c.c \Big ] \Bigg \}, \label{E2}
\end{equation}
where c.c is the complex conjugate and $\rho_{ji}$ and $\kappa_{ji}$ are calculated from Eqs.~(\ref{rho_qp}) and (\ref{kappa_qp}) for some quasiparticle distribution $\{n_{i}\}$. We note that $\lambda$ does not appear in this equation as the two contributions in Eq.~(\ref{H2_LM}) cancel when we take the average. The total energy of the system is given to quadratic order by
\begin{equation}
E_{i}\{n_{i}\} = H_{0}\{n_{i}\} + E_{2}\{n_{i}\},
\label{E_total}
\end{equation}
where $H_{0}$ depends on $\{n_{i}\}$ via its dependence on $N_{0} = N - \langle \hat{N}_{ex} \rangle =  N - \sum_{i \neq 0} \rho_{ii}$.

The excitation energies correspond to the change in the energy of the system when a single quasiparticle is created, and can therefore be calculated from the expression
\begin{equation}
\begin{array}{c@{\hspace{1cm}}c}
{\displaystyle \epsilon_{p} = E_{i}\big [ \{n_{i} \}, n_{p}+1 \big ]  - E_{i}\big [ \{n_{i}  \}, n_{p} \big ],  } &{\displaystyle  (i \neq p). }
\end{array}
\end{equation}
Linearizing this equation and using Eqs.~(\ref{H0}) and (\ref{E2}), we obtain
\begin{equation}
\epsilon_{p} = \sum_{ij \neq 0} \left [ {\cal L}_{ij} \Delta \rho_{ji}(p) + \frac{1}{2}{\cal M}^{*}_{ij} \Delta \kappa_{ji}(p) + \frac{1}{2}{\cal M}_{ij} \Delta \kappa^{*}_{ji}(p) \right ],
\label{qp_energy}
\end{equation}
with ${\cal L}_{ij}$ and ${\cal M}_{ij}$ as in Eq.~(\ref{LM_zeroT}) and $\Delta \rho_{ij}(p)$ and $\Delta \kappa_{ij}(p)$ given by Eq.~(\ref{deltarhokappa_qp}). This result is of course the same expression for the energy of a quasiparticle as can be obtained directly from the BdG equations [c.f. Eq.~(\ref{qp_energy_matrix})]. The method of derivation given here is more useful, however, when we come to consider the contribution of non-quadratic terms.

We note that the parameter $\lambda$ (contained within ${\cal L}_{ij}$), which is absent in the expression for $E_{2}$ of Eq.~(\ref{E2}), is introduced into the quasiparticle energy as the contribution from the change in $H_{0}$ when $N_{0}$ changes. This term therefore takes into account the fact that the creation of an excitation requires the removal of particles from the condensate. As a result the quasiparticle energies are measured relative to the condensate. 

\subsection{The Position Representation} \label{position}

The BdG equations are often quoted in the position representation using the contact potential and we will therefore rewrite some of the previous equations in this more familiar form. In so doing we will address the issue of the orthogonality of the excitations to the condensate which arises in this representation.

The spatial representation of the quasiparticle transformation coefficients $U_{ij}$ and $V_{ij}$ of Eq.~(\ref{qp}) is given by the functions $u_{i}({\bf r})$ and $v_{i}({\bf r})$, defined by
\begin{equation}
\begin{array}{c@{\hspace{1cm}}c}
{\displaystyle u_{i}({\bf r}) = \sum_{j \neq 0} U_{ij} \zeta_{j}({\bf r}), } & {\displaystyle v^{*}_{i}({\bf r}) = \sum_{j \neq 0} V^{*}_{ij} \zeta_{j}({\bf r}). }
\end{array}
\label{uv_position}
\end{equation}
The orthonormality and symmetry conditions of Eq.~(\ref{orthosym1}), which ensure that the quasiparticle transformation is canonical, become the integral relations
\begin{equation}
\begin{array}{lcr}
{\displaystyle \int \! d^{3}{\bf r} \, \Big \{ u_{i}({\bf r})u^{*}_{j}({\bf r}) - v_{i}({\bf r})v^{*}_{j}({\bf r}) \Big \} } &=& {\displaystyle \delta_{ij}, } \nonumber \\[2mm]
{\displaystyle \int \! d^{3}{\bf r} \, \Big \{ u_{i}({\bf r})v_{j}({\bf r}) - u_{j}({\bf r})v_{i}({\bf r}) \Big \} } &=& 0.
\end{array}
\label{orthosym2}
\end{equation}

Using the contact potential, the BdG equations can be written as 
\begin{equation}
\begin{array}{lcr}
{\displaystyle {\cal L}({\bf r})u_{j}({\bf r}) + {\cal M}({\bf r})v_{j}({\bf r}) } &=& {\displaystyle \epsilon_{j}u_{j}({\bf r}) + c_{j}\zeta_{0}({\bf r}), \nonumber } \nonumber \\[2mm]
{\displaystyle {\cal L}({\bf r})v_{j}({\bf r}) + {\cal M}^{*}({\bf r})u_{j}({\bf r}) } &=& {\displaystyle -\epsilon_{j}v_{j}({\bf r}) + c_{j}\zeta^{*}_{0}({\bf r}), }
\end{array}
\label{BdGposition}
\end{equation}
where
\begin{equation}
\begin{array}{c@{\hspace{1cm}}c}
{\displaystyle {\cal L}({\bf r}) = \hat{H}^{sp} -\lambda + 2N_{0}U_{0}|\zeta_{0}({\bf r})|^{2}, } & {\displaystyle {\cal M}({\bf r}) = N_{0}U_{0}\zeta^{2}_{0}({\bf r}). }
\end{array}
\label{LM_r}
\end{equation}
and
\begin{equation}
c_{j} = \int \! d^{3}{\bf r} \, U_{0} N_{0}| \zeta_{0}({\bf r})|^{2} \, \Big [ \zeta^{*}_{0}({\bf r})u_{j}({\bf r}) + \zeta_{0}({\bf r})v_{j}({\bf r}) \Big ].
\label{cj_orthogonality}
\end{equation}
The parameters $c_{j}$ ensure that the solutions to Eq.~(\ref{BdGposition}) with $\epsilon_{j} \neq 0$ are orthogonal to the condensate \cite{Gardiner_orthogonal}. We note that the equations also have a solution at zero energy with $u_{j}({\bf r}) = \zeta_{0}({\bf r})$ and $v_{j}({\bf r}) = -\zeta^{*}_{0}({\bf r})$ because the theory is gapless. This does not correspond to a solution of Eq.~(\ref{BdGmatrix}), however, because it is not orthogonal to the condensate.

The BdG equations are usually written in the position representation without the parameters $c_{j}$ on the right hand side, i.e. as
\begin{equation}
\begin{array}{lcr@{\hspace{-.1mm}}l}
{\displaystyle {\cal L}({\bf r})\tilde{u}_{j}({\bf r}) + {\cal M}({\bf r})\tilde{v}_{j}({\bf r}) } &=& {\displaystyle \epsilon_{j}\tilde{u}_{j}({\bf r}), } & \nonumber \\[2mm]
{\displaystyle {\cal L}({\bf r})\tilde{v}_{j}({\bf r}) + {\cal M}^{*}({\bf r})\tilde{u}_{j}({\bf r}) } &=& {\displaystyle  -\epsilon_{j}\tilde{v}_{j}({\bf r})} &.
\end{array}
\label{BdGposition2}
\end{equation}
These equations have the same eigenvalue solutions as Eq.~(\ref{BdGposition}) and they also have a solution with zero energy of the same form as that given above. Substituting this into the orthogonality and symmetry conditions of Eq.~(\ref{orthosym2}) shows that the solutions for $\epsilon_{j} \neq 0$ are orthogonal to the condensate in the generalized sense $\int \! d^{3}{\bf r} \Big [ \zeta^{*}_{0}({\bf r})\tilde{u}_{j}({\bf r}) + \zeta_{0}({\bf r})\tilde{v}_{j}({\bf r}) \Big ] = 0$. The integrals $\int \! d^{3}{\bf r} \, \zeta^{*}_{0}({\bf r})\tilde{u}_{j}({\bf r})$ and $\int \! d^{3}{\bf r} \, \zeta_{0}({\bf r})\tilde{v}_{j}({\bf r})$ are not separately zero, however, and so the quasiparticle functions $\tilde{u}_{j}({\bf r})$ and $\tilde{v}_{j}({\bf r})$ are not individually orthogonal to the condensate. Consequently, they do not correspond to the functions $u_{j}({\bf r})$ and $v_{j}({\bf r})$ defined in Eq.~(\ref{uv_position}). These functions can be obtained from the solutions of Eq.~(\ref{BdGposition2}) by removing the projection onto the condensate\footnote{This procedure also works for the generalized BdG equations which are used in the HFB-Popov and gapless HFB theories. The method fails for the full HFB theory, however, because this is not gapless (and so the condensate does not provide a zero energy solution of the equations).}
\be
\ba{c@{\hspace{1cm}}c}
{\disp u_{j}({\bf r}) = \tilde{u}_{j}({\bf r}) - c_{j}\zeta_{0}({\bf r})/\epsilon_{j},} &
v_{j}({\bf r}) = {\disp \tilde{v}_{j}({\bf r}) + c_{j}\zeta^{*}_{0}({\bf r})/\epsilon_{j},}
\ea
\label{ortho_subtract}
\ee
for $\epsilon_{j} \neq 0$. The significance of this result numerically is that it allows the BdG equations to be solved using a basis set which consists of the eigenstates of $\hat{H}^{sp}$ (which are not orthogonal to the condensate).

We will complete this section by giving the (local) spatial representations of the one-body density matrix $\rho_{ij}$ and the anomalous average $\kappa_{ij}$ which were defined in Eq.~(\ref{rhokappa}). Denoting these by $\rho_{ex}({\bf r})$ and $\kappa({\bf r})$ respectively, we have
\begin{equation}
\begin{array}{c@{\hspace{1cm}}c}
{\displaystyle \rho_{ex}({\bf r}) = \sum_{ij \neq 0} \zeta^{*}_{j}({\bf r})\zeta_{i}({\bf r})\rho_{ij},  } & {\displaystyle \kappa({\bf r}) = \sum_{ij \neq 0} \zeta_{i}({\bf r})\zeta_{j}({\bf r})\kappa_{ij} }.
\end{array}
\label{rhokappa_r}
\end{equation}
Using Eqs.~(\ref{rho_qp}), (\ref{kappa_qp}) and (\ref{uv_position}) these can be written in terms of the quasiparticle amplitudes as
\begin{eqnarray}
\rho_{ex}({\bf r}) &=& \sum_{p \neq 0} (|u_{p}({\bf r})|^{2} + |v_{p}({\bf r})|^{2})n_{p} + |v_{p}({\bf r})|^{2}, \label{rho_r} \\
\kappa({\bf r}) &=& \sum_{p \neq 0} u_{p}({\bf r})v^{*}_{p}({\bf r})(2n_{p} + 1).
\label{kappa_r}
\end{eqnarray}
The change in these quantities when $n_{p} \rightarrow n_{p}+1$ will be denoted by $\Delta \rho_{p}({\bf r})$ and $\Delta \kappa_{p}({\bf r})$.

\subsection{Thermal Averages} \label{Thermal}

All the averages $\langle \ldots \rangle$ which have appeared to this point have referred to ordinary quantum expectation values in pure quasiparticle number states. For comparison with experiment, however, we are more interested in thermal averages, which  should strictly be calculated via the canonical partition function ${\cal Z}_{c} = \sum_{\{ n_{i}\}} e^{-\beta E_{i}\{ n_{i}\}}$ with $E_{i}\{ n_{i}\}$ as in Eq.~(\ref{E_total}). To a good approximation, however, they can be obtained simply by replacing the condensate and quasiparticle populations with their thermal averages, in which case the BdG equations and quasiparticle energies become dependent on temperature (via their dependence on $N_{0}$).

The quasiparticle populations can be taken to have the usual Bose-Einstein distribution
\begin{equation}
n_{p}(T) = \frac{1}{z^{-1}e^{\beta \epsilon_{p}}-1}
\label{BE_distribution2}
\end{equation}
where the fugacity is defined by\footnote{The fugacity is defined here by $z = e^{\beta (\mu - \lambda)}$ rather than by $z = e^{\beta \mu}$ because the quasiparticle energies are measured relative to the condensate.} $z = e^{\beta (\mu - \lambda)}$ and is calculated from the condition that the thermal population satisfies $\langle \hat{N}_{ex} \rangle = N-N_{0}(T)$. The condensate population $N_{0}(T)$ should strictly be determined from the requirement that the system minimizes its free energy ${\cal F} = -k_{B}T\log{\cal Z}_{c}$. The fugacity only differs significantly from unity, however, when the condensate population is of order $1$, and so it should be sufficient simply to use the noninteracting gas result $N_{0}(T) = z/(1-z)$.

We note that the Bose-Einstein distribution makes use of the fact that the condensate eigenvalue and the chemical potential are not identical. In the limit that $N_{0} \rightarrow \infty$ the difference disappears and Eq.~(\ref{BE_distribution2}) reduces to the Planck distribution ($z=1$). The distinction is important for finite systems, however, especially near the critical point where the Planck distribution can lead to the number of excited particles being greater than $N$.

\section{Beyond the Quadratic Hamiltonian} \label{C4}

In this section we will describe how the properties of a Bose condensed gas can be calculated beyond the approximation of the quadratic Hamiltonian. The non-quadratic terms are expected to be small and we will therefore treat them using first and second order perturbation theory. This allows a calculation of the quasiparticle shifts and widths at zero and finite temperature. In addition, the GPE is upgraded to include the effect of the noncondensate on the energy and shape of the condensate.

\subsection{The Generalized GPE} \label{finiteT_condensate}

The GPE of Eq.~(\ref{zeroT_GPE}) was obtained by requiring that the condensate wave function minimize the energy functional $H_{0}$. This is a special case of the more general method of this section and is sufficient if we are only interested in the quadratic Hamiltonian. For calculations at higher order, however, we should include the effect of the noncondensate on the condensate atoms. This can be achieved by incorporating the quadratic Hamiltonian in the energy functional for the condensate and leads to a generalized GPE.

The condensate wave function is determined in general by the requirement that the system minimizes its free energy ${\cal F} = -k_{B}T\log{\cal Z}_{c}$, where ${\cal Z}_{c}$ is the canonical partition function defined above. From the definition of ${\cal F}$ and ${\cal Z}_{c}$ we have
\begin{equation}
\frac{d {\cal F}}{d\zeta^{*}_{0}} =  \frac{1}{{\cal Z}_{c}} \sum_{\{ n_{i}\}} \frac{d E_{i}}{d\zeta^{*}_{0}}e^{-\beta E_{i}} = \bigg \langle \! \bigg  \langle \frac{d E_{i}}{d\zeta^{*}_{0}} \bigg \rangle \! \bigg \rangle,
\end{equation}
where $E_{i}$ is the energy of the system for some particular quasiparticle distribution and $\langle \! \langle \ldots \rangle \! \rangle$ denotes a thermal average. To obtain the GPE of Eq.~(\ref{zeroT_GPE}) we calculated $E_{i}$ using only $H_{0}$, but at this order the more accurate result of Eq.~(\ref{E_total}) is required. Using the expression for $E_{2}$ from Eq.~(\ref{E2}) we obtain the generalized GPE
\begin{equation}
H^{sp}_{k0} + N_{0}\Vs{k}{0}{0}{0} + \sum_{ij \neq 0} \Big [ 2\Vs{k}{i}{j}{0}\rho_{ji} + \Vs{k}{0}{i}{j}\kappa_{ji} \Big ] = \lambda_{G} \delta_{k0},
\label{finiteT_GPE}
\end{equation}
where $\rho_{ij}$ and $\kappa_{ij}$ are to be interpreted here as thermal averages. We have denoted the generalized condensate eigenvalue by $\lambda_{G}$ to distinguish it from the parameter $\lambda$ which was introduced in Eq.~(\ref{condensate_eigenvalue}). Since $\lambda_{G}$ must be real, we have
\begin{equation}
\sum_{ij \neq 0} \Vs{0}{0}{i}{j} \kappa_{ji} = \sum_{ij \neq 0} \Vs{i}{j}{0}{0} \kappa^{*}_{ji}.
\label{kappa_real}
\end{equation}
If we assume a contact potential and use the fact that the condensate wave function can be chosen to be real, then this result shows that the anomalous average $\kappa({\bf r})$ of Eq.~(\ref{rhokappa_r}) can also be chosen to be real.

If we rewrite Eq.~(\ref{finiteT_GPE}) in the position representation using the contact potential approximation, then it takes the more familiar form
\begin{equation}
- \frac{\hbar^2}{2m}\nabla^{2}\zeta_{0}({\bf r}) + V_{\mbox{\scriptsize Trap}}({\bf r})\zeta_{0}({\bf r}) + N_{0}U_{0}|\zeta_{0}({\bf r})|^{2}\zeta_{0}({\bf r}) + 2U_{0}\rho_{ex}({\bf r})\zeta_{0}({\bf r}) + U_{0}\kappa({\bf r}) \zeta^{*}_{0}({\bf r}) = \lambda_{G} \, \zeta_{0}({\bf r}).
 \label{finiteT_GPE_position}
\end{equation}
For a trapped gas, a convenient choice for the basis states $\zeta_{i}({\bf r})$ is therefore provided by the solutions to the equation\footnote{To show that this gives wave functions orthogonal to the condensate we need to use the fact that $\zeta_{0}({\bf r})$, $\kappa({\bf r})$ and $\zeta_{i}({\bf r})$ can be chosen to be real.}
\begin{equation}
- \frac{\hbar^2}{2m}\nabla^{2}\zeta_{i}({\bf r}) + \bigg [ V_{\mbox{\scriptsize Trap}}({\bf r}) + N_{0}U_{0}|\zeta_{0}({\bf r})|^{2} \bigg ] \zeta_{i}({\bf r}) + 2U_{0}\rho_{ex}({\bf r})\zeta_{i}({\bf r}) + U_{0}\kappa({\bf r}) \zeta^{*}_{i}({\bf r}) = \epsilon^{\scriptscriptstyle B}_{i} \,\zeta_{i}({\bf r}),
\label{trap_basis2}
\end{equation}
which replaces Eq.~(\ref{trap_basis}).

The use of the generalized GPE leads to a change in the energy of the condensate from $\lambda$ to $\lambda_{G}$ and also to a change in its shape (this effect is absent in the homogeneous limit).\footnote{We have not introduced an additional notation to describe the change in shape, so in the rest of this section the index $0$ which appears in matrix elements is to be interpreted as referring to a solution of the generalized GPE of Eq.~(\ref{finiteT_GPE}) rather than the ordinary GPE of Eq.~(\ref{zeroT_GPE}).} These changes in the condensate affect the excited atoms and produce shifts in the quasiparticle energies. These effects are discussed further in Secs.~\ref{quartic} and \ref{cubic}.

As a result of the additional terms in Eq.~(\ref{finiteT_GPE}) relative to the ordinary GPE, the linear Hamiltonian $\hat{H}_{1}$ of Eq.~(\ref{H1}) no longer vanishes. The remainder (which we will call $\Delta \hat{H}_{1}$) is
\begin{equation}
\Delta \hat{H}_{1} = - \sqrt{N_{0}} \sum_{ijk \neq 0} \bigg \{ \Big [ 2\Vs{k}{i}{j}{0} \rho_{ji} + \Vs{k}{0}{i}{j}\kappa_{ji} \Big ] \hat{a}^{\dagger}_{k} + h.c. \bigg \}.
\label{delta_H1}
\end{equation}
$\Delta \hat{H}_{1}$ has the same characteristic size as $\hat{H}_{3}$ ($\sim \sqrt{N_{0}}$) and it can therefore be incorporated naturally into a redefinition of the cubic Hamiltonian (see Sec.~\ref{cubic}).

\subsection{The Cubic and Quartic Hamiltonians} \label{cubic_quartic}

The conventional treatment of the non-quadratic terms for trapped gases is based on factorization approximations which reduce them to linear or quadratic forms respectively \cite{Griffin}. These approximations are obtained by pairing operators in all possible ways and then replacing these pairs by their expectation values in a quasiparticle state. Thus we have for example
\begin{eqnarray}
\hat{a}^{\dagger}_{i}\hat{a}_{j}\hat{a}_{k} &\not \rightarrow& \rho_{ji}\hat{a}_{k} + \rho_{ki}\hat{a}_{j} + \kappa_{kj}\hat{a}^{\dagger}_{i}, \label{factor3} \\
\hat{a}^{\dagger}_{i}\hat{a}^{\dagger}_{j}\hat{a}_{k}\hat{a}_{m} &\not \rightarrow& \rho_{ki}\hat{a}^{\dagger}_{j}\hat{a}_{m} + \rho_{mi}\hat{a}^{\dagger}_{j}\hat{a}_{k} + \rho_{kj}\hat{a}^{\dagger}_{i}\hat{a}_{m} \nonumber \\
&+& \rho_{mj}\hat{a}^{\dagger}_{i}\hat{a}_{k} + \kappa_{mk}\hat{a}^{\dagger}_{i}\hat{a}^{\dagger}_{j} + \kappa^{*}_{ij}\hat{a}_{k}\hat{a}_{m} \label{factor4} \\
&-& (\rho_{ki}\rho_{mj} + \rho_{mi}\rho_{kj} + \kappa^{*}_{ij}\kappa_{mk}), \nonumber
\end{eqnarray}
where we have used the symbol $\not \rightarrow$ to indicate that we will {\em not} be using these approximations. Equations~(\ref{factor3}) and (\ref{factor4}) lead to a modified quadratic Hamiltonian which can be diagonalized exactly by a (self-consistent) quasiparticle transformation [with $\rho_{ij}$ and $\kappa_{ij}$ calculated from Eqs.~(\ref{rho_qp}) and (\ref{kappa_qp})].

Equation~(\ref{factor4}) is usually justified using Wick's theorem which gives \cite{Blaizot_Ripka}
\begin{equation}
\langle \! \langle\hat{a}^{\dagger}_{i}\hat{a}^{\dagger}_{j}\hat{a}_{k}\hat{a}_{m} \rangle \! \rangle = \rho_{ki}\rho_{mj} + \rho_{kj}\rho_{mi} + \kappa^{*}_{ij}\kappa_{mk},
\label{Wick}
\end{equation}
while Eq.~(\ref{factor3}) is justified by analogy. In addition, these approximations lead to the same quadratic Hamiltonian as a variational approach \cite{Blaizot_Ripka}. This is not entirely satisfactory, however, and instead we will simply apply perturbation theory to $\hat{H}_{3}$ and $\hat{H}_{4}$ on the assumption that they are small compared to the quadratic Hamiltonian. This assumption is based on the fact that these terms contain smaller powers of the condensate population, and its validity will be confirmed by an explicit calculation of the energy shifts in the homogeneous limit (see Sec.~\ref{C6}).

Since the condensate wave function now obeys the generalized GPE, the perturbing Hamiltonian includes the contribution from $\Delta \hat{H}_{1}$ and is given by
\begin{equation}
\hat{H}_{\mbox{{\scriptsize Pert}}} = \hat{H}_{3} + \Delta \hat{H}_{1} + \hat{H}_{4}.
\label{H_pert}
\end{equation}
We will treat this Hamiltonian using first and second order perturbation theory in a quasiparticle basis. Only $\hat{H}_{4}$ gives a non-zero contribution in first order perturbation theory, while at second order we only need to consider $\hat{H}_{3} + \Delta \hat{H}_{1}$ by virtue of the extra factor of $\sqrt{N_{0}}$ which this contains relative to $\hat{H}_{4}$ [c.f. Eqs.~(\ref{H3}), (\ref{H4}) and (\ref{delta_H1})]. For the same reason, the first order calculation with $\hat{H}_{4}$ is of the same order of magnitude as the second order calculation with $\hat{H}_{3} + \Delta \hat{H}_{1}$, so both must be taken into account if the calculation is to be consistent.

In the following sections we will give the formal expressions resulting from these perturbative calculations. These expressions can be evaluated analytically in the homogeneous limit (see Sec.~\ref{C6}) but otherwise can be incorporated into existing numerical codes \cite{Rusch2}. We also show that if the first order perturbation theory on $\hat{H}_{4}$ is made self-consistent, then it leads to the same results as Eq.~(\ref{factor4}). The factorization approximation on $\hat{H}_{4}$ is therefore justified as long as a first order perturbative treatment is valid. However, the factorization approximation on the product of three operators is \emph{not} justified and amounts to neglecting a number of terms which are obtained from second order perturbation theory. This leads to a gap in the excitation spectrum and a failure of infra-red divergences to cancel in physical quantities. In contrast, the perturbative treatment of the cubic Hamiltonian leads to a finite, gapless theory.

\subsection{First Order Perturbation Theory} \label{quartic}

In first order perturbation theory, the energy shift to a quasiparticle number state $\keta{s} = |n_{1}, n_{2}, n_{3} \ldots \rangle$ is
\begin{equation}
E_{\mbox{{\scriptsize Pert}}}(s,1) = \braa{s} \hat{H}_{\mbox{{\scriptsize Pert}}} \keta{s},
\end{equation}
where the perturbing Hamiltonian $\hat{H}_{\mbox{{\scriptsize Pert}}}$ is given in Eq.~(\ref{H_pert}).
Only the fourth order Hamiltonian of Eq.~(\ref{H4}) contributes in this expression and so we will denote this first order energy shift as $E_{4}$.

To evaluate $E_{4}$ we need to calculate the expectation value of the product of four operators which appears in Eq.~(\ref{H4}). This is straightforward and the result is
\begin{equation}
\langle \hat{a}^{\dagger}_{i}\hat{a}^{\dagger}_{j}\hat{a}_{k}\hat{a}_{m} \rangle = \rho_{ki}\rho_{mj} + \rho_{mi}\rho_{kj} + \kappa^{*}_{ij}\kappa_{mk},
\label{average4}
\end{equation}
where we have left out terms which are negligible in the thermodynamic limit (the full expression is given in Ref.~\cite{thesis}). This result is consistent with Wick's theorem [c.f. Eq.~(\ref{Wick})], although in that case $\rho_{ij}$ and $\kappa_{ij}$ are defined in terms of thermal averages whereas Eq.~(\ref{average4}) has been derived for a pure state. Using Eq.~(\ref{average4}) we find that $E_{4}$ is given by
\begin{equation}
E_{4} = \langle \hat{H}_{4} \rangle = \frac{1}{2} \sum_{ijkm \neq 0}  \Vs{i}{j}{k}{m} \left [ \rho_{ki}\rho_{mj} + \rho_{mi}\rho_{kj} + \kappa^{*}_{ij}\kappa_{mk} \right ].
\label{E4}
\end{equation}

Equation~(\ref{E4}) gives the contribution to the total energy of the system from first order perturbation theory on the non-quadratic Hamiltonian. For comparison with experiment, however, we are more interested in calculating the frequencies at which the system responds to external perturbations. These correspond to the quasiparticle energies, i.e. the energy required to create a quasiparticle. We are therefore interested in the change in $E_{4}$ as $n_{p} \rightarrow n_{p}+1$, which we denote by $\Delta E_{4}(p)$
\begin{equation}
\Delta E_{4}(p) = E_{4}(n_{1},n_{2} \ldots n_{p}+1 \ldots)-E_{4}(n_{1},n_{2} \ldots n_{p} \ldots).
\end{equation}
The dependence of $E_{4}$ on the quasiparticle populations is contained in $\rho_{ij}$ and $\kappa_{ij}$ so we can obtain an expression for $\Delta E_{4}(p)$ simply by writing $\rho_{ij} \longrightarrow \rho_{ij} + \Delta \rho_{ij}(p)$ and $\kappa_{ij} \longrightarrow \kappa_{ij} + \Delta \kappa_{ij}(p)$, where $\Delta \rho_{ij}(p)$ and $\Delta \kappa_{ij}(p)$ are defined in Eq.~(\ref{deltarhokappa_qp}). This gives
\begin{equation}
\Delta E_{4}(p) = \sum_{ijkm \neq 0} \left [ 2\Vs{k}{i}{j}{m}\rho_{mk}\Delta \rho_{ji}(p) +\frac{1}{2}\Vs{i}{j}{k}{m}\kappa_{km}\Delta \kappa^{*}_{ji}(p) +  \frac{1}{2}\Vs{k}{m}{i}{j}\kappa^{*}_{km}\Delta \kappa_{ji}(p)  \right ].
\label{delta_E4}
\end{equation}
In the position representation (using the contact potential approximation) this becomes
\begin{equation}
\Delta E_{4}(p) = U_{0} \int \! d^{3}{\bf r} \left [ 2\rho_{ex}({\bf r}) \Delta \rho_{p}({\bf r}) + \kappa({\bf r})\Delta \kappa_{p}({\bf r}) \right ],
\label{delta_E4_position}
\end{equation}
where we have used the fact that $\kappa({\bf r})$ can be chosen to be real.

We must also take into account, however, that the total number of particles is fixed, so a change in the quasiparticle distribution must be accompanied by a change in the condensate population. This means that the creation of a quasiparticle via $n_{p} \rightarrow n_{p}+1$ also leads to a change in the energy of the quadratic Hamiltonian which is
\begin{eqnarray}
\Delta (H_{0}+ E_{2})  &=& \bigg \{ H^{sp}_{00} + N_{0}\Vs{0}{0}{0}{0} +  \sum_{ij \neq 0} \left [ 2\Vs{0}{i}{j}{0}\rho_{ji} + \Vs{0}{0}{i}{j}\kappa_{ji} \right ] \bigg \} \Delta N_{0} \nonumber \\
&=&  -\sum_{ij \neq 0} \lambda_{G} \delta_{ij} \Delta \rho_{ji}(p),
\label{delta_Equad}
\end{eqnarray}
where we have used Eqs.~(\ref{finiteT_GPE}) and (\ref{kappa_real}) and the fact that $N_{0} = N - \langle \hat{N}_{ex} \rangle$. However, the contribution to $\lambda_{G}$ from the ordinary GPE is included in the BdG equations of Eq.~(\ref{BdGmatrix}) and is diagonalized exactly. The perturbative shift in the quasiparticle energy as a result of Eq.~(\ref{delta_Equad}) is therefore
\begin{equation}
\Delta E_{\lambda}(p) = -\sum_{ij \neq 0} \big [ \lambda_{G}-\lambda \big ] \delta_{ij} \Delta \rho_{ji}(p) = -\big [ \lambda_{G}-\lambda \big ]\Delta \langle \hat{N}_{ex} \rangle (p),
\label{delta_Elambda}
\end{equation}
where $\Delta \langle \hat{N}_{ex} \rangle(p)$ is the change in the population of the noncondensate when a quasiparticle is created in mode $p$.

We note that for a trapped gas $\lambda_{G}-\lambda$ is not simply given by the final two terms of Eq.~(\ref{finiteT_GPE}). The reason is that in this case the solution of the generalized GPE has a different shape from the solution of the ordinary GPE. Thus the condensate interaction terms $N_{0}\Vs{0}{0}{0}{0}$ are numerically different in the two equations, even though they have the same functional form. In the homogeneous limit this issue does not arise and we can write an explicit expression for $\Delta E_{\lambda}(p)$ [see Sec.~\ref{hardsphere_higher}]. We note also that the energy shift $\Delta E_{\lambda}(p)$ arises naturally as a consequence of the constraint on $N$ which means that an excitation can only be created by removing particles from the condensate. However, the same shift also appears in a  broken symmetry (and hence grand canonical) approach because the chemical potential is taken to be $\lambda_{G}$ rather than $\lambda$ for the higher order calculation.

The energy shifts described by $\Delta E_{4}(p)$ and $\Delta E_{\lambda}(p)$ are quadratic forms in the quasiparticle transformation coefficients $U_{pi}$ and $V_{pi}$ for the particular mode $p$ under consideration [this follows from the fact that $\Delta \rho_{ji}(p)$ and $\Delta \kappa_{ji}(p)$ are quadratic forms, c.f. Eq.~(\ref{deltarhokappa_qp})]. They can therefore be written straightforwardly as a modification to the matrices ${\cal L}$ and ${\cal M}$ which appear in the BdG equations of Eq.~(\ref{BdGmatrix}) [c.f. Eqs.~(\ref{qp_energy}), (\ref{delta_E4}) and (\ref{delta_Elambda})]. The new matrices are defined by
\begin{eqnarray}
{\cal L}_{ij} &=& H^{sp}_{ij} - \lambda_{G} \delta_{ij} + 2N_{0}\Vs{0}{i}{j}{0} + \sum_{km \neq 0} 2\Vs{k}{i}{j}{m}\rho_{mk}, \hspace{1cm}  \label{L_finiteT} \\
{\cal M}_{ij} &=& N_{0}\Vs{i}{j}{0}{0} + \sum_{km \neq 0} \Vs{i}{j}{k}{m}\kappa_{km}.
\label{M_finiteT}
\end{eqnarray}

Writing the energy shifts in the form of Eqs.~(\ref{L_finiteT}) and (\ref{M_finiteT}) allows the perturbative calculation to be made self-consistent if an exact diagonalization of the new BdG equations is performed. When we refer to ordinary perturbation theory we will therefore mean the simple evaluation of the expressions for $\Delta E_{4}(p)$ and $\Delta E_{\lambda}(p)$ of Eqs.~(\ref{delta_E4}) and (\ref{delta_Elambda}) using quasiparticle energies and transformation coefficients calculated from the quadratic BdG equations [i.e. with ${\cal L}$ and ${\cal M}$ given by Eq.~(\ref{LM_zeroT})]. Self-consistent perturbation theory, on the other hand, refers to the exact diagonalization of the generalized BdG equations which are obtained from the matrices ${\cal L}$ and ${\cal M}$ of Eqs.~(\ref{L_finiteT}) and (\ref{M_finiteT}) (and also including the corrections from $\hat{H}_{3}$ which will be discussed in the next section).

Comparison of Eqs.~(\ref{LM_zeroT}), (\ref{L_finiteT}) and (\ref{M_finiteT}) shows that there are three types of correction introduced by $\Delta E_{4}(p)$ and $\Delta E_{\lambda}(p)$. First, the condensate energy $\lambda$ which appears in ${\cal L}$ is upgraded to the value appropriate to the generalized GPE. Second, direct and exchange collisions between noncondensate atoms are included via the term $\sum_{km \neq 0} 2\Vs{k}{i}{j}{m}\rho_{mk}$ which appears in ${\cal L}_{ij}$. Finally, the anomalous average $\kappa_{ij}$ is introduced into the off-diagonal terms ${\cal M}_{ij}$. We will show in Sec.~\ref{C5} that the interpretation of this correction is that it upgrades the condensate-condensate interactions which appear in the leading order contribution to ${\cal M}_{ij}$ so that these are described by a many-body T-matrix.

We note that the changes to the coefficients ${\cal L}_{ij}$ and ${\cal M}_{ij}$ which are introduced by $\Delta E_{4}(p)$ are exactly the same as would have been obtained by using the factorization approximation of Eq.~(\ref{factor4}) on the operators appearing in $\hat{H}_{4}$. Thus this approximation on the product of four operators is equivalent to self-consistent first order perturbation theory and is therefore justified to the order of this calculation. This result is not too surprising given that Wick's theorem is concerned with operator averages and hence is closely related to first order perturbation theory.

Although the expressions for $\Delta E_{4}(p)$ and $\Delta E_{\lambda}(p)$ have been derived for pure quasiparticle number states, they can be straightforwardly reinterpreted as thermal averages along the lines of the discussion in Sec.~\ref{Thermal}. The energy shifts therefore become functions of temperature via their dependence on the quasiparticle populations and transformation coefficients.

\subsection{Second Order Perturbation Theory} \label{cubic}

The expression for the energy shift of a quasiparticle number state $\keta{s}$ from second order perturbation theory is
\begin{equation}
E_{\mbox{{\scriptsize Pert}}}(s,2) = \sum_{r \neq s} \frac{|\braa{r} \hat{H}_{\mbox{{\scriptsize Pert}}} \keta{s}|^{2}}{E_{s}-E_{r}},
\end{equation}
where $E_{s}$ and $E_{r}$ are energies of the {\em system} calculated from the quadratic Hamiltonian via Eq.~(\ref{E_total}). As mentioned earlier, we can neglect the contribution from $\hat{H}_{4}$ at this order of perturbation theory and consider only the modified cubic Hamiltonian defined by $\hat{H}_{3}' = \hat{H}_{3} + \Delta \hat{H}_{1}$. For this reason we will denote the second order shift in the energy as $E_{3}$.

It is convenient to rewrite the Hamiltonian $\hat{H}_{3}'$ in the quasiparticle basis. Using Eqs.~(\ref{H3}), (\ref{qp_inverse}) and (\ref{delta_H1}) we obtain the result
\begin{equation}
\hat{H}_{3}' = \hat{H}_{3} + \Delta \hat{H}_{1} = \Bigg \{ \sum_{ijk \neq 0} \left [ A_{ijk} \hat{\beta}_{i}\hat{\beta}_{j}\hat{\beta}_{k} + B_{ijk}\hat{\beta}^{\dagger}_{i}\hat{\beta}_{j}\hat{\beta}_{k} \right ] + \sum_{i \neq 0} C_{i}\hat{\beta}_{i} \Bigg \} + h.c. \, ,
\label{H3_qp}
\end{equation}
where the coefficients $A_{ijk}$, $B_{ijk}$ and $C_{i}$ are given by
\begin{equation}
\begin{array}{lcr}
A_{ijk} &=& {\displaystyle \sqrt{N_{0}} \sum_{mnq \neq 0} \Vs{q}{0}{m}{n} V_{iq}U_{jn}U_{km} } \\
&& {\displaystyle + \Vs{m}{n}{q}{0} U_{iq}V_{jn}V_{km}, }
\end{array}
\label{A}
\end{equation}

\begin{eqnarray}
B_{ijk} &=& \sqrt{N_{0}} \sum_{mnq \neq 0} \Vs{q}{0}{m}{n} \left [ U^{*}_{iq}U_{jn}U_{km} + V^{*}_{in}V_{jq}U_{km} + V^{*}_{im}U_{jn}V_{kq}  \right ] \hspace{15mm} \nonumber \\
&& \hspace{15mm} + \Vs{m}{n}{q}{0} \left [ U^{*}_{in}U_{jq}V_{km} + U^{*}_{im}V_{jn}U_{kq} + V^{*}_{iq}V_{jn}V_{km}  \right ],
\label{B}
\end{eqnarray}

\begin{equation}
\begin{array}{lcr}
{\displaystyle C_{i} + \sum_{q \neq 0}(B_{qqi} + B_{qiq})n_{q} } &=& {\displaystyle 0.}
\end{array}
\label{C}
\end{equation}
These coefficients can be rewritten straightforwardly in the position representation [where they have the form of integrals over the quasiparticle functions $u({\bf r})$ and $v({\bf r})$, c.f. Eqs.~(\ref{Abar_position})-(\ref{Cbar_position})] using Eq.~(\ref{uv_position}). We have given the linear coefficients $C_{i}$ in the particular combination of Eq.~(\ref{C}) because they appear in the energy in this form [see Eq.~(\ref{E3})]. Linear terms in the Hamiltonian arise both from $\Delta \hat{H}_{1}$ and from the use of quasiparticle commutation relations when $\hat{H}_{3}$ is written in normal order. The fact that the right hand side of Eq.~(\ref{C}) is zero demonstrates that $\Delta \hat{H}_{1}$ cancels naturally with part of the contribution to the energy from $\hat{H}_{3}$. In fact, $\Delta \hat{H}_{1}$ removes that part of $\hat{H}_{3}$ which is obtained from the factorization approximation as we will discuss near the end of this section.

The modified cubic Hamiltonian gives a contribution to the energy which is
\begin{eqnarray}
E_{3} &=& \frac{1}{6} \sum_{ijk \neq 0} \frac{|A^{P}_{ijk}|^{2}}{\epsilon_{i} + \epsilon_{j} + \epsilon_{k}} \left [ n_{i}n_{j}n_{k} - (n_{i}+1)(n_{j}+1)(n_{k}+1) \right ] \nonumber \\
&+& \frac{1}{2} \sum_{ijk \neq 0} \frac{|B_{ijk} + B_{ikj}|^{2}}{\epsilon_{j} + \epsilon_{k} - \epsilon_{i}} \left [ (n_{i}+1)n_{j}n_{k} - n_{i}(n_{j}+1)(n_{k}+1) \right ] \nonumber \\
&-& \sum_{i \neq 0}\frac{1}{\epsilon_{i}}|C_{i} + \sum_{j}(B_{jji} + B_{jij})n_{j}|^{2},
\label{E3}
\end{eqnarray}
where we have again left out terms which are negligible in the thermodynamic limit (the full expression is given in Ref~\cite{thesis}). The coefficient $A^{P}_{ijk}$ is defined as a sum over the permutations of the three indices in $A_{ijk}$, i.e.  $A^{P}_{ijk} = A_{ijk} + A_{ikj} + A_{jik} + A_{jki} + A_{kij} + A_{kji}$. We note that the expression for $E_{3}$ contains quasiparticle population terms in exactly the factors we would expect for the Bose enhancement of scattering into and out of the states involved. 

Equation~(\ref{E3}) gives the contribution to the total energy of the \emph{system} from second order perturbation theory. The contribution to the energy of a \emph{quasiparticle} corresponds to the change in $E_{3}$ when a quasiparticle is created in some particular mode $p$. We are therefore interested in the change in $E_{3}$ as $n_{p} \rightarrow n_{p}+1$, which we denote by $\Delta E_{3}(p)$
\begin{equation}
\Delta E_{3}(p) = E_{3}(n_{1},n_{2} \ldots n_{p}+1 \ldots)-E_{3}(n_{1},n_{2} \ldots n_{p} \ldots).
\end{equation}
Evaluating $\Delta E_{3}(p)$ using Eq.~(\ref{E3}) gives
\begin{eqnarray}
\Delta E_{3} (p) &=& - \sum_{ij \neq 0} \frac{|A^{P}_{pij}|^{2}}{2 \,( \epsilon_{p} + \epsilon_{i} + \epsilon_{j} )} \left [ 1 + n_{i} + n_{j} \right ] \nonumber \\
&&  +\sum_{ij \neq 0} \frac{|B_{pij} + B_{pji}|^{2}}{2 \,( \epsilon_{p} -\epsilon_{i} - \epsilon_{j})} \left [1 + n_{i} + n_{j} \right ] \label{delta_E3} \\
&& + \sum_{ij \neq 0} \frac{|B_{ijp} + B_{ipj}|^{2}}{\epsilon_{p} -\epsilon_{i}+ \epsilon_{j}} \left [n_{j} - n_{i} \right ]. \nonumber
\end{eqnarray}
Although this expression has been derived for pure states, we will now reinterpret it as a thermal average along the lines of the discussion in Sec.~\ref{Thermal}. As before, this has the effect that the energy shift becomes dependent on temperature.

We will show in Sec.~\ref{C5} that the physical interpretation of $\Delta E_{3}(p)$ is that it introduces a T-matrix into the description of condensate--noncondensate collisions in the diagonal terms of the BdG equations. We note here, however, that the first term in Eq.~(\ref{delta_E3}) corresponds to the simultaneous annihilation or creation of three quasiparticles, while the second describes Beliaev processes in which a single quasiparticle spontaneously decomposes into two others \cite{Beliaev}. These processes can occur at zero temperature and are therefore dominant in the low-temperature regime. The final term corresponds to Landau processes in which two quasiparticles collide and coalesce to form a single quasiparticle (these are essentially the reverse of Beliaev processes). Landau processes can not occur at zero temperature (because there are no excited quasiparticles) but they dominate at high temperature. If an energy matching condition is satisfied, then the Beliaev and Landau processes can occur in a real rather than a virtual sense and lead to a finite lifetime for the quasiparticles. In the homogeneous limit these lifetimes can be calculated analytically (Sec.~\ref{Widths}) and give the same result as the application of Fermi's Golden Rule to the Hamiltonian $\hat{H}_{3}$. In a trap the situation is more complicated because there may be no exact energy matches, and only a few states which are close enough to be strongly coupled. Nonetheless, $\Delta E_{3}(p)$ can be used for a trapped gas to calculate the time evolution of a quasiparticle and determine its lifetime if this is meaningful. Such calculations are currently underway and will be described in a forthcoming publication \cite{Rusch2}.

The expression for $\Delta E_{3}(p)$ of Eq.~(\ref{delta_E3}) is a quadratic form in the quasiparticle transformation coefficients $U_{pi}$ and $V_{pi}$ for the particular mode $p$ under consideration. This can be seen from the fact that each of the indices in the coefficients $A^{P}_{pij}$ and $B_{pij}$ corresponds to a linear dependence on a quasiparticle transformation matrix $U$ or $V$. Although Eq.~(\ref{delta_E3}) is all that is required for ordinary perturbation theory, if we wish to calculate $\Delta E_{3}(p)$ self-consistently we need to rewrite it in a form which explicitly separates out the quadratic dependence on $U_{pi}$ and $V_{pi}$. The result is
\begin{equation}
\Delta E_{3}(p) =
\left (
\begin{array}{cc}
\vec{u}^{*}_{p} & -\vec{v}^{*}_{p}
\end{array}
\right )
\left (
\begin{array}{cc}
\Delta {\cal L}(\epsilon_{p}) & \Delta {\cal M}(\epsilon_{p}) \\
-\Delta {\cal M}^{*} (-\epsilon_{p}) & -\Delta {\cal L}^{*} (-\epsilon_{p})
\end{array}
\right )
\left (
\begin{array}{c}
\vec{u}_{p} \\
\vec{v}_{p}
\end{array}
\right ),
\label{delta_E3_matrix}
\end{equation}
where the matrices $\Delta {\cal L}(\epsilon_{p})$ and $\Delta {\cal M}(\epsilon_{p})$ depend on the energy of the quasiparticle mode $\epsilon_{p}$ (and on temperature) and are defined by
\begin{eqnarray}
\Delta {\cal L}_{ij}(\epsilon_{p}) &=&  \sum_{km \neq 0} \frac{( 1 + n_{k} + n_{m}) }{2} \left [  \frac{\bar{A}_{ikm}\bar{A}^{*}_{jkm}}{\epsilon_{p}-(\epsilon_{k} + \epsilon_{m})} - \frac{\bar{B}^{*}_{ikm}\bar{B}_{jkm}}{\epsilon_{p} + \epsilon_{k} + \epsilon_{m}}\right ] \nonumber \\
&& \hspace{4.5cm} + \frac{(n_{m}-n_{k})\,\bar{C}^{*}_{ikm}\bar{C}_{jkm}}{\epsilon_{p} + \epsilon_{m} - \epsilon_{k}}\, , \label{delta_E3_L} \\
\Delta {\cal M}_{ij}(\epsilon_{p}) &=&  \sum_{km \neq 0} \frac{( 1 + n_{k} + n_{m}) }{2} \left [  \frac{\bar{A}_{ikm}\bar{B}^{*}_{jkm}}{\epsilon_{p}-(\epsilon_{k} + \epsilon_{m})} - \frac{\bar{B}^{*}_{ikm}\bar{A}_{jkm}}{\epsilon_{p} + \epsilon_{k} + \epsilon_{m}}\right ] \nonumber \\ 
&& \hspace{4.5cm} + \frac{(n_{m}-n_{k})\,\bar{C}^{*}_{ikm}\bar{C}^{*}_{jmk}}{\epsilon_{p} + \epsilon_{m} - \epsilon_{k}}\, ,
\label{delta_E3_M}
\end{eqnarray}
where
\begin{eqnarray}
\bar{A}_{ikm} &=& 2\sqrt{N_{0}}\sum_{qr \neq 0} \Vs{i}{0}{q}{r}U_{kq}U_{mr} + \Vs{i}{r}{q}{0}[U_{kq}V_{mr}+ V_{kr}U_{mq}], \hspace{15mm} \label{A_bar} \\
\bar{B}_{ikm} &=& 2\sqrt{N_{0}}\sum_{qr \neq 0} \Vs{q}{r}{i}{0}V_{kq}V_{mr} + \Vs{r}{0}{q}{i}[U_{kq}V_{mr}+ V_{kr}U_{mq}],  \label{B_bar} \\
\bar{C}_{ikm} &=& 2\sqrt{N_{0}}\sum_{qr \neq 0} \Vs{q}{r}{i}{0}U^{*}_{kr}V_{mq} + \Vs{q}{0}{i}{r}[U^{*}_{kq}U_{mr}+ V^{*}_{kr}V_{mq}]. \label{C_bar}
\end{eqnarray}
These coefficients can be written in the position representation using the contact potential (in which form they may be more convenient for calculations) as
\begin{eqnarray}
\bar{A}_{ikm} &=& 2\sqrt{N_{0}} U_{0}\!\int \! d^{3}{\bf r} \, \zeta^{*}_{i}({\bf r}) \bigg \{ \zeta^{*}_{0}({\bf r})u_{k}({\bf r})u_{m}({\bf r}) + \zeta_{0}({\bf r})\Big [ u_{k}({\bf r})v_{m}({\bf r}) + v_{k}({\bf r})u_{m}({\bf r})\Big ] \bigg \},  \label{Abar_position}\\
\bar{B}_{ikm} &=& 2\sqrt{N_{0}} U_{0}\!\int \! d^{3}{\bf r} \, \zeta_{i}({\bf r})\bigg \{ \zeta_{0}({\bf r})v_{k}({\bf r})v_{m}({\bf r}) + \zeta^{*}_{0}({\bf r})\Big [ u_{k}({\bf r})v_{m}({\bf r}) + v_{k}({\bf r})u_{m}({\bf r})\Big ] \bigg \},  \\
\bar{C}_{ikm} &=& 2\sqrt{N_{0}} U_{0}\!\int \! d^{3}{\bf r} \, \zeta_{i}({\bf r})\bigg \{ \zeta_{0}({\bf r})u^{*}_{k}({\bf r})v_{m}({\bf r}) + \zeta^{*}_{0}({\bf r})\Big [ u^{*}_{k}({\bf r})u_{m}({\bf r}) + v^{*}_{k}({\bf r})v_{m}({\bf r})\Big ] \bigg \}. \label{Cbar_position}
\end{eqnarray}
The quantities $\Delta {\cal L}_{ij}(\epsilon_{p})$ and $\Delta {\cal M}_{ij}(\epsilon_{p})$ satisfy $\Delta {\cal L}^{*}_{ji}(\epsilon_{p}) = \Delta {\cal L}_{ij}(\epsilon_{p})$ and $\Delta {\cal M}_{ji}(\epsilon_{p}) = \Delta {\cal M}_{ij}(-\epsilon_{p})$.

Since $\Delta E_{3}(p)$ simply modifies the matrices ${\cal L}$ and ${\cal M}$ which appear in the BdG equations, its calculation can be made self-consistent by including $\Delta {\cal L}(\epsilon_{p})$ and $\Delta {\cal M}(\epsilon_{p})$ in the BdG equations and solving them exactly. If we also include the effects of $\Delta E_{4}(p)$ and $\Delta E_{\lambda}(p)$ self-consistently, then the generalized BdG equations are
\begin{equation}
\left (
\begin{array}{cc}
{\cal L}(\epsilon_{p}) & {\cal M}(\epsilon_{p}) \\
-{\cal M}^{*} (-\epsilon_{p}) & -{\cal L}^{*} (-\epsilon_{p})
\end{array}
\right )
\left (
\begin{array}{c}
\vec{u}_{p} \\
\vec{v}_{p}
\end{array}
\right )
= \epsilon_{p}
\left (
\begin{array}{c}
\vec{u}_{p} \\
\vec{v}_{p}
\end{array}
\right ),
\label{BdGmatrix2}
\end{equation}
where ${\cal L}(\epsilon_{p})$ and ${\cal M}(\epsilon_{p})$ are the matrices with elements
\begin{equation}
{\cal L}_{ij}(\epsilon_{p}) = {\cal L}_{ij} + \Delta {\cal L}_{ij}(\epsilon_{p}), \hspace*{15mm} {\cal M}_{ij}(\epsilon_{p}) = {\cal M}_{ij} + \Delta {\cal M}_{ij}(\epsilon_{p}), \label{LM_finiteT2}
\end{equation}
and ${\cal L}_{ij}$ and ${\cal M}_{ij}$ are defined in Eqs.~(\ref{L_finiteT}) and (\ref{M_finiteT}), while $\Delta {\cal L}_{ij}(\epsilon_{p})$ and $\Delta {\cal M}_{ij}(\epsilon_{p})$ are defined above. Eq.~(\ref{BdGmatrix2}) defines the final form of the BdG equations to the order of the calculation presented in this paper.

We note that the inclusion of $\Delta E_{3}(p)$ modifies the structure of the BdG equations, as can be seen from a comparison of Eqs.~(\ref{BdGmatrix}) and (\ref{BdGmatrix2}). There are two important changes. The first is that the matrix which is to be diagonalized now depends on the energy of the quasiparticle mode under consideration. This means that a single matrix diagonalization no longer yields the whole quasiparticle spectrum. The second change is that (for $\epsilon_{p} \neq 0$) the diagonal elements are no longer proportional to (the complex conjugates of) each other, and similarly for the off-diagonal elements. However, this proportionality is a consequence of any quadratic Hamiltonian, i.e. any Hamiltonian of the form of Eq.~(\ref{H2_LM}) with arbitrary coefficients ${\cal L}_{ij}$ and ${\cal M}_{ij}$. Thus the full effect of $\Delta E_{3}(p)$ can not be reproduced by any quadratic Hamiltonian. This is the reason why a variational approach to the problem of the dilute Bose gas only reproduces the HFB theory and does not include the effect of $\Delta E_{3}(p)$ (see Sec.~\ref{HFB}) \cite{Blaizot_Ripka}.

A further feature of $\Delta E_{3}(p)$ is that it is intrinsically non-local. Thus even if a contact potential is used to describe particle interactions, the position representation of Eq.~(\ref{delta_E3_matrix}) still involves integrals over two spatial coordinates. In contrast, the energies obtained from the quadratic theory or from $\Delta E_{4}(p)$ depend only on a single spatial integral if the contact potential is used.

As we discussed in Sec.~\ref{cubic_quartic}, the conventional treatment of the cubic Hamiltonian is based on the factorization approximation. A comparison of Eqs.~(\ref{H3}), (\ref{delta_H1}) and (\ref{factor3}) shows that this leads to a modified cubic Hamiltonian $\hat{H}_{3}' = \hat{H}_{3} + \Delta \hat{H}_{1}$ which is identically zero, so $E_{3}$ and $\Delta E_{3}(p)$ are completely neglected. This means that the factorization approximation only takes into account that part of $\hat{H}_{3}$ which cancels with $\Delta \hat{H}_{1}$. This term appears as a result of the change in shape of the condensate when we use the generalized rather than the ordinary GPE. Thus the factorization approximation takes condensate shape effects into account but neglects all the Beliaev and Landau processes which can occur in the noncondensate. These give a contribution to the energy which is of the same order of magnitude as $\Delta E_{4}(p)$, however, so the factorization approximation results in an {\em inconsistent} treatment of the non-quadratic Hamiltonian. Indeed we will show in Sec.~\ref{C6} that $\Delta E_{3}(p)$ is required to remove infra-red divergences in the theory and leads to a gapless spectrum. The factorization approximation on the product of three operators is therefore invalid and is the cause of many of the difficulties encountered in extending the Bogoliubov theory of BEC to higher order.

\subsection{HFB and HFB-Popov} \label{HFB}

The generalized HFB theory includes the effect of the non-quadratic Hamiltonian via the factorization approximations of Eqs.~(\ref{factor3}) and ~(\ref{factor4}). As we have seen, these approximations correctly include the contribution from $\Delta E_{4}$ but neglect the terms in $\Delta E_{3}$. The generalized HFB theory therefore involves solving the BdG equations in the form of Eq.~(\ref{BdGmatrix}) but with the matrices ${\cal L}$ and ${\cal M}$ taken from Eqs.~(\ref{L_finiteT}) and (\ref{M_finiteT}). The condensate wave function and eigenvalue are determined from the generalized GPE of Eq.~(\ref{finiteT_GPE}).

The HFB theory can also be obtained using a variational approach \cite{Blaizot_Ripka}. In this case the Hamiltonian is assumed to have the quadratic form of Eq.~(\ref{H2_LM}), but the coefficients ${\cal L}_{ij}$ and ${\cal M}_{ij}$ are treated as variational parameters which are used to minimize the free energy of the system. This procedure leads to the same coefficients as are obtained by the factorization approximations and hence this approach reproduces the HFB theory. The variational calculation does not include the effect of $\Delta E_{3}(p)$ because, as we mentioned in Sec.~\ref{cubic}, this leads to an expression for the energy of a quasiparticle which does not correspond to any quadratic Hamiltonian.

The full HFB theory is not used in practice, however, because it does not predict a gapless spectrum and it suffers from infra-red divergences (see Sec.~\ref{C6}). Instead, the Popov approximation is used in which the contribution from the anomalous average is neglected in both the generalized GPE and the BdG equations. In the position representation (using the contact potential), the BdG equations therefore have the form of Eq.~(\ref{BdGposition}) with ${\cal L}({\bf r})$ and ${\cal M}({\bf r})$ given by\footnote{The equations are usually given without the coefficients $c_{j}$, i.e. as in Eq.~(\ref{BdGposition2}). Excitations orthogonal to the condensate can be obtained in this case using the prescription of Eq.~(\ref{ortho_subtract}).}
\begin{equation}
\begin{array}{c@{\hspace{1cm}}c}
{\displaystyle {\cal L}({\bf r}) = \hat{H}^{sp} -\lambda_{G} + 2U_{0}[N_{0}|\zeta_{0}({\bf r})|^{2} + \rho_{ex}({\bf r})], } & {\displaystyle {\cal M}({\bf r}) = N_{0}U_{0}\zeta^{2}_{0}({\bf r}). }
\end{array}
\label{LM_r_Popov}
\end{equation}
The HFB-Popov theory is gapless and it is also free of infra-red divergences as we will show in Sec.~\ref{C6}. It is therefore preferable to the full HFB approach and for this reason it is the basis of recent numerical calculations at finite temperature \cite{Hutchinson_Popov,Dodd_Popov}.

An interesting feature of the Popov approximation is that the shifts in the quasiparticle energies depend predominately on the spatial variation of the noncondensate density in the region of the condensate. This can easily be seen if we write the expression for $\Delta E_{\lambda}(p)$ of Eq.~(\ref{delta_Elambda}) in the position representation using the contact potential approximation. If we ignore the fact that the condensate shape changes in going from the ordinary to the generalized GPE, then $\Delta E_{\lambda}(p)$ is given by
\begin{equation}
\Delta E_{\lambda}(p) \sim - U_{0} \int \! d^{3}{\bf r} \, 2\rho_{ex}({\bf r})|\zeta_{0}({\bf r})|^{2} \int \! d^{3}{\bf r'}\Delta \rho_{p}({\bf r'}).
\end{equation}
Comparison with the expression for $\Delta E_{4}(p)$ of Eq.~(\ref{delta_E4_position}) (with $\kappa({\bf r}) \rightarrow 0$) shows that if $\rho_{ex}({\bf r})$ is independent of ${\bf r}$ then the combination $\Delta E_{4}(p) + \Delta E_{\lambda}(p)$ is zero. A consequence of this is that in the homogeneous limit there is no change in the excitation spectrum within the Popov approximation [c.f. Eq.~(\ref{deltaE4_Elambda_homogeneous})].

\section{Ultra-Violet Divergences and the T-Matrix} \label{C5}

In the previous sections we have used the bare interatomic potential $V({\bf r})$ to describe particle interactions. It is well-known, however, that at low
temperatures the scattering of neutral atoms in a three dimensional gas can be
characterized by the {\em s}-wave scattering length
$a$. This parameter is usually introduced by replacing $V({\bf r})$ with the contact potential of Eq.~(\ref{contactpotential}), although this leads to the appearance of ultra-violet divergences in the theory. This is not surprising when we consider that a contact potential can scatter high-energy atoms as effectively as low-energy ones. Of course this is physically
unrealistic, and in reality the momentum transfer between atoms will vanish at
large momenta ($ k > 1/a$). The contact potential is a low-energy
approximation and care must be taken to ensure that high-energy states are dealt with correctly.

The purpose of this section is to enable the contact potential to be used to describe particle interactions while introducing the required ultra-violet renormalization in a rigorous manner. This is achieved using the fact that
the contact potential is really an approximation to the low-energy limit of the
two-body T-matrix ($T_{\mbox{{\scriptsize 2b}}}$) which describes particle scattering in a vacuum. We will therefore rewrite interaction matrix elements in terms of this T-matrix and show that the difference between $T_{\mbox{{\scriptsize 2b}}}$ and the interatomic potential $V({\bf r})$ provides the necessary ultra-violet renormalization.

We note, however, that the procedure we will describe is of general validity and can be used for situations where the contact potential approximation to the T-matrix does not apply. In particular, the expressions we will obtain can be used to describe a charged Bose gas or neutral atoms in two dimensions and the only change is that a different substitution for the T-matrix is required. For this reason we have written all equations in terms of $T_{\mbox{{\scriptsize 2b}}}$, and only replaced this with $U_{0}\,\delta({\bf r})$ when we have specialized to the case of neutral atoms in three dimensions.

As well as the removal of ultra-violet divergences, there are a number of additional reasons why it is important to rewrite matrix elements in terms of
$T_{\mbox{{\scriptsize 2b}}}$. The first is the fact that the details of the
interatomic potential are often not very well-known, while the low-energy T-matrix is fairly universal and can be characterized by a single parameter, namely the {\em s}-wave scattering length $a$. The second is that the contact potential is very convenient for numerical calculations since it leads to a considerable simplification of the equations. A third reason is that, for singular potentials (such as a hard sphere), the matrix elements $\Vs{i}{j}{k}{m}$ are actually very large or poorly
defined for low-energy states. This means that the Hamiltonian written in terms of $V({\bf r})$ is not convenient for numerical calculations. In contrast, the T-matrix elements describing scattering off such potentials are finite and well-defined.

Perhaps the most important reason for the introduction of the T-matrix, however, is the fact that our perturbative analysis can not be expected to be valid if interactions are characterized by the actual interatomic potential $V({\bf r})$. The reason is that the physical interpretation of higher order terms [specifically $\kappa_{ij}$ and $\Delta E_{3}(p)$] is that they introduce T-matrix corrections into the description of particle collisions, as we will show in Sec.~\ref{Interpret}. However, a perturbative treatment of two-body collisions (i.e. the Born approximation) is known to fail at low-energy. This is apparent from the fact that this approach gives $T_{\mbox{{\scriptsize 2b}}} \approx V({\bf r})$, whereas the low-energy T-matrix is actually independent of the details of $V({\bf r})$ (as in the contact potential for example). It is therefore necessary to take into account those terms which upgrade $V({\bf r}) \rightarrow T_{\mbox{{\scriptsize 2b}}}$ to all orders and regroup terms so that the Hamiltonian is written in terms of $T_{\mbox{{\scriptsize 2b}}}$. We will show that perturbation theory then corresponds to a Born expansion of the difference between scattering in the presence of a condensate (described by the many-body T-matrix $T_{\mbox{{\scriptsize mb}}}$) and scattering in a vacuum. These effects are small for a dilute gas away from the critical point and therefore a perturbative treatment should be valid.

The structure of this section is as follows: in Secs.~\ref{T2_section} and \ref{Tmb_section}, we give a brief discussion of the two-body and many-body T-matrices. In Sec.~\ref{Interpret} we discuss the physical interpretation of $\kappa_{ij}$ and $\Delta E_{3}(p)$ in terms of these quantities. In Sec.~\ref{UVrenorm} we show that the Hamiltonian can be rewritten in terms of $T_{\mbox{{\scriptsize 2b}}}$, and that this leads automatically to an ultra-violet renormalization of the theory. Some of the mathematical details of the argument are given in Appendix~\ref{Tupper_app}.

\subsection{The Two-Body T-Matrix} \label{T2_section}

The two-body T-matrix is defined as a function of the complex parameter $z$ by the Lippmann-Schwinger equation
\begin{equation}
T_{\mbox{\scriptsize 2b}}(z) = V + \sum_{pq}
V \ket{p}{q}^{sp}\frac{1}{z -
(\epsilon^{sp}_{p} + \epsilon^{sp}_{q})} \, ^{sp}\bra{p}{q} T_{\mbox{\scriptsize 2b}}(z),
\label{T2} 
\end{equation}
where $\epsilon^{sp}_{p}$ and $\epsilon^{sp}_{q}$ are particle energies (eigenvalues of $\hat{H}^{sp}$). The kets $\ket{p}{q}^{sp}$ are the corresponding two particle eigenstates and describe the intermediate states in the collision of two atoms. The summation over these states includes the noninteracting ground state.\footnote{The labels of the intermediate states in Eq.~(\ref{T2}) correspond to single-particle wave functions $\zeta^{sp}_{i}({\bf r})$. For a trapped gas these are different from the basis functions $\zeta_{i}({\bf r})$ we have used in the theory so far because they are not orthogonal to the condensate. This does not create any difficulties, however, since Eq.~(\ref{T2}) simply serves to define an operator whose matrix elements become an input to the theory.}

In the homogeneous limit, the scattering between states with relative momentum $\hbar{\bf k}$ and $\hbar{\bf k'}$ is independent of the centre-of-mass momentum $\hbar {\bf K}$ and is described by the matrix element $T_{\mbox{\scriptsize 2b}}({\bf k'},{\bf k},z) = \bra{{\bf k'}\,}{{\bf K}} T_{\mbox{\scriptsize 2b}} (z) \ket{{\bf k}\,}{{\bf K}}$. Physical scattering events correspond to on-shell matrix elements for which $k = k'$ and $z = \lim_{\eta \rightarrow 0} \epsilon^{sp}_{k} + \imath \eta$ (here $k = |{\bf k}|$, $\epsilon^{sp}_{k} = \hbar^{2}k^{2}/2\tilde{m}$, $\tilde{m}=m/2$ is the reduced mass, and we have redefined $z$ to include the centre-of-mass energy). For neutral atoms in three dimensions, these matrix elements can be calculated in the low-energy limit $ka \ll 1$, with the result $\lim_{\stackrel{{\scriptstyle |{\bf k}| = |{\bf k'}|}}{\eta \rightarrow 0}} T_{\mbox{\scriptsize 2b}}({\bf k'},{\bf k},\epsilon^{sp}_{k} + \imath \eta)= U_{0} + \mbox{O}(ka)$. This implies that the spatial representation of the low-energy, on-shell T-matrix is given by the contact potential of Eq.~(\ref{contactpotential}). Measurements of the scattering length thus correspond to measurements of this T-matrix.

We note that the contact potential is only a valid approximation at low-energy and the full T-matrix vanishes for $k > 1/a$. The simplest way to include this in the theory is to use
the contact potential together with a momentum space cut-off
around $k = 1/a$. In fact our results do not depend on the position of this
cut-off so that it may ultimately be taken to infinity. This result (which is
discussed in Sec.~\ref{UVrenorm}) demonstrates that there is no ultra-violet
divergence in the theory if the contact potential is used as an approximation
to $T_{\mbox{\scriptsize 2b}}$ rather than $V({\bf r})$.

Although the relationship between the two-body T-matrix and the contact potential has only been discussed for on-shell elements in the homogeneous
limit, we will assume that the result can also be used to calculate
off-shell matrix elements, and in addition that it can be applied to a trapped
gas. In the latter case, this assumption relies on the validity of a local density approximation for two-body scattering in a trap and is discussed in more detail in Ref.~\cite{thesis}.

\subsection{The Many-Body T-Matrix} \label{Tmb_section}

In a Bose condensed system, particle collisions occur in the presence of the condensate and are therefore described by a many-body T-matrix. In this section we will define this T-matrix, discuss the physical effects it describes and relate it to the two-body T-matrix.

The many-body T-matrix $T_{\mbox{{\scriptsize mb}}}(z)$ appropriate to this paper is defined by the Lippmann-Schwinger equation
\begin{equation}
T_{\mbox{{\scriptsize mb}}}(z) = V + \sum_{pq \neq 0}
\frac{ V \ket{p}{q}^{sp} \, (1 + n_{p} + n_{q}) \, ^{sp}\bra{p}{q} T_{\mbox{{\scriptsize
mb}}}(z)}{z-(\epsilon_{p} + \epsilon_{q})}, \label{T_many}
\end{equation}
where $z$ is a complex parameter,\footnote{Although the many-body T-matrix is defined for a general complex $z$, this parameter is physically interpreted as the energy of a collision. We note therefore that the zero of energy in Eq.~(\ref{T_many}) is not the same as in Eq.~(\ref{T2}). The many-body T-matrix is defined in terms of quasiparticle energies so $z$ is measured relative to the condensate in Eq.~(\ref{T_many}). The two-body T-matrix is written in terms of particle energies so in that case $z$ is measured relative to the energy of a stationary particle at the bottom of the trapping potential.} and $\epsilon_{p}$ and $n_{p}$ are quasiparticle energies and populations. We note, however, that the intermediate states $\ket{p}{q}^{sp}$ correspond to \emph{particle} wave functions (see below). In the homogeneous limit this does not pose any difficulties because a single momentum can label both particle and quasiparticle states. For a trapped gas, however, the definition of Eq.~(\ref{T_many}) only applies in the high-energy limit where quasiparticle and particle wave functions are identical.

Comparison of Eqs.~(\ref{T2}) and (\ref{T_many}) shows that the many-body T-matrix takes into account two effects of the medium in which collisions occur. First, the intermediate states may be occupied so there is Bose enhancement of scattering into and out of these levels described by the factor $1 + n_{p} + n_{q} = (n_{p}+1)(n_{p}+1) - n_{p}n_{q}$. Second, the existence of a condensate means that the energies and populations of the intermediate states correspond to quasiparticles rather than particles. 

The full quasiparticle character of the intermediate states is not taken into account in the T-matrix of Eq.~(\ref{T_many}), however, because the transformation coefficients $U_{ij}$ and $V_{ij}$ do not explicitly appear. Thus the wave functions of the intermediate states are treated in the perturbative (particle) approximation $U_{ij} \rightarrow \delta_{ij}$, $V_{ij} \rightarrow 0$. A more general T-matrix which includes the full quasiparticle wave functions has been discussed by Bijlsma and Stoof \cite{Bijlsma_Stoof}. It is the simpler expression of Eq.~(\ref{T_many}) which appears naturally in the theory, however, at least to the order of this calculation [c.f. Eq.~(\ref{M_Tmany})].

It is convenient to rewrite $T_{\mbox{{\scriptsize mb}}}$ in terms of $T_{\mbox{{\scriptsize 2b}}}$ because both these quantities are well-defined for singular potentials. Using their respective definitions, it is straightforward to show that they are related by
\begin{eqnarray}
T_{\mbox{{\scriptsize mb}}}(z)  &=&  T_{\mbox{{\scriptsize 2b}}}(\bar{z}) + \sum_{pq \neq 0}
\frac{ T_{\mbox{{\scriptsize 2b}}}(\bar{z}) \ket{p}{q}^{sp}\, (1 + n_{p} + n_{q}) \, ^{sp}\bra{p}{q} T_{\mbox{{\scriptsize
mb}}}(z)}{z-(\epsilon_{p} + \epsilon_{q})}  \nonumber \\
&& \hspace{13mm} - \sum_{pq} T_{\mbox{\scriptsize
2b}}(\bar{z}) \ket{p}{q}^{sp} \frac{1}{\bar{z} - (\epsilon^{sp}_{p} + \epsilon^{sp}_{q})} \, ^{sp}\bra{p}{q} T_{\mbox{\scriptsize
mb}}(z).
\label{T_many_T2}
\end{eqnarray}
This result greatly simplifies the calculation of $T_{\mbox{{\scriptsize mb}}}$ in terms of the \emph{s}-wave scattering length. In addition, although a Born expansion of $T_{\mbox{{\scriptsize 2b}}}$ and $T_{\mbox{{\scriptsize mb}}}$ in terms of $V({\bf r})$ will fail at low energies, an expansion of $T_{\mbox{{\scriptsize mb}}}$ in terms of $T_{\mbox{{\scriptsize 2b}}}$ based on Eq.~(\ref{T_many_T2}) should be a good approximation for a dilute gas away from the critical point.

\subsection{Interpretation of $\kappa_{ij}$ and $\Delta E_{3}$}
\label{Interpret}

In this section we will discuss the physical interpretation of $\kappa_{ij}$ and
$\Delta E_{3}(p)$ and show that they introduce the many-body T-matrix defined above into the description of particle collisions. Our discussion is based predominately on the homogeneous limit, although the results also apply to a trapped gas at high energy.

\subsubsection{Interpretation of $\kappa_{ij}$} \label{Interpret_K}

The anomalous average $\kappa_{ij}$ is defined in terms of the quasiparticle transformation coefficients by Eq.~(\ref{kappa_qp}), and as far as the numerical implementation of the theory is concerned it is simply calculated directly from this definition. For the purposes of interpretation, however, it is useful to express it in terms of quantities which have a more direct physical significance such as interaction matrix elements and quasiparticle energies and populations. This can be achieved in the high-energy limit ($i,j \rightarrow \infty$) using Eq.~(\ref{UV_high}) which gives
\begin{equation}
\kappa_{ij} \sim \frac{(1 + n_{i} + n_{j}){\cal M}_{ij}}{-(\epsilon_{i}+ \epsilon_{j})},
\label{kappa_Tmany_pert}
\end{equation}
In the quadratic theory, ${\cal M}_{ij}$ is given by Eq.~(\ref{LM_zeroT}), so comparison with Eqs.~(\ref{finiteT_GPE}) and (\ref{M_finiteT}) shows that $\kappa_{ij}$ introduces the second order Born approximation to the many-body T-matrix into condensate-condensate interactions.

In fact, self-consistent perturbation theory leads to the introduction of this T-matrix to all orders in its Lippmann-Schwinger definition. This result can be demonstrated in the homogeneous limit where the BdG equations which determine $\kappa_{ij}$ can be solved analytically. In this case it is easy to show that Eq.~(\ref{kappa_Tmany_pert}) is exact for all $i,j$, provided only that ${\cal M}_{ij}$ does not depend explicitly on $\epsilon_{p}$ (which is the case if we neglect $\Delta E_{3}(p)$ or approximate it by setting $\epsilon_{p} \rightarrow 0$). In particular, the result applies if $\Delta E_{4}$ is calculated self-consistently, in which case ${\cal M}_{ij}$ is given by Eq.~(\ref{M_finiteT}). This leads directly to the results
\begin{equation}
{\cal M}_{ij} = N_{0} \bra{i}{j} T_{\mbox{{\scriptsize mb}}}(z=0) \ket{0}{0},
\label{M_Tmany}
\end{equation}
\begin{equation}
\kappa_{ij} = \frac{(n_{i} + n_{j} + 1)}{-(\epsilon_{i}+\epsilon_{j})}N_{0}\bra{i}{j} T_{\mbox{{\scriptsize mb}}}(z=0) \ket{0}{0},
\label{kappa_solution}
\end{equation}
with $T_{\mbox{{\scriptsize mb}}}(z)$ defined as in Eq.~(\ref{T_many}).

The interpretation of the anomalous average is therefore that it introduces the many-body T-matrix into the description of condensate-condensate collisions in both the generalized GPE and the off-diagonal elements of the BdG equations. Self-consistent perturbation theory leads to the introduction of this T-matrix to all orders in the Lippmann-Schwinger equation, while ordinary perturbation theory introduces the second order contribution in a Born expansion.

\subsubsection{Interpretation of $\Delta E_{3}$} \label{Interpret_E3}

Although the general expression for $\Delta E_{3}$ is rather complicated, the physical interpretation of this term can be seen by considering its form when the coefficients $\bar{A}_{ijk}$, $\bar{B}_{ijk}$ and $\bar{C}_{ijk}$ of Eqs.~(\ref{A_bar})-(\ref{C_bar}) are treated in the particle (high-energy or perturbative) approximation $U_{ij} \rightarrow \delta_{ij}$, $V_{ij} \rightarrow 0$. In this case Eqs.~(\ref{delta_E3_L}) and (\ref{delta_E3_M}) become
\begin{equation}
\Delta {\cal L}_{ij}(\epsilon_{p}) = \sum_{km \neq 0} \left [
\frac{2N_{0}\Vs{0}{i}{k}{m}(1+n_{k}+n_{m})\Vs{k}{m}{j}{0}}{\epsilon_{p}-(\epsilon_{k}+\epsilon_{m})}
\right ] + \left [ \frac{4N_{0}
\Vs{i}{m}{k}{0}(n_{m}-n_{k})\Vs{k}{0}{j}{m}}{\epsilon_{p}+\epsilon_{m}-\epsilon_{k}}
\right ],
\label{delta_L_pert} 
\end{equation}
and
\begin{equation}
\Delta {\cal M}_{ij}(\epsilon_{p}) =  \sum_{km \neq 0}
\frac{4N_{0}\Vs{i}{m}{k}{0}(n_{m}-n_{k})\Vs{j}{k}{m}{0}}{\epsilon_{p} + \epsilon_{m}-\epsilon_{k}}.
\label{delta_M_pert} 
\end{equation}

Comparison with Eq.~(\ref{T_many}) shows that the first term in $\Delta {\cal L}_{ij}(\epsilon_{p})$ is the second order contribution to $T_{\mbox{{\scriptsize mb}}}$ for the expression $2N_{0}\Vs{0}{i}{j}{0}$,
which describes condensate-excited state collisions in ${\cal L}_{ij}$ [c.f. Eq.~(\ref{LM_zeroT})]. The T-matrix is introduced with the on-shell value of $z$ since in the limit $\epsilon_{p} \rightarrow \infty$ the quasiparticle energy is given by $\epsilon_{p} \rightarrow {\cal L}_{pp}$ [c.f. Eq.~(\ref{qpenergy_HF})].

\begin{figure}
\centerline{\psfig{file=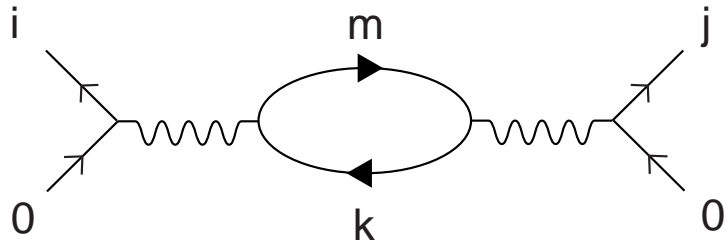,width=10cm}}
\caption{\footnotesize One of the polarization diagrams which contributes to $\Delta {\cal M}_{ij}(\epsilon_{p})$ of Eq.~(\ref{delta_M_pert}).\label{polar_fig}}
\end{figure}

The second term in the equation for $\Delta {\cal L}_{ij}(\epsilon_{p})$ is similar in structure to $\Delta {\cal M}_{ij}(\epsilon_{p})$. Each of these expressions contains the effect of four distinct scattering processes because they are written in terms of symmetrized matrix elements. The simplest contribution to $\Delta {\cal M}_{ij}(\epsilon_{p})$ is illustrated in Fig~\ref{polar_fig}, and is a modification to the processes of pair production out of the condensate described by the leading order expression for ${\cal M}_{ij}$ of Eq.~(\ref{LM_zeroT}). Figure~\ref{polar_fig} shows that $\Delta {\cal M}_{ij}(\epsilon_{p})$ describes interactions between condensate atoms which occur via the exchange of a particle from the noncondensate. These interactions can also be thought of in terms of polarization of the medium in which collisions occur; the first condensate atom causes a density fluctuation in the noncondensate which then affects the motion of a second condensate atom. 

Thus, just as $\kappa_{ij}$ introduces the many-body T-matrix in the off-diagonal terms of the BdG equations, $\Delta E_{3}(p)$ introduces these corrections into the diagonal terms and in addition introduces more complicated polarization effects. An important difference, however, is that whereas self-consistent perturbation theory on $\kappa_{ij}$ introduces $T_{\mbox{{\scriptsize mb}}}$ to all orders, $\Delta E_{3}$ only introduces it to second order in a Born expansion. This result is not altered by making the perturbation theory self-consistent, because the result of Eq.~(\ref{delta_L_pert}) was derived by considering the high-energy limit $U_{ij} \rightarrow \delta_{ij}$, $V_{ij} \rightarrow 0$. This limit is independent of the form of the interaction potential and is therefore unchanged as the perturbation theory is made self-consistent. The terms which introduce the third order contributions to $T_{\mbox{{\scriptsize mb}}}$ in condensate-excited state collisions are in fact beyond the order of this calculation and come from the use of third order perturbation theory. The fact that $\Delta E_{3}$ only introduces $T_{\mbox{{\scriptsize mb}}}$ to second order has implications for the gapless nature of the theory if self-consistent perturbation theory is used and is discussed further in Sec.~\ref{C6}.

The anomalous average and $\Delta E_{3}$ are the terms in the GPE and BdG equations which contain ultra-violet divergences if the contact potential is used as an approximation for $V({\bf r})$. The fact that both these quantities introduce T-matrix corrections suggests that there is an intimate connection between ultra-violet divergences and the T-matrix and provides a simple explanation of why such divergences occur. As we discussed in Sec.~\ref{T2_section}, the contact potential is really an approximation to $T_{\mbox{{\scriptsize 2b}}}$ rather than to $V({\bf r})$. Some of the corrections introduced by $\kappa_{ij}$ and $\Delta E_{3}$ exist even in the extremely dilute limit and correspond to the introduction of $T_{\mbox{{\scriptsize 2b}}}$. It is therefore incorrect to use the contact potential together with the full expressions for $\kappa_{ij}$ and $\Delta E_{3}$ as this amounts to double counting two-body effects. We will show in the next section that the correct use of the contact potential as an approximation for $T_{\mbox{\scriptsize 2b}}$ avoids this double counting and leads directly to a renormalization of both $\kappa_{ij}$ and $\Delta E_{3}$.

\subsection{Ultra-Violet Renormalization} \label{UVrenorm}

In this section we show how all interaction matrix elements can be rewritten in terms of the two-body T-matrix and give explicit expressions for the ultra-violet renormalization of $\kappa_{ij}$ and $\Delta E_{3}(p)$. The mathematical details of the argument are given in Appendix~\ref{Tupper_app}.

The introduction of the T-matrix occurs in two stages. In the first, we return to the original Hamiltonian of Eq.~(\ref{Hamiltonian}) and divide the single-particle basis states $\zeta_{i}({\bf r})$ into two subspaces, a low-energy region (L) and a high-energy region (H). In the low-energy subspace interactions between particles may be significant and the quasiparticle techniques we have discussed in the previous sections are needed. In the high-energy subspace, however, the effect of interactions is small and the characteristic time scale of evolution is fast compared to the low-energy states. It is therefore possible to integrate the equations of motion for high-energy operators and express them in terms of operators which act only on low-lying states. This results in the replacement of the bare interaction potential between two low-energy particles with an effective interaction which has the form of a restricted two-body T-matrix, the restriction being that the intermediate states must be of high energy.

The initial Hamiltonian of Eq.~(\ref{Hamiltonian}) is therefore replaced with an effective low-energy Hamiltonian which is [c.f. Eq.~(\ref{H_eff_app})]
\begin{equation}
\hat{H}_{\mbox{{\scriptsize eff}}} = \sum^{L}_{ij} H^{sp}_{ij}\hat{a}^{\dagger}_{i}\hat{a}_{j} + \frac{1}{2} \sum^{L}_{ijkm}  \bra{i}{j} T_{\mbox{{\scriptsize H}}} \ket{k}{m} \hat{a}^{\dagger}_{i}\hat{a}^{\dagger}_{j}\hat{a}_{k}\hat{a}_{m},
\label{H_eff}
\end{equation}
where $\sum^{L}$ means a sum over low indices. The restricted two-body T-matrix $T_{\mbox{{\scriptsize H}}}$ which describes low-energy interactions is defined by the Lippmann-Schwinger equation [c.f. Eq.~(\ref{TH_app})]
\begin{equation}
T_{\mbox{{\scriptsize H}}} = V +  \sum^{H}_{pq} V\ket{p}{q}^{sp} \frac{1}{-(\epsilon^{sp}_{p} + \epsilon^{sp}_{q})} \, ^{sp}\bra{p}{q} T_{\mbox{{\scriptsize H}}},
\label{TH}
\end{equation}
which is the usual expression for the two-body T-matrix [c.f. Eq.~(\ref{T2})] except for the fact that the intermediate states are restricted to the high-energy subspace. All our previous results can therefore be taken over by making the substitutions $\Vs{i}{j}{k}{m} \rightarrow \bra{i}{j} T_{\mbox{{\scriptsize H}}} \ket{k}{m}$ and $\sum \rightarrow \sum^{L}$ (matrix elements involving $T_{\mbox{{\scriptsize H}}}$ are to be interpreted as being symmetrized). Since the Hamiltonian now has an explicit high-energy cut-off, the problem of ultra-violet divergences is replaced by the problem that physical quantities may depend on the cut-off frequency.

The second stage of the procedure is to replace the restricted T-matrix $T_{\mbox{{\scriptsize H}}}$ with the full two-body T-matrix of Eq.~(\ref{T2}). This is achieved using the fact that these quantities are related by the expression [c.f. Eq.~(\ref{TH_T2_app})]
\begin{equation}
T_{\mbox{{\scriptsize 2b}}}(z) = T_{\mbox{{\scriptsize H}}} +  \sum^{L}_{pq} T_{\mbox{{\scriptsize H}}} \ket{p}{q}^{sp} \frac{1}{z-(\epsilon^{sp}_{p}+\epsilon^{sp}_{q})} \, ^{sp} \bra{p}{q} T_{\mbox{{\scriptsize 2b}}}(z).
\label{TH_T2}
\end{equation}
The importance of this result is that $T_{\mbox{{\scriptsize H}}}$ should be approximately equal to $T_{\mbox{{\scriptsize 2b}}}$ for systems where the density and temperature are low enough that the boundary between high and low-energy states can be taken fairly low. In this case the second term in Eq.~(\ref{TH_T2}) should be small compared with the first, and a Born expansion of $T_{\mbox{{\scriptsize H}}}$ in terms of $T_{\mbox{{\scriptsize 2b}}}$ should be valid. This is in contrast to the fact that a Born expansion of either $T_{\mbox{{\scriptsize H}}}$ or $T_{\mbox{{\scriptsize 2b}}}$ in terms of $V({\bf r})$ is expected to fail at low energy and is essentially the reason for the introduction of $T_{\mbox{{\scriptsize H}}}$.\footnote{Difficulties will arise, however, if $T_{\mbox{{\scriptsize 2b}}}(z)$ depends strongly on $z$ at low energy, or (equivalently) if $T_{\mbox{{\scriptsize H}}}$ is sensitive to the position of the cut-off. In this case the Born approximation [which in leading order is simply $T_{\mbox{{\scriptsize H}}} \approx T_{\mbox{{\scriptsize 2b}}}(z)$] will fail. This is not a problem in three dimensions where $T_{\mbox{{\scriptsize 2b}}}(z)$ tends to a constant (the contact potential) at low energy, but it will be important in two dimensions where $T_{\mbox{{\scriptsize 2b}}}(z) \sim -1/\ln(z)$ as $z \rightarrow 0$ \cite{Fisher_Hohenberg}. In this case we will have to work explicitly with $T_{\mbox{{\scriptsize H}}}$.} We show in Appendix~\ref{Tupper_app}, that at $T=0$ the Born expansion of Eq.~(\ref{TH_T2}) is valid in three dimensions if the dilute gas criterion $na^{3} \ll 1$ is satisfied. At finite temperature the requirement becomes $a/\lambda_{dB} \ll 1$ where $\lambda_{dB}$ is the de~Broglie wave length. This condition is satisfied for a dilute gas even at temperatures above the critical point.

We therefore introduce the full two-body T-matrix by writing Eq.~(\ref{TH_T2}) in a second order Born approximation as
\begin{equation}
T_{\mbox{{\scriptsize H}}} \approx T_{\mbox{{\scriptsize 2b}}}(z) -  \sum^{L}_{pq} T_{\mbox{{\scriptsize 2b}}}(z) \ket{p}{q}^{sp} \frac{1}{z-(\epsilon^{sp}_{p}+\epsilon^{sp}_{q})} \, ^{sp} \bra{p}{q} T_{\mbox{{\scriptsize 2b}}}(z).
\label{TH_T2_Born}
\end{equation}
The contribution from the first term amounts to replacing $T_{\mbox{{\scriptsize H}}}$ with $T_{\mbox{{\scriptsize 2b}}}(z)$ in all interaction matrix elements. This is all that is required (to the order of this calculation) in $\hat{H}_{3}$ and $\hat{H}_{4}$, but in $H_{0}$ and $\hat{H}_{2}$ the second term must also be included. This has the effect of renormalizing both $\kappa_{ij}$ and $\Delta E_{3}(p)$ so that they are finite even if a contact potential is used for $T_{\mbox{{\scriptsize 2b}}}$. The renormalization therefore simply amounts to the addition and subtraction of the second term in Eq.~(\ref{TH_T2_Born}). The addition of this contribution gives a better approximation to $T_{\mbox{{\scriptsize 2b}}}$ in leading order matrix elements, while the subtraction renormalizes higher order terms and ensures that two-body effects are not double counted. Since the theory is now well-behaved in the high-energy limit, all summations are convergent and they may therefore be extended to infinity. This means that physical quantities do not depend on the exact position chosen for the cut-off between high and low-energy states.

Although this procedure is valid mathematically for any (small) value of $z$, the physically appropriate values are those which are introduced naturally by $\kappa_{ij}$ and $\Delta E_{3}(p)$ in the two-body perturbative limit. In this case we have $\epsilon_{p} \rightarrow \epsilon^{sp}_{p}-\epsilon^{sp}_{0}$ [c.f. Eqs.~(\ref{condensate_eigenvalue}) and (\ref{qpenergy_HF})], so Eq.~(\ref{kappa_Tmany_pert}) shows that in the matrix elements $\bra{i}{j} T_{\mbox{{\scriptsize 2b}}}(z) \ket{0}{0}$ we should set $z = 2\epsilon^{sp}_{0}$, while Eq.~(\ref{delta_L_pert}) shows that for the matrix elements $\bra{0}{i} T_{\mbox{{\scriptsize 2b}}}(z) \ket{j}{0}$ we should set $z = \epsilon^{sp}_{p} + \epsilon^{sp}_{0}$. In the highest order matrix elements (those introduced by $\hat{H}_{3}$ and $\hat{H}_{4}$), the values of $z$ should be taken to be the two-body limit of the energy of the corresponding collision process. This means that in terms involving $\kappa_{ij}$ we have $z = 2\epsilon^{sp}_{0}$, because $\kappa_{ij}$ always introduces the T-matrix into collisions of the form $\bra{i}{j} T_{\mbox{{\scriptsize 2b}}}(z) \ket{0}{0}$. Terms involving $\rho_{ij}$ are more complicated and involve a range of different energies. The appropriate values of $z$ can be found by decomposing $\rho_{ij}$ into quasiparticles via Eq.~(\ref{rho_qp}), and considering the two-body limit of the various different quasiparticle collisions which then appear. In three dimensions, however, the T-matrix is independent of $z$ at low energy and is simply given by the contact potential.

We have now succeeded in rewriting all interaction matrix elements in terms of the two-body T-matrix. For neutral atoms in three dimensions this can be replaced by the contact potential in the homogeneous limit. For a trapped gas the situation is more complicated, however, because the T-matrix defined in Eq.~(\ref{T2}) is not the same in a trap as it is in the homogeneous limit. Nonetheless it is this T-matrix which appears in the theory so in principle we should calculate all matrix elements of $T_{\mbox{{\scriptsize 2b}}}(z)$ explicitly from their definition. This is hardly practical, however, and we will assume instead that the T-matrix in a trap can also be replaced by a contact potential. This assumption relies on the validity of the local density approximation and should be valid if the dominant contribution to particle scattering involves intermediate states of high energy which are insensitive to the details of the trapping potential. This issue is discussed further in Ref.~\cite{thesis}.

\subsubsection{The GPE and BdG equations}

Following the procedure discussed above, the generalized GPE of Eq.~(\ref{finiteT_GPE}) becomes
\begin{equation}
H^{sp}_{k0} + N_{0}\bra{k}{0} T_{\mbox{\scriptsize 2b}}(2\epsilon^{sp}_{0}) \ket{0}{0} + \sum_{ij \neq 0} \left [ 2\bra{k}{i} T_{\mbox{\scriptsize 2b}}(z) \ket{j}{0}\rho_{ji} + \bra{k}{0} T_{\mbox{\scriptsize 2b}}(2\epsilon^{sp}_{0}) \ket{i}{j}\kappa^{\scriptscriptstyle R}_{ji} \right ] = \lambda_{G} \, \delta_{k0},
\label{finiteT_GPE_renorm}
\end{equation}
where the values of $z$ in the term involving $\rho_{ji}$ are calculated as discussed above.
The renormalized anomalous average $\kappa^{\scriptscriptstyle R}_{ij}$ is defined by
\begin{equation}
\sum_{ij \neq 0} \ket{i}{j} \kappa^{\scriptscriptstyle R}_{ij} \equiv \sum_{ij \neq 0} \ket{i}{j} \kappa_{ij} - \sum_{ij} \ket{i}{j}^{sp} \kappa^{sp}_{ij},
\label{kappa_renorm}
\end{equation}
where $\kappa^{sp}_{ij}$ is the two-body, perturbative limit of $\kappa_{ij}$ [c.f. Eq.~(\ref{kappa_Tmany_pert})]
\begin{equation}
 \kappa^{sp}_{ij} = \frac{ N_{0} \, ^{sp}\bra{i}{j} T_{\mbox{\scriptsize
2b}}(2\epsilon^{sp}_{0}) \ket{0}{0}}{2\epsilon^{sp}_{0}-(\epsilon^{sp}_{i} + \epsilon^{sp}_{j})}.
\label{kappa_sp}
\end{equation}
In the position representation using the contact potential, Eq.~(\ref{kappa_renorm}) becomes
\begin{equation}
\kappa^{\scriptscriptstyle R}({\bf r}) = \kappa({\bf r}) - \kappa^{sp}({\bf r}),
\label{kappa_renorm_r}
\end{equation}
with
\begin{equation}
\kappa^{sp}({\bf r}) = N_{0}U_{0}\sum_{ij} \zeta^{sp}_{i}({\bf r})\zeta^{sp}_{j}({\bf r}) \frac{\int d^{3}{\bf r'} [ \zeta^{sp*}_{i}({\bf r'})\zeta^{sp*}_{j}({\bf r'})\zeta^{2}_{0}({\bf r'}) ]}{2\epsilon^{sp}_{0}-(\epsilon^{sp}_{i}+\epsilon^{sp}_{j})}.
\label{kappa_sp_r}
\end{equation}

The renormalized form of the coefficients ${\cal L}_{ij}(\epsilon_{p})$ and ${\cal M}_{ij}(\epsilon_{p})$ of Eq.~(\ref{LM_finiteT2}) is
\begin{eqnarray}
{\cal L}_{ij}(\epsilon_{p}) &=& H^{sp}_{ij} - \lambda_{G} \delta_{ij} + 2N_{0}\bra{0}{i} T_{\mbox{{\scriptsize 2b}}}(\epsilon^{sp}_{0} + \epsilon^{sp}_{p}) \ket{j}{0} + \sum_{km \neq 0} 2\bra{k}{i} T_{\mbox{{\scriptsize 2b}}}(z) \ket{j}{m}\rho_{mk} + \Delta {\cal L}^{\scriptscriptstyle R}_{ij}(\epsilon_{p}), \hspace{1cm} \label{L_T2} \\
{\cal M}_{ij}(\epsilon_{p}) &=& N_{0} \bra{i}{j} T_{\mbox{\scriptsize 2b}}(2\epsilon^{sp}_{0}) \ket{0}{0} + \sum_{km \neq 0} \bra{i}{j} T_{\mbox{\scriptsize 2b}}(2\epsilon^{sp}_{0}) \ket{k}{m} \kappa^{\scriptscriptstyle R}_{km} + \Delta {\cal M}_{ij}(\epsilon_{p}),
\label{M_T2}
\end{eqnarray}
where $\Delta {\cal L}^{\scriptscriptstyle R}_{ij}(\epsilon_{p})$ is calculated from the renormalized form of $\Delta E_{3}(p)$ (see below) and the condensate eigenvalue $\lambda_{G}$ is calculated from the renormalized GPE of Eq.~(\ref{finiteT_GPE_renorm}). The values of $z$ in the term involving $\rho_{mk}$ are calculated as discussed above (Sec.~\ref{UVrenorm}).
The renormalization of $\Delta E_{3}(p)$ comes from the use of the second order expression of Eq.~(\ref{TH_T2_Born}) in the matrix element $2N_{0}\bra{0}{i} T_{\mbox{\scriptsize 2b}}(\epsilon^{sp}_{0} + \epsilon^{sp}_{p}) \ket{j}{0}$ which appears in ${\cal L}_{ij}(\epsilon_{p})$.
The new form of $\Delta E_{3}(p)$ is
\begin{equation}
\Delta E^{\scriptscriptstyle R}_{3}(p) = 
\Delta E_{3}(T_{\mbox{{\scriptsize 2b}}},p) - \Delta E^{sp}_{3}(p),
\label{delta_E3_renorm}
\end{equation}
where
\begin{equation}
\Delta E^{sp}_{3}(p) = 2 N_{0} \sum_{ij \neq 0} \sum_{km} \frac{\bra{0}{i} T_{\mbox{\scriptsize 2b}}(\epsilon^{sp}_{0} + \epsilon^{sp}_{p}) \ket{k}{m}^{sp \: sp}\bra{k}{m}
T_{\mbox{\scriptsize 2b}}(\epsilon^{sp}_{0} + \epsilon^{sp}_{p}) \ket{j}{0} }{(\epsilon^{sp}_{0} + \epsilon^{sp}_{p}) - (\epsilon^{sp}_{k} + \epsilon^{sp}_{m}) } \Delta \rho_{ji}(p).
\label{delta_E3_sp}
\end{equation}
Here $\Delta E_{3}(T_{\mbox{\scriptsize 2b}},p)$ stands for the expression of Eq.~(\ref{delta_E3}) with the coefficients $A_{ijk}$ and $B_{ijk}$ given from Eqs.~(\ref{A}) and (\ref{B}) but with $V({\bf r})$ replaced by $T_{\mbox{\scriptsize 2b}}(\epsilon^{sp}_{0} + \epsilon^{sp}_{p})$ in all matrix elements. $\Delta E^{sp}_{3}(p)$ corresponds to the perturbative, two-body limit of $\Delta E_{3}(p)$ and the expression for $\Delta E^{\scriptscriptstyle R}_{3}(p)$ is finite even if the contact potential approximation is used for $T_{\mbox{\scriptsize 2b}}$. 

The physical interpretation of the renormalized quantities $\kappa^{\scriptscriptstyle R}_{ij}$ and $\Delta E^{\scriptscriptstyle R}_{3}(p)$ is that they describe the difference between scattering in a medium and scattering in a vacuum. They therefore upgrade the two-body T-matrix which now appears in leading order matrix elements and replace it with the many-body T-matrix. The use of ordinary perturbation theory on these terms corresponds to a second order Born approximation of $T_{\mbox{{\scriptsize mb}}}$ in terms of $T_{\mbox{{\scriptsize 2b}}}$ [c.f. Eq.~(\ref{T_many_T2})]. In contrast to a Born expansion in $V({\bf r})$, this is expected to be valid for a low temperature, dilute gas away from the critical point. As before, self-consistent perturbation theory introduces $T_{\mbox{\scriptsize mb}}$ to all orders in condensate-condensate collisions but still only to second order in condensate-excited state collisions. Since the expansion is now in terms of $T_{\mbox{{\scriptsize 2b}}}$, however, the difference between these is of higher order than the calculation of this paper. Thus, now that matrix elements are written in terms of the two-body T-matrix, a perturbative treatment of the non-quadratic terms in the Hamiltonian should be justified, at least away from the critical point. The validity of the theory in this form is discussed further in Sec.~\ref{validity}.

Although we have given explicit formulae for the renormalization of $\kappa_{ij}$ and $\Delta E_{3}$, it is often sufficient simply to neglect the zero-temperature contribution to these quantities. This part of the expressions contains all the ultra-violet divergence because the quasiparticle populations which appear at finite temperature vanish exponentially at high energy. Of course this procedure is not exact as it neglects the difference between the many-body and two-body T-matrices at zero temperature. For dilute gases these corrections are of order $(na^{3})^{1/2}$ and so can justifiably be ignored for the trapped gases of current experiments.

\subsubsection{The ground state energy}

In the previous section we have shown that the use of the two-body T-matrix to describe particle interactions leads to finite expressions in the GPE and BdG equations. It remains to show
that the total energy of the system (and in particular the ground state energy) is also finite as this will
establish that the theory is completely free of ultra-violet divergences.
The renormalization is rather more subtle
in this case, however, and the contribution from the non-quadratic Hamiltonian is logarithmically divergent if the contact potential is used.

The total energy of the system is given to the order of this calculation by
\begin{equation}
E = H_{0} + E_{2} + E_{3} + E_{4}.
\end{equation}
Writing this in terms of the two-body T-matrix and introducing the appropriate ultra-violet renormalization gives [c.f. Eqs.~(\ref{H0}), (\ref{E2}), (\ref{E3}) and (\ref{E4})]
\begin{eqnarray}
E &=& N_{0} \Big [ H^{sp}_{00} + \frac{N_{0}}{2} \bra{0}{0} T_{\mbox{\scriptsize 2b}} \ket{0}{0} \Big ] \nonumber \\
&&+ \sum_{ij \neq 0} \Big [ H^{sp}_{ij} + 2N_{0}\bra{0}{i} T_{\mbox{\scriptsize 2b}} \ket{j}{0} \Big ] \rho_{ji} + \sum_{ij \neq 0}  \frac{N_{0}}{2}\bra{0}{0} T_{\mbox{\scriptsize 2b}} \ket{i}{j}\kappa^{\scriptscriptstyle R}_{ji} + \frac{N_{0}}{2}\bra{i}{j} T_{\mbox{\scriptsize 2b}} \ket{0}{0}\kappa^{*}_{ji} \nonumber \\
&&+ E_{3}(V \rightarrow T_{\mbox{\scriptsize 2b}}) - 2 N_{0} \sum_{ij \neq 0} \sum_{km} \frac{\bra{0}{i} T_{\mbox{\scriptsize 2b}} \ket{k}{m}^{sp \: sp}\bra{k}{m}
T_{\mbox{\scriptsize 2b}} \ket{j}{0} }{(\epsilon^{sp}_{0} + \epsilon^{sp}_{j}) - (\epsilon^{sp}_{k} + \epsilon^{sp}_{m}) } \rho_{ji} \label{E_total_T}\\
&&+ \frac{1}{2} \sum_{ijkm \neq 0}  \bra{i}{j} T_{\mbox{\scriptsize 2b}}\ket{k}{m} \left [ \rho_{ki}\rho_{mj} + \rho_{mi}\rho_{kj} + \kappa^{{\scriptscriptstyle R} *}_{ij}\kappa^{\scriptscriptstyle R}_{mk} \right ], \nonumber
\end{eqnarray}
where for simplicity we have neglected the $z$ dependence of the T-matrix.
We note that only one of the linear contributions from $\kappa_{ij}$ in this equation is renormalized and the other still contains an ultra-violet divergence if we use a contact potential. The single-particle energy $\sum_{ij \neq 0} H^{sp}_{ij}\rho_{ji}$ is also
divergent, however, and it turns out that the two divergences cancel exactly.

The ground state energy is obtained by using the zero-temperature expressions for $\rho_{ij}$, $\kappa_{ij}$ and $E_{3}$. The renormalization leads to a finite contribution at quadratic order and in the homogeneous limit we obtain the well-known result \cite{Beliaev,Lee_Huang_Yang}
\begin{equation} 
\frac{E_{g}}{N} = \frac{nU_{0}}{2} \left[ 1 +
\frac{128}{15\sqrt{\pi}}(na^{3})^{1/2} \right],
\end{equation}
where $n$ is the total number density. At next order we need to consider the effect of $E_{3}$ and
$E_{4}$. The renormalization of $E_{4}$ leads to a finite contribution of order $na^{3}$, while the renormalization of $E_{3}$ removes its linear divergence, but leaves a logarithmic term. This can be dealt with using the fact that the full two-body T-matrix has a high-energy cut-off around $k \sim 1/a$. Truncating the summations for $E_{3}$ in this region leads to an expression for the ground state energy which has been calculated by Hugenholtz and Pines \cite{Hugenholtz_Pines} and Wu \cite{Wu} 
\begin{equation} 
\frac{E_{g}}{N} = \frac{nU_{0}}{2}
\bigg [ 1 + \frac{128}{15\sqrt{\pi}}(na^{3})^{1/2} + \,
8 \, (\frac{4\pi}{3}-\sqrt{3})(na^{3})\ln(na^{3})  +
\mbox{O}(na^{3}) \bigg]. \label{gs_energy}
\end{equation}
The term of order $na^{3}$ in this expression depends on the exact position of the cut-off and hence on the particular form of the
interatomic potential. Thus to the order of the calculation we have presented,
the contribution to the ground state energy at order $na^{3}$ is the only
quantity which can not be calculated solely in terms of the {\em s}-wave scattering length.

\section{Infra-red divergences and a gapless spectrum} \label{C6}

As mentioned in the introduction, two difficulties commonly associated with attempts to extend the Bogoliubov
method beyond the quadratic Hamiltonian are the presence of a gap in the
excitation spectrum at low energy and the appearance of infra-red
divergences in the higher order energy shifts. Genuine infinities do not appear in a trapped gas because this has a natural low energy cut-off set by the trap
frequency. Divergences may re-appear, however, as the trap is opened and the system approaches the
homogeneous limit. Thus for a trapped gas the
problem of infra-red divergences corresponds to an anomalous dependence of
physical quantities on the trap frequency. A discussion of
infra-red divergences and a gapless spectrum is therefore a discussion of the
homogeneous limit, and if the theory is well-defined in this case we can assume that this is also true in a trap.

In this section we will show that the theory developed in this paper is finite and gapless in the homogeneous limit. We will restrict our attention to neutral atoms in three dimensions for which the contact potential can be used as an approximation to the two-body T-matrix. Details of the relevant calculations can be found in Ref.~\cite{thesis}. Our notation follows that of Refs.~\cite{Mohling_Sirlin,Mohling_Morita}.

\subsection{Quadratic theory} \label{quadratic_homogeneous}

In the homogeneous limit, the GPE of Eq.~(\ref{zeroT_GPE}) [with $V({\bf r}) \rightarrow T_{\mbox{\scriptsize 2b}}(z) \rightarrow U_{0}\,\delta({\bf r})$] leads to a condensate wave function and eigenvalue given by $\zeta_{0} = 1/\sqrt{\Omega}$ and $\lambda = n_{0}U_{0}$ where $n_{0} = N_{0}/\Omega$ and $\Omega$ is the volume of the system. The wave functions $\zeta_{i}({\bf r})$ which describe the noncondensate can be chosen to be plane waves $\zeta_{i}({\bf r}) = 1/\sqrt{\Omega}e^{\imath {\bf k_{i}}.{\bf r}}$ since these form a complete set orthogonal to the condensate. The coefficients ${\cal L}_{ij}$ and ${\cal M}_{ij}$ of Eq.~(\ref{LM_zeroT}) which define the BdG equations are therefore
\begin{equation}
\begin{array}{c@{\hspace{1cm}}c}
{\displaystyle {\cal L}_{ij} = \left [ \frac{\hbar^{2}k^{2}_{i}}{2m} + n_{0}U_{0} \right ] \delta_{ij}, } &
{\displaystyle {\cal M}_{ij} = n_{0}U_{0} \, \delta_{i(-j)}, }
\end{array}
\label{LM_homogeneous}
\end{equation}
where $k_{i} = |{\bf k}_{i}|$ and the label $-j$ refers to the state with wave vector ${\bf k_{-j}} = -{\bf k_{j}}$.

The quasiparticle transformation of Eq.~(\ref{qp}) reduces to
\be
\hat{\beta}_{k} = u_{k}\hat{a}_{k} - v_{k}\hat{a}^{\dagger}_{-k},
\ee
and the solution of the BdG equations gives
\begin{equation}
\begin{array}{c@{\hspace{1cm}}c}
{\displaystyle u_{k} = \frac{1}{\sqrt{1-\alpha^{2}_{k}}}\, , } & {\displaystyle v_{k} = \frac{-\alpha_{k}}{\sqrt{1-\alpha^{2}_{k}}}\, , }
\end{array}
\label{UV_homogeneous}
\end{equation}
where
\begin{equation}
\alpha_{k} = 1+y^{2}_{k}-y_{k}(2 + y^{2}_{k})^{1/2},
\label{alpha_yk}
\end{equation}
and the dimensionless wave vector $y_{k}$ is defined by 
\begin{equation}  
\begin{array}{ccr}
{\displaystyle y_{k} = \frac{k}{k_{0}}, }&\hspace{5mm}&  {\displaystyle \left(
\frac{\hbar^{2}k^{2}_{0}}{2m} \right) = n_{0}U_{0} \Rightarrow k^{2}_{0}
= 8 \pi a n_{0}. }\\
\end{array} 
\end{equation}
The quasiparticle energies are given by the usual Bogoliubov expression
\begin{equation}
\epsilon_{k} = \left [ (c\hbar k)^{2} + \left ( \frac{\hbar^{2} k^{2}}{2m} \right )^{2} \right ]^{1/2} \hspace{-3mm} = \,  n_{0}U_{0} \, y_{k}(2 + y^{2}_{k})^{1/2} \equiv n_{0}U_{0} \, z_{k},
\label{qpenergy_homogeneous}
\end{equation}
where $c = (n_{0}U_{0}/m)^{1/2}$ is the speed of sound, and the final term on the right hand side simply serves to define the dimensionless energy $z_{k}$.

The expression for the quasiparticle energies of Eq.~(\ref{qpenergy_homogeneous}) must be supplemented by an expression for the condensate density in terms of the total number of particles. This can be obtained by solving the nonlinear equation $n = n_{0}-\rho_{ex}$, where the noncondensate density $\rho_{ex}$ is obtained from the expression [c.f. Eq.~(\ref{rho_r})]
\begin{equation}
\rho_{ex}  =  \frac{k^{3}_{0}}{(2\pi)^{3}}\int \! d^{3}y_{k} \left \{ \frac{n_{k}+(n_{-k}+1)\,\alpha^{2}_{k}}{1-\alpha^{2}_{k}} \right \}.
\label{rho_homogeneous}
\end{equation}
For a dilute gas at zero temperature, the condensate population is therefore given to order $(na^{3})^{1/2}$ by the well-known result \cite{Bogoliubov}
\begin{equation}
n_{0}(T=0) = n\left [ 1-\frac{8}{3} \left ( \frac{na^{3}}{\pi} \right )^{1/2} \right ].
\label{zeroT_depletion}
\end{equation}

The renormalized anomalous average of Eq.~(\ref{kappa_renorm_r}) is given by the expression [c.f. Eq.~(\ref{kappa_r})]
\begin{equation}
\kappa^{\scriptscriptstyle R} =  -\frac{k^{3}_{0}}{(2\pi)^{3}} \int \! d^{3}y_{k} \left \{ \frac{(n_{k}+n_{-k}+1)\,\alpha_{k}}{1-\alpha^{2}_{k}} -\frac{1}{2y^{2}_{k}} \right \}
\label{kappaR_homogeneous}
\end{equation}
where the last term corresponds precisely to the ultra-violet renormalization of Eq.~(\ref{kappa_sp_r}). $\kappa^{\scriptscriptstyle R}$ is finite at any temperature and at $T=0$ we have the result $\kappa^{\scriptscriptstyle R} = 3 \rho_{ex} = k^{3}_{0}/(2\sqrt{2}\pi^{2})$.

\subsection{Higher Order Terms} \label{hardsphere_higher}

For calculations beyond quadratic order, the properties of the condensate are calculated from the generalized GPE of Eq.~(\ref{finiteT_GPE_renorm}) rather than the ordinary GPE. The condensate wave function is unchanged, but the eigenvalue becomes
\begin{equation}
\lambda_{G} = n_{0}U_{0} + 2U_{0}\rho_{ex} + U_{0}\kappa^{\scriptscriptstyle R}.
\label{lambda_G_homogeneous}
\end{equation}
Evaluating this at $T=0$, we obtain the well-known result \cite{Beliaev}
\begin{equation}
\lambda_{G}(T=0) = nU_{0} \Big [1 + \frac{32}{3\sqrt{\pi}} (na^{3})^{1/2} \Big ].
\end{equation}

The discussion of Sec.~\ref{C4} shows that the total contribution to the quasiparticle energy from non-quadratic terms can be written as
\begin{equation} 
\Delta E(k) = \Delta E_{3}(k) + \Delta E_{4}(k) + \Delta E_{\lambda}(k).
\label{E_shift_homogeneous}
\end{equation}
The energy shifts $\Delta E_{4}(k)$ and $\Delta E_{\lambda}(k)$ are defined in Eqs.~(\ref{delta_E4}) and (\ref{delta_Elambda}). Using these, together with Eq.~(\ref{lambda_G_homogeneous}), we obtain
\begin{equation}
\Delta E_{4}(k) + \Delta E_{\lambda}(k) =  -U_{0} \kappa^{\scriptscriptstyle R}\frac{(1+\alpha_{k})^{2}}{1-\alpha^{2}_{k}}\, .
\label{deltaE4_Elambda_homogeneous}
\end{equation}
This result demonstrates that in the homogeneous limit there is no shift in the quasiparticle excitation spectrum in the HFB-Popov approximation, where both $\Delta E_{3}(k)$ and $\kappa^{\scriptscriptstyle R}$ are neglected. This result is expected from our discussion of Sec.~\ref{HFB} where we showed that the energy shifts in the Popov approximation depend on the variation of the noncondensate density in the region of the condensate.

The energy shift $\Delta E_{3}(k)$ is defined in Eq.~(\ref{delta_E3}) and can be written in the homogeneous limit as 
\begin{equation}
\Delta E_{3}(k) = -\frac{U_{0}k^{3}_{0}}{4\pi^{3}(1-\alpha^{2}_{k})} \, \left [ \Delta
E^{a}_{3}(k) + \Delta E^{b}_{3}(k) + \Delta E^{c}_{3}(k) \right ],
\label{delta_E3_homogeneous}
\end{equation}
where
\begin{equation} 
\Delta E^{a}_{3}(k) =
 \int \! d^{3}y_{j}  \left \{
\frac{(1+n_{i}+n_{j}).[1-\alpha_{i}-\alpha_{j}+\alpha_{i}\alpha_{k}+\alpha_{j}\alpha_{k}-\alpha_{i}\alpha_{j}\alpha_{k}]^{2}}{(z_{i}+z_{j}-z_{k}).(1-\alpha^{2}_{i}).(1-\alpha^{2}_{j})} +  \frac{1}{(y^{2}_{k}-y^{2}_{i}-y^{2}_{j})} 
\right \},
\label{delta_E3a}
\end{equation} 
\begin{equation}
\Delta E^{b}_{3}(k) =
\int \! d^{3}y_{j}  \left \{
\frac{(1+n_{-i}+n_{-j}).[\alpha_{i}+\alpha_{j}+\alpha_{k}-\alpha_{i}\alpha_{j}-\alpha_{i}\alpha_{k}-\alpha_{j}\alpha_{k}]^{2}}{(z_{i}+z_{j}+z_{k}).(1-\alpha^{2}_{i}).(1-\alpha^{2}_{j})} + 
\frac{\alpha^{2}_{k}}{(y^{2}_{k}-y^{2}_{i}-y^{2}_{j})}  \right \},  
\label{delta_E3b} 
\end{equation} 
\begin{eqnarray} 
\Delta E^{c}_{3}(k) &=&
-2\int \! d^{3}y_{j} \left \{
\frac{(n_{i}-n_{-j}).[1-\alpha_{j}-\alpha_{k}+\alpha_{i}\alpha_{j}+\alpha_{i}\alpha_{k}-\alpha_{i}\alpha_{j}\alpha_{k}]^{2}}{(z_{i}-z_{j}-z_{k}).(1-\alpha^{2}_{i}).(1-\alpha^{2}_{j})}
\right \}, \label{delta_E3c} \\[5mm]
&& \hspace{6cm} \big (\mbox{{\bf i} } = \mbox{ {\bf k} } - \mbox{ {\bf
j}}\big ) \, . \nonumber 
\end{eqnarray}
The final terms of Eqs.~(\ref{delta_E3a}) and (\ref{delta_E3b}) correspond to the ultra-violet renormalization of $\Delta E_{3}(k)$ given in Eq.~(\ref{delta_E3_sp}). In the above equations a mixed notation of $\alpha$'s, $z$'s and $y$'s has been used in order to keep the expressions as compact as possible. The evaluation of the integrals is greatly simplified using the following relationships
between these quantities
\begin{equation}
\begin{array}{c@{\hspace{1cm}}c} 
{\displaystyle \alpha_{k} =
[1+z^{2}_{k}]^{1/2} - z_{k}, } &  {\displaystyle \alpha^{-1}_{k} =
[1+z^{2}_{k}]^{1/2} + z_{k}, }
\\ {\displaystyle 1-\alpha^{2}_{k} =
2z_{k}\alpha_{k},} & {\displaystyle (1-\alpha_{k})^{2} = 2y^{2}_{k}\alpha_{k}.}
\end{array}
\label{azy}
\end{equation}

The integrals of Eqs.~(\ref{delta_E3a})-(\ref{delta_E3c}) contain energy denominators which vanish when a decay process is energetically allowed. The integrals
must therefore be calculated by inserting a small imaginary part $\imath \epsilon$ and using the result 
\begin{equation} 
\lim_{\epsilon
\rightarrow 0} \, \frac{1}{x+\imath \epsilon} = {\cal P} \left ( \frac{1}{x} \right ) -\imath \pi
\delta(x),
\label{principal}
\end{equation} 
where ${\cal P}$ means `the principal part'. The first contribution describes the energy shifts of the excitations, while the imaginary part corresponds to their decay and describes the quasiparticle lifetimes. These two contributions are discussed in the following subsections.

\subsubsection{Energy shifts} 

The energy shifts at low momentum can be calculated by introducing an expansion in powers of $z_{k}$ [we could equally well use an expansion in powers of $y_{k}$ since for $y_{k} \ll 1$ we have $z_{k} = \sqrt{2}y_{k} + \mbox{O}(y^{3}_{k})$]. This expansion has the form
\begin{equation}
\Delta E_{3,4,\lambda}(k) = \frac{A}{z_{k}} + B + Cz_{k} + \mbox{O}(z^{2}_{k}).
\label{zk_expansion}
\end{equation}
The origin of
the term in $1/z_{k}$ is the prefactor $1/(1-\alpha^{2}_{k})$, which appears in
each of $\Delta E_{3}(k)$, $\Delta E_{4}(k)$ and $\Delta E_{\lambda}(k)$ [c.f. Eqs.~(\ref{deltaE4_Elambda_homogeneous}) and (\ref{delta_E3_homogeneous})] and which has the expansion
\begin{equation} 
\frac{1}{1-\alpha^{2}_{k}} = \frac{1}{2z_{k}} + \frac{1}{2} +
\frac{z_{k}}{4} + \mbox{O}(z^{3}_{k}).  
\end{equation}
The presence of terms
proportional to $1/z_{k}$ means that there is an infra-red divergence in both
$\Delta E_{3}(k)$ and $\Delta E_{4}(k) + \Delta E_{\lambda}(k)$. To prove that there are no infra-red divergences in the overall energy shift, we therefore need to show that when we add these quantities the resulting coefficient $A$ is zero. To show in addition that the theory is gapless we need to show that the sum of the $B$ coefficients is also zero. The total $C$ coefficient then describes the energy shift of the low-momentum states and corresponds to a modification of the speed of sound. This should be small if the use of perturbation theory on the non-quadratic Hamiltonian is justified.

The expansion of $\Delta E_{4}(k) + \Delta E_{\lambda}(k)$ can be obtained directly from Eq.~(\ref{deltaE4_Elambda_homogeneous}) and is
\begin{equation}
\Delta E_{4}(k) + \Delta E_{\lambda}(k) = -\frac{2U_{0}\kappa^{\scriptscriptstyle R}}{z_{k}} -\frac{U_{0} \kappa^{\scriptscriptstyle R}z_{k}}{2} + \mbox{O}(z^{3}_{k}).
\label{delta_E4lambda}
\end{equation}
This result demonstrates the presence of infra-red divergences in the full HFB theory (where $\Delta E_{3}(k)$ is neglected) which we mentioned earlier.
The calculation of $\Delta E_{3}(k)$ is more difficult and the details are given in Ref.~\cite{thesis}. The leading order contribution is simply obtained by setting ${\bf i} = {\bf j}$ in Eqs.~(\ref{delta_E3a})--(\ref{delta_E3c}). This gives
\begin{equation} 
\Delta E^{a}_{3}(k) = \Delta E^{b}_{3}(k) = \lim_{Y \rightarrow \infty}
-\frac{U_{0}k^{3}_{0}}{(2\pi)^{3}z_{k}} \left [ \int^{Y}_{0} \! \! \! d^{3} y_{j} \left(
\frac{1+n_{j}+n_{-j}}{2z_{j}} \right) - 2\pi Y \right],
\end{equation}
while $\Delta E^{c}_{3}(k) = 0$.
Comparison
with Eq.~(\ref{kappaR_homogeneous}) [using the fact that $\alpha_{k}/(1-\alpha^{2}_{k}) =
1/(2z_{k})$ from Eq.~(\ref{azy})] gives
\begin{equation} 
\Delta E_{3}(k) = \Delta E^{a}_{3}(k) + \Delta E^{b}_{3}(k) =
\frac{2U_{0}\kappa^{\scriptscriptstyle R}}{z_{k}} + \ldots
\label{delta_E3_expand}
\end{equation}

Equations~(\ref{delta_E4lambda}) and (\ref{delta_E3_expand}) show that the
infra-red divergence in $\Delta E_{3}(k)$ cancels exactly with that in $\Delta
E_{4}(k)+\Delta E_{\lambda}(k)$. We note that this occurs without the need to assume any particular functional form for the quasiparticle populations. Thus the infra-red divergences cancel for non-equilibrium quasiparticle distributions as well as for thermal distributions of any temperature. This general result is what is needed to show that the theory is well-defined, since the cancellation of divergences should be an intrinsic property of the theory and not of a particular distribution.

To show in addition that the theory is gapless
requires a calculation of $\Delta E_{3}(k)$ to order $(z_{k})^{0}$.
The result is that this term
vanishes, and since there is also no contribution at this order from $\Delta E_{4}(k)+\Delta E_{\lambda}(k)$ of Eq.~(\ref{delta_E4lambda}) this means that the theory is gapless. We have proved this result for any quasiparticle distribution which depends only on the modulus of the wave vector and which varies at worst as $n_{k} \sim 1/y_{k}$ as $y_{k} \rightarrow 0$. Since the low-momentum excitations are phonons with a linear dispersion relation, this includes the Planck distribution at finite temperature. The overall shift to the quasiparticle energies from the non-quadratic Hamiltonian is of order $z_{k}$, and the theory is therefore finite and gapless.

Thus the effect of the non-quadratic terms on the phonon spectrum is simply to modify the speed of sound. At zero temperature we have calculated the energy shift to be
\be
\Delta E(k,T=0) = \frac{7U_{0}k^{3}_{0}}{12\sqrt{2}\pi^{2}} \, z_{k} + \mbox{O}(z^{2}_{k}).
\label{shift0}
\ee
This shift is
small compared to the leading order expression $\epsilon_{k} = n_{0}U_{0}z_{k}$ (justifying the use of perturbation theory) if the dilute gas criterion
$na^{3} \ll 1$ is satisfied.
At high temperature ($k_{B}T/n_{0}U_{0} \gg 1$), the energy shift has been calculated by Fedichev and Shlyapnikov \cite{Fedichev_Shlyapnikov} and is
\begin{equation}
\Delta E(k,T) \approx -7 n_{0}U_{0}z_{k} \left ( \frac{k_{B}T}{n_{0}U_{0}} \right ) (n_{0}a^{3})^{1/2},
\label{soundT}
\end{equation}
which is small if $(k_{B}T/n_{0}U_{0}).(n_{0}a^{3})^{1/2} \ll 1$. This is therefore the condition for the validity of the theory in the high temperature regime and is discussed further in Sec.~\ref{validity}.

\subsubsection{Decay rates} \label{Widths}

$\Delta E_{3}(k)$ develops an imaginary part $\Delta E^{\mbox{{\scriptsize I}}}_{3}(k)$ whenever a real decay process is energetically allowed. The corresponding decay rate $\gamma(k) = -2 \Delta E^{\mbox{{\scriptsize I}}}_{3}(k)/\hbar$ can be calculated by replacing the energy denominators in Eqs.~(\ref{delta_E3a}) and (\ref{delta_E3c}) with $-\imath \pi \delta (E)$, where $\delta (E)$ is an energy conserving delta function [c.f. Eq.~(\ref{principal})]. This leads to the same results as the application of Fermi's Golden Rule to the cubic Hamiltonian for the case that the decaying mode has a large population.

Integrating over the energy conserving delta function, we obtain an expression for the decay rate in the form $\gamma(k) = \gamma_{B}(k) + \gamma_{L}(k)$, where the Beliaev decay rate $\gamma_{B}(k)$ is given by
\begin{eqnarray} 
\gamma_{B}(k)  &=& \frac{\hbar k^{3}_{0} a}{8my_{k}z_{k}\alpha_{k}}
\int^{z_{k}}_{0} \! \! dz_{j} \,
\frac{(1+n_{i}+n_{j}).[1-\alpha_{i}-\alpha_{j}+\alpha_{i}\alpha_{k}+\alpha_{j}\alpha_{k}-\alpha_{i}\alpha_{j}\alpha_{k}]^{2}}{\alpha_{i}\alpha_{j}(1+z^{2}_{i})^{1/2}(1+z^{2}_{j})^{1/2}}, \nonumber \\
&& \mbox{\hspace{7cm}} (z_{k} = z_{i}+z_{j}),
\label{Beliaev}
\end{eqnarray}
and the Landau decay rate $\gamma_{L}(k)$ is given by
\begin{eqnarray} 
\gamma_{L}(k) &=&  \frac{\hbar k^{3}_{0} a}{8my_{k}z_{k}\alpha_{k}} \int^{\infty}_{0} \! \! dz_{j} \,
\frac{(n_{j}-n_{i}).[1-\alpha_{j}-\alpha_{k}+\alpha_{i}\alpha_{k}+\alpha_{i}\alpha_{j}-\alpha_{i}\alpha_{j}\alpha_{k}]^{2}}{\alpha_{i}\alpha_{j}(1+z^{2}_{i})^{1/2}(1+z^{2}_{j})^{1/2}},
\nonumber \\
&& \mbox{\hspace{7cm}} (z_{k} = z_{i}-z_{j}).
\label{Landau}
\end{eqnarray}
Beliaev damping corresponds to processes in which the decaying quasiparticle decomposes into two of lower energy, while Landau damping is essentially the reverse process in which two quasiparticles collide and annihilate to form a third of higher energy.

Using Eqs.~(\ref{Beliaev}) and (\ref{Landau}) all the standard results for the decay rates of a homogeneous Bose gas can be obtained. For example, at $T=0$ the decay rate is entirely due to Beliaev processes and for the phonon part of the spectrum it is given by \cite{Beliaev,Lee_Yang,Mohling_Sirlin}
\begin{equation}
\begin{array}{c@{\hspace{1cm}}c}
{\displaystyle \gamma = \frac{3\hbar k^{5}}{320 \pi m n_{0}}\, , }& {\displaystyle (T=0\mbox{, }k/k_{0} \ll 1). }
\end{array}
\label{Beliaev0_low}
\end{equation}
At high temperatures Landau processes dominate and the phonon decay rate is given by \cite{Fedichev_Shlyapnikov,Shi,Giorgini,Pitaevskii_Landaudamping}
\begin{equation}
\gamma =  \frac{3 \pi^{3/2}}{2}.\frac{\epsilon_{k}}{\hbar}.\left ( \frac{k_{B}T}{n_{0}U_{0}} \right )(n_{0}a^{3})^{1/2}, \hspace{1cm} (k_{B}T/n_{0}U_{0} \gg 1\mbox{, }k/k_{0} \ll 1). 
\label{Landau_highT}
\end{equation}

\subsection{Physical Discussion} \label{shifts_physics}

We have shown that the inclusion of $\Delta E_{3}(k)$ leads to the cancellation of infra-red divergences and to a gapless spectrum. We will now discuss the physical reasons why these difficulties can arise in the theory of BEC and why individual terms in the perturbation theory contain divergences.

The excitation spectrum is calculated as the energy required
to create a quasiparticle. This involves a change in the
number of excited atoms which is $\Delta \langle \hat{N}_{ex} \rangle =
(1+\alpha^{2}_{k})/(1-\alpha^{2}_{k}) \sim \lim_{k \rightarrow 0} 1/(\sqrt{2}y_{k})$. A single low-energy quasiparticle therefore contains a very large number of particles and this is the physical origin of infra-red divergences in individual terms of the perturbation theory. A consequence of this is that care must be taken to ensure that the total number of particles is fixed so that the additional excited particles are taken from the condensate. This is the origin of the contribution from $\Delta E_{\lambda}(k)$ which is crucial in producing a well-defined theory.

The fact that low-energy quasiparticles consist of a large number of particles can also be seen as the reason why the low-momentum excitations of the system have the phonon dispersion relation $\epsilon_{k} = c \hbar k$. This involves a non-perturbative change in the excitation spectrum from the quadratic form appropriate to particles, and occurs because the energy of a quasiparticle corresponds roughly to the energy of a particle multiplied by the number of particles it contains. Since the latter quantity scales as $1/k$ as $k \rightarrow 0$ this gives $\epsilon_{k} \sim (\hbar^{2} k^{2}/2m ).(k_{0}/\sqrt{2}k) = c \hbar k/2$, which only differs from the actual value by a factor of two.

The mathematical origin of infra-red divergences is the fact that the quasiparticle
functions $u_{k}$ and $v_{k}$ scale as $1/k$ for $k \rightarrow 0$. The energy of a quasiparticle
is a quadratic form in the $u$'s and $v$'s and the only quadratic
combination which is finite as $k \rightarrow 0$ is
$u^{2}_{k}-v^{2}_{k} = 1$. Solving the BdG equations ensures that the quasiparticle energies depend only on this particular combination and hence are well-defined. If we modify these equations, however, and treat the changes via ordinary perturbation
theory, then we will get energy shifts which are still quadratic forms in the $u$'s and $v$'s but
which are not necessarily proportional to $u^{2}-v^{2}$ [c.f. Eq.~(\ref{deltaE4_Elambda_homogeneous})]. In this case there may be an infra-red divergence simply because of the low-momentum behaviour of the $u$'s and $v$'s and not because of the particular change to the BdG equations. Explicit calculations as in the previous section are required to show whether or not infra-red divergences cancel in physical quantities.

If we use self-consistent perturbation theory, however, and diagonalize the BdG equations exactly then these infra-red divergences will disappear, although the remnant of the problem will be seen in the energy spectrum having a gap at low energy (as in the HFB theory). Self-consistency therefore leads to finite energy shifts, but these will certainly be large compared with the quadratic theory results if the approach is not gapless. Any such theory is therefore inconsistent because a perturbative treatment of the non-quadratic Hamiltonian requires that the energy shifts are small compared to the leading order terms.

The appearance of infra-red divergences (and hence a gap in the excitation spectrum) can also be seen as a consequence of an inconsistent treatment of interactions involving the condensate. If $\Delta E_{3}(k)$ is neglected (as it is in the HFB theory), then the many-body T-matrix is introduced into pair excitation processes of the form $\Vs{i}{j}{0}{0}$ but not into the Hartree-Fock interactions $\Vs{0}{i}{j}{0}$. These terms are of the same order, however, as each involves two condensate labels and two excited state labels. The different treatment of these interactions in the HFB theory is therefore inconsistent and is responsible for the gap in the excitation spectrum. In contrast, the inclusion of $\Delta E_{3}$ leads to a consistent theory and a gapless spectrum. We showed in Sec.~\ref{C5}, however, that $\Delta E_{3}$ only introduces the T-matrix to second order in the Hartree-Fock terms, whereas self-consistent perturbation theory with the anomalous average gives the T-matrix to all orders in pair excitation processes. This means that the theory developed in this paper may not be gapless if self-consistent perturbation theory is used. Since we have proved that ordinary perturbation theory gives a gapless spectrum, however, the size of any gap will be beyond the order of this calculation. In contrast, the size of the gap in the HFB theory is proportional to $\kappa^{{\scriptscriptstyle R}}$ and hence is of the same order of magnitude as the calculation which is being performed.

The above argument also shows why the Popov approximation produces a gapless theory. There are essentially two consistent ways to treat condensate interactions; either we include the effects of the medium in all collisions or we neglect them completely. The first approach corresponds to the theory discussed in this paper and requires the inclusion of $\Delta E_{3}(k)$. The second approach corresponds to the Popov approximation in which both $\Delta E_{3}(k)$ and $\kappa^{{\scriptscriptstyle R}}$ are neglected. In this case all interactions are described by the two-body T-matrix (contact potential). The gapless nature of the Popov approximation is shown by Eq.~(\ref{delta_E4lambda}) which also demonstrates that there are no infra-red divergences in this approach. Thus the HFB-Popov theory is preferable to the full HFB approach and for this reason it has been used in recent calculations for trapped gases \cite{Hutchinson_Popov,Dodd_Popov}. Nonetheless, the neglect of many-body effects can only be justified at near zero temperature and the full second order theory of this paper will be needed whenever there is significant depletion of the condensate.

\section{Validity} \label{validity}

The principle assumption we have made in this paper is that the existence of a condensate means that the non-quadratic terms in the Hamiltonian can be treated using perturbation theory. We argued in
Sec.~\ref{C5}, however, that this can only be the case if interactions are described by the two-body T-matrix rather than the interaction potential. The discussion in this section will assume that this has already been done.

The use of perturbation theory requires that the effect of non-quadratic terms is small, and this can be verified after any calculation. A necessary condition is that the total energy shifts do not contain infra-red divergences and vanish at low momentum. We have shown in Sec.~\ref{C6} that this is indeed the case and the validity of the theory therefore depends on the size of the resulting shifts relative to the leading order terms. A comparison of Eqs.~(\ref{qpenergy_homogeneous}), (\ref{shift0}) and (\ref{soundT}) shows that the validity of the theory in the homogeneous limit can be summarized by
\begin{eqnarray}
(na^{3})^{1/2} \ll 1,  &\hspace{1cm}& (T = 0),  \label{validity0} \\[2mm]
\left ( \frac{k_{B}T}{n_{0}U_{0}} \right )(n_{0}a^{3})^{1/2} \ll 1, &\hspace{1cm}& (k_{B}T/n_{0}U_{0} \gg 1). \label{validityT}
\end{eqnarray}
These same conditions are obtained by requiring that the quasiparticles are well-defined (i.e. that their decay rates are small compared to their frequencies) as can be seen from Eqs.~(\ref{Beliaev0_low}) and (\ref{Landau_highT}).

The condition of Eq.~(\ref{validity0}) for the validity of the theory at zero temperature is simply the criterion for a dilute gas. The finite temperature condition of Eq.~(\ref{validityT}) is closely related to the Ginzburg criterion for the validity of mean-field theory \cite{Huang}. The fact that this expression scales with the condensate density as $(n_{0})^{-1/2}$ means that the theory must fail in the region of the critical point where $n_{0} \rightarrow 0$. The boundary of the critical region can be estimated by setting $(k_{B}T/n_{0}U_{0})(n_{0}a^{3})^{1/2} \sim 1$. If we write this in terms of the de~Broglie wave length of the atoms and use the noninteracting gas result for the condensate density near the critical point, then it becomes $t^{1/2} \sim a^{1/2}/(n^{1/2}\lambda^{2}_{dB})$, where $t$ is the reduced temperature defined by $t = (T_{c}-T)/T_{c}$ for $T < T_{c}$. Since $n\lambda^{3}_{dB} \sim 1$ near the critical point, this can be simplified to $t \sim a/\lambda_{dB}$. The critical region is therefore very small for a dilute gas for which $a/\lambda_{dB} \ll 1$. We note that the failure of the theory at the critical point is a result of the growth of long wavelength fluctuations which destroy the mean field of the condensate. This `infra-red' behaviour produces a genuine breakdown in perturbation theory and leads to the occurrence of critical phenomena. It is completely distinct, however, from the problem of infra-red divergences in individual terms of the perturbative expressions with which we have been concerned in this paper (and which we have shown cancel in physical quantities). 

Equations~(\ref{validity0}) and (\ref{validityT}) show that the expansion parameter in the homogeneous, thermodynamic limit is proportional to $a^{3/2}$ at $T=0$ and $a^{1/2}$ at finite temperature. These are non-analytic functions of $a$ which means that the perturbation expansion is at best only asymptotically convergent. This is demonstrated both by the presence of a logarithmic term in the expression for the ground state energy of Eq.~(\ref{gs_energy}), and by the fact that a condensate can not exist with a negative scattering length \cite{Lifshitz_Pitaevskii}. In a trap, however, the finite size of the system means that a condensate can exist in the negative scattering length case, provided that its population is not too large \cite{Ruprecht,Dodd_negative}. This leads to the possibility that the expansion is convergent in a trap, at least for small condensate populations.

This result indicates that the trap potential can have a profound effect on the nature of BEC and the validity of mean-field theory. In particular, it is possible that the theory developed in this paper can be meaningfully applied to a trapped gas even in the region of the phase transition, at least for some range of values of the {\em s}-wave scattering length. The reason is that the finite length scale in a trap suppresses the long wavelength fluctuations which
are responsible for destroying the condensate mean field. One obvious consequence of this is that BEC is possible in one and two dimensions in a trap, whereas in the homogeneous, thermodynamic limit it can not occur in fewer than three dimensions.

A criterion for the validity of the theory which can be applied to a trapped gas is provided by the anomalous average $\kappa_{ij}$. This is an important quantity because it vanishes in the absence of interactions and is only large when there is a significant noncondensate fraction. The anomalous average introduces many-body effects into particle collisions and the requirement that these are small compared to two-body effects can be written as
\begin{equation}
\frac{\sum_{ij \neq 0}
 \bra{0}{0} T_{\mbox{\scriptsize 2b}} \ket{i}{j}\kappa^{\scriptscriptstyle R}_{ji} }{N_{0}\bra{0}{0} T_{\mbox{\scriptsize 2b}} \ket{0}{0}} \ll 1.
\end{equation}
In the homogeneous limit this becomes $\kappa^{\scriptscriptstyle R}/n_{0} \ll 1$ and leads directly to the conditions of Eqs.~(\ref{validity0}) and (\ref{validityT}). The fact that the finite temperature criterion can not be satisfied close to the phase transition can be seen from the fact that the many-body T-matrix vanishes at the critical point for a homogeneous gas \cite{Shi,Bijlsma_Stoof}. In this case the effects of the medium on the collision of two particles are as large as the two-body effects and the perturbative expansion breaks down. In a trap, however, the many-body T-matrix remains finite in the region of the phase transition and only decreases by of order $10 \%$ \cite{Proukakis_gapless,Hutchinson_private}. This is a further indication that the theory may remain valid through the phase transition in a trap.

Another simple test of the theory in a harmonic trap is
provided by the calculation of the Kohn modes \cite{Dobson}. These are excitations which consist of a centre of mass oscillation of all the atoms about the trap minimum, and they are exact solutions
of the equations of motion for an interacting system. We therefore expect there to be solutions of the BdG equations with frequencies of exactly
one in the trap units appropriate to the axis of oscillation.

The Kohn modes are recovered exactly in the quadratic
theory because this includes the full dynamics of the condensate. They are not obtained exactly in higher order theories, however, because the motion of the thermal
cloud is treated approximately. This can be seen from the fact that first
order perturbation theory is used on the quartic Hamiltonian $\hat{H}_{4}$, corresponding to the assumption that the noncondensate is static. Some of the dynamics of the
thermal cloud are included in this paper, however, because second order perturbation theory is used on the cubic Hamiltonian $\hat{H}_{3}$. Nonetheless we do
not expect to recover the Kohn modes exactly. This fact can be turned into a useful estimate of the error in the theory, however, since our method is based on a systematic treatment of the original Hamiltonian, so the degree to which an exact result is violated provides a measure of the importance of higher order
terms.

Finally, we mention that although our discussion of validity has focused on the regime below the critical temperature, the leading order predictions of the theory above $T_{c}$ should also be correct. The reason is that in the zero-condensate limit the theory reduces to the Hartree-Fock approximation for the noncondensate, which should be reasonably accurate for a dilute gas above the critical point. In addition, the leading order properties of the system at high temperature are simply those of a noninteracting gas, and this contribution is treated exactly in the quadratic Hamiltonian. For this reason numerical simulations for trapped gases reproduce the noninteracting gas results for $T \gg T_{c}$ \cite{Hutchinson_Popov,Dodd_Popov,Proukakis_gapless}.

\section{Gapless HFB} \label{C7}

In this section we describe and motivate a gapless extension of the conventional HFB theory which has recently been developed \cite{Proukakis_gapless,Hutchinson_gapless}. This theory has the advantage that it is significantly easier to compute with than the full second order theory of this paper, and it can therefore be used as a first approximation when many-body effects are not negligible (so that the HFB-Popov approximation can not be applied).

In Sec.~\ref{C6} we argued that the physical reason for the appearance of a gap in the HFB theory is the inconsistent treatment of interactions involving the condensate. In particular, the anomalous average introduces the many-body T-matrix into the off-diagonal terms of the BdG equations, but if $\Delta E_{3}$ is neglected then the diagonal terms only contain the two-body T-matrix. A simple way to solve this problem is therefore to insert the many-body T-matrix `by hand' into the diagonal terms by replacing the matrix elements $\bra{0}{i} T_{\mbox{\scriptsize 2b}} \ket{j}{0}$ with $\bra{0}{i} T_{\mbox{\scriptsize mb}} \ket{j}{0}$. This is the basis of the gapless HFB theory and it is also equivalent to the effective Hamiltonian approach of Bijlsma and Stoof \cite{Bijlsma_Stoof}.

If the two-body T-matrix is approximated by a contact potential then the low-energy limit of the many-body T-matrix will also have this form. The diagonal matrix elements can therefore be written as
\begin{equation}
\bra{0}{i} T_{\mbox{\scriptsize mb}} \ket{j}{0} = \int \! \! d^{3}{\bf r} \, \zeta^{*}_{i}({\bf r}) \zeta_{j}({\bf r}) T_{\mbox{\scriptsize mb}}({\bf r})|\zeta_{0}({\bf r})|^{2},
\label{Tmany_diag_offdiag}
\end{equation}
where we have denoted the T-matrix by $T_{\mbox{\scriptsize mb}}({\bf r})$, since many-body effects are spatially dependent for a trapped gas.
The interaction strength $T_{\mbox{\scriptsize mb}}({\bf r})$ can be found from consideration of the off-diagonal matrix elements\footnote{We are assuming that the dependence of the T-matrix on the energy of the collision (the parameter $z$) can be neglected at low-energy. In reality the many-body T-matrix does have a momentum dependence and at high-energy it reduces to the two-body T-matrix. This ultimately leads to two versions of the gapless HFB theory as discussed later in this section.} $\bra{i}{j} T_{\mbox{\scriptsize mb}} \ket{0}{0}$. Writing Eq.~(\ref{T_many_T2}) in the position representation and using Eqs.~(\ref{kappa_solution}), (\ref{kappa_renorm}) and (\ref{kappa_sp}) we have\footnote{We note that $\kappa^{\scriptscriptstyle R}$ is defined here using $T_{\mbox{\scriptsize mb}}$ rather than $T_{\mbox{\scriptsize 2b}}$ in Eq.~(\ref{kappa_sp}). This definition is more appropriate to self-consistent perturbation theory, but the two are equivalent to the order of this paper.}
\begin{eqnarray}
N_{0} \int \! \! d^{3}{\bf r} \, \zeta^{*}_{i}({\bf r}) \zeta^{*}_{j}({\bf r}) T_{\mbox{\scriptsize mb}}({\bf r})\zeta^{2}_{0}({\bf r}) \! &=& \! \int \! \! d^{3}{\bf r} \, \zeta^{*}_{i}({\bf r}) \zeta^{*}_{j}({\bf r}) \, \left [ N_{0}U_{0}\zeta^{2}_{0}({\bf r}) + U_{0}\kappa^{\scriptscriptstyle R}({\bf r}) \right ] \hspace*{10mm} \label{Tmb_r_offdiag} \\[2mm]
&=& \! N_{0} \int \! \! d^{3}{\bf r} \, \zeta^{*}_{i}({\bf r}) \zeta^{*}_{j}({\bf r}) \, U_{0} \left [ 1 + \frac{\kappa^{\scriptscriptstyle R}({\bf r})}{N_{0} \zeta^{2}_{0}({\bf r})} \right ]\zeta^{2}_{0}({\bf r}). \nonumber 
\end{eqnarray}
We can therefore obtain a gapless HFB theory by starting with the generalized HFB approach (i.e. $\kappa^{\scriptscriptstyle R}({\bf r}) \neq 0$, $\Delta E_{3}=0$) and introducing the many-body T-matrix to all orders in the diagonal elements of the BdG equations by replacing $U_{0}$ with $T_{\mbox{\scriptsize mb}}({\bf r})$, defined by
\begin{equation}
T_{\mbox{\scriptsize mb}}({\bf r}) = U_{0} \left [ 1 + \frac{\kappa^{\scriptscriptstyle R}({\bf r})}{N_{0} \zeta^{2}_{0}({\bf r})} \right ].
\label{Tmb_r}
\end{equation}
The result of Eq.~(\ref{kappa_real}) shows that $T_{\mbox{\scriptsize mb}}({\bf r})$ is real and therefore the quasiparticles do not decay in this approach.

The discussion to this point has focused on the condensate-excited state collisions $\bra{0}{i} T_{\mbox{\scriptsize mb}} \ket{j}{0}$ which are the leading order terms in the diagonal elements of the BdG equations. The introduction of the many-body T-matrix in these interactions is mandatory if we wish to have a gapless theory. However, there are further interaction terms in the BdG and GPE equations which only contain the two-body T-matrix at the order of this calculation. In particular, collisions of noncondensate atoms are described in the BdG equations by the term $\sum_{km \neq 0} 2\bra{k}{i} T_{\mbox{{\scriptsize 2b}}} \ket{j}{m}\rho_{mk}$ which appears in Eq.~(\ref{L_T2}), while the effect of condensate-excited state collisions on the condensate is described by the expression $\sum_{ij \neq 0} 2\bra{k}{i} T_{\mbox{\scriptsize 2b}} \ket{j}{0}\rho_{ji}$ which appears in the generalized GPE of Eq.~(\ref{finiteT_GPE_renorm}).\footnote{The interaction strength which appears in terms involving the anomalous average must remain written in terms of $T_{\mbox{\scriptsize 2b}}$, since this is what is required to introduce the many-body T-matrix into condensate-condensate collisions [c.f. Eq.~(\ref{Tmb_r_offdiag})].} The theory remains gapless if these interactions are also upgraded to the many-body T-matrix and we therefore obtain two versions of the theory depending on the treatment of these terms. In the first approach (GHFB1) these collisions remain described by the two-body T-matrix $U_{0}$, while in the second approach (GHFB2) many-body effects are introduced by replacing $U_{0}$ with $T_{\mbox{\scriptsize mb}}({\bf r})$.

The two gapless HFB theories can therefore be summarized by the following equations\footnote{To obtain excitations orthogonal to the condensate we should really write these equations in the form of Eq.~(\ref{BdGposition}) with $U_{0} \rightarrow T_{\mbox{\scriptsize mb}}({\bf r})$ in Eq.~(\ref{cj_orthogonality}). Since the theory is gapless, however, we can simply apply the method of Eq.~(\ref{ortho_subtract}).}
\begin{equation}
- \frac{\hbar^2}{2m}\nabla^{2}\zeta_{0}({\bf r}) + V_{\mbox{\scriptsize Trap}}({\bf r}) +  N_{0}T_{\mbox{\scriptsize mb}}({\bf r}) |\zeta_{0}({\bf r})|^{2}\zeta_{0}({\bf r}) + 2 U_{\mbox{\scriptsize ex}}({\bf r}) \rho_{ex}({\bf r})\zeta_{0}({\bf r}) = \lambda_{G} \zeta_{0}({\bf r}),
\label{GHFB0}
\end{equation}
\begin{equation}
\begin{array}{lcr@{\hspace{-.1mm}}l}
{\displaystyle {\cal L}({\bf r})u_{j}({\bf r}) + N_{0}T_{\mbox{\scriptsize mb}}({\bf r}) \zeta^{2}_{0}({\bf r})v_{j}({\bf r}) } &=& {\displaystyle \epsilon_{j}u_{j}({\bf r}), } & \nonumber \\[2mm]
{\displaystyle {\cal L}({\bf r})v_{j}({\bf r}) + N_{0}T_{\mbox{\scriptsize mb}}({\bf r})  \zeta^{*2}_{0}({\bf r})u_{j}({\bf r}) } &=& {\displaystyle  -\epsilon_{j}v_{j}({\bf r})} &,
\end{array}
\label{GHFB_position}
\end{equation}
where ${\cal L}({\bf r})$ is defined by
\begin{equation}
{\cal L} ({\bf r}) = - \frac{\hbar^2}{2m}\nabla^{2} + V_{\mbox{\scriptsize Trap}}({\bf r}) - \lambda_{G} + 2N_{0}T_{\mbox{\scriptsize mb}}({\bf r}) |\zeta_{0}({\bf r})|^{2} + 2 U_{\mbox{\scriptsize ex}}({\bf r}) \rho_{ex}({\bf r}),
\end{equation}
and we have used the fact that $T_{\mbox{\scriptsize mb}}({\bf r})$ is real.
The GHFB1 theory corresponds to setting $U_{\mbox{\scriptsize ex}}({\bf r}) = U_{0}$ while the GHFB2 theory is defined by $U_{\mbox{\scriptsize ex}}({\bf r}) = T_{\mbox{\scriptsize mb}}({\bf r})$. We note that there may be numerical difficulties in the implementation of the GHFB2 theory because the BdG equations contain the term $2U_{0} \left [ 1 \hspace{-2.5pt} + \hspace{-2.5pt} \kappa^{\scriptscriptstyle R}({\bf r})/N_{0} \, \zeta^{2}_{0}({\bf r}) \right ] \rho_{ex}({\bf r})$, which may not be well-defined at the edge of the condensate. There is no such problem in the GHFB1 theory because in that case $T_{\mbox{\scriptsize mb}}({\bf r})$ always appears multiplied by $\zeta^{2}_{0}({\bf r})$.

The additional many-body effects introduced in the GHFB2 theory can be justified on the physical grounds that all low-energy collisions should be described by the same effective interaction. Nonetheless, the many-body T-matrix depends on the relative momentum of the colliding particles and $T_{\mbox{\scriptsize mb}}({\bf r})$ is only a good approximation at low energy. The GHFB2 theory is therefore only appropriate if most of the collisions involving the noncondensate occur at low energy. If the dominant collisions occur at an energy which is large compared to the mean-field interaction, then the GHFB1 theory should be used because in this limit the many-body T-matrix reduces to the two-body T-matrix. The two theories therefore correspond to the two extreme assumptions about the energy regime of noncondensate collisions.

We will now discuss the relationship between the gapless HFB approach and the full second order theory of this paper. The essential feature of the GHFB theories is the introduction of the many-body T-matrix into condensate-excited state collisions in the diagonal terms of the BdG equations. We showed in Sec.~\ref{C5} that such T-matrix corrections are introduced by $\Delta E_{3}$, albeit only to second order. $\Delta E_{3}$ also introduces polarization effects, however, and these are not contained in the many-body T-matrix. The GHFB1 theory can therefore be seen as an approximation to the full theory developed in this paper which neglects these polarization terms, but includes higher order contributions to the T-matrix. The additional many-body effects in the GHFB2 theory are also of higher order and will be introduced in a full theory by higher order perturbation theory as discussed in Ref.~\cite{thesis}. The higher order terms in both the GHFB theories should not have a significant effect, however, and if they are important then a systematic calculation beyond the order of this paper will be required. The gapless HFB approach is only useful therefore if the dominant contribution from $\Delta E_{3}$ corresponds to many-body T-matrix corrections, and if terms beyond the order of this calculation can be neglected.

The quasiparticle energies predicted by the gapless HFB theories can easily be calculated in the homogeneous limit. In this case, the total energy shift from non-quadratic terms can be written as
\begin{equation} 
\Delta E(k) = \Delta E_{\mbox{\scriptsize GHFB}}(k) + \Delta E_{4}(k) + \Delta E_{\lambda}(k),
\label{E_shift_gapless}
\end{equation}
where $\Delta E_{4}(k) + \Delta E_{\lambda}(k)$ is given in Eq.~(\ref{deltaE4_Elambda_homogeneous}) and the contribution from the terms introduced in the GHFB theories is
\begin{equation}
\Delta E_{\mbox{\scriptsize GHFB}}(k) = 2U_{0}\kappa^{\scriptscriptstyle R}\frac{1+\alpha^{2}_{k}}{1-\alpha^{2}_{k}}\, ,
\label{delta_EGHFB}
\end{equation}
with $\alpha_{k}$ and $\kappa^{\scriptscriptstyle R}$ defined in Eqs.~(\ref{alpha_yk}) and (\ref{kappaR_homogeneous}) respectively. We note that the two gapless HFB theories give exactly the same energy shifts in the homogeneous limit because the additional many-body effects introduced in GHFB2 cancel with each other.

Substituting Eqs.~(\ref{deltaE4_Elambda_homogeneous}) and (\ref{delta_EGHFB}) into Eq.~(\ref{E_shift_gapless}) gives
\begin{equation}
\Delta E(k) = U_{0}\kappa^{\scriptscriptstyle R}\frac{(1-\alpha_{k})^{2}}{1-\alpha^{2}_{k}}.
\end{equation}
In the low-momentum limit $k \rightarrow 0$ this becomes
\begin{equation}
\Delta E(k) =  \frac{U_{0}\kappa^{\scriptscriptstyle R} \, z_{k}}{2} + \mbox{O}(z^{3}_{k}).
\end{equation}
Evaluating $\kappa^{\scriptscriptstyle R}$ using Eq.~(\ref{kappaR_homogeneous}), we find that the zero and finite-temperature energy shifts for the phonon spectrum ($k \ll k_{0}$) are
\be
\Delta E(k,T=0) = \frac{U_{0}k^{3}_{0}}{4\sqrt{2}\pi^{2}} \, z_{k},
\ee
\be
\Delta E(k,T) = -2\pi^{1/2}n_{0}U_{0}z_{k}\left ( \frac{k_{B}T}{n_{0}U_{0}} \right )(n_{0}a^{3})^{1/2}, \hspace{1cm} (k_{B}T/n_{0}U_{0} \gg 1). 
\ee
Comparison with Eqs.~(\ref{shift0}) and (\ref{soundT}) shows that these results agree with those from the full theory to within factors of order unity.

For trapped gases, the gapless HFB theories have been implemented numerically in both one and three dimensions \cite{Proukakis_gapless,Hutchinson_gapless}. In one dimension, the predictions of GHFB1 for the quasiparticle energies did not differ significantly from those of the HFB-Popov theory, although there was a noticeable difference in GHFB2 \cite{Proukakis_gapless}. This is somewhat surprising in the light of the homogeneous limit results (where all the effects occur in the GHFB1 theory), and the fact that the additional many-body corrections introduced in GHFB2 are expected to be of higher order. In three dimensions, the GHFB2 theory gives good agreement with the experimental results for one of the low-lying excitations at temperatures approaching $T_{c}$, although the agreement for another is worse than the HFB-Popov predictions \cite{Hutchinson_gapless}.

\section{Conclusions}

In this paper we have developed a gapless theory of BEC which goes beyond the quadratic theory of Bogoliubov in a consistent manner, and which can be applied to trapped gases as well as to the homogeneous limit. The theory is based on rewriting the many-body Hamiltonian in a form which is approximately quadratic, and using first and second order perturbation theory to treat the non-quadratic terms. Infra-divergences can appear in individual terms of the perturbation expansion, but the total contribution is finite and the change in the excitation spectrum is small for a dilute gas away from the critical point, justifying the use of perturbation theory. The problem of ultra-violet divergences is dealt with by using the contact potential as an approximation to the two-body T-matrix rather than the bare interaction potential. This leads naturally to a renormalization of the theory at high energy.

The numerical implementation of the theory for trapped gases is currently in progress \cite{Rusch2}, and it is hoped that this will lead to good agreement with experimental results for the energies and lifetimes of the excitations at temperatures approaching the phase transition. The theory can also be applied to the study of charged Bose gases and neutral atoms in two dimensions, and work along these lines is also in progress \cite{Lee}.

A further area of interest for future research is the study of the properties of BEC near the phase transition. We showed in Sec.~\ref{validity} that the theory developed in this paper fails in the homogeneous limit near the critical point, but may remain valid in this region for a trapped gas. Whether or not this is the case can be established using the requirement that the perturbative shifts predicted by the theory are small. The signature for the failure of the theory is the appearance of large shifts for low-energy states, and this should occur as the trap is opened and the system behaves in a quasi-homogeneous manner.

The failure of perturbation theory is associated with the emergence of critical phenomena, and is a consequence of the growth of long wavelength fluctuations on the condensate. This can only be studied using an approach which is free of the spurious infra-red divergences which arise from an inconsistent treatment of the non-quadratic Hamiltonian. It is therefore important to apply the theory developed in this paper to the region of the phase transition and to study the cross-over between mean-field and critical behaviour for a Bose gas.

\section{Acknowledgements}

I am especially grateful to Professor Keith Burnett for his supervision of this work and his many invaluable contributions to it. I would also like to thank Professor Crispin Gardiner for a number of useful discussions, and Matthew Davis for proof-reading an early version of the manuscript. I thank the UK EPSRC and St. John's College, Oxford for financial assistance.

\appendix

\section{Effective Low-Energy Hamiltonian} \label{Tupper_app}

In this appendix we will show for a Bose gas at low temperature that the high-energy states of the system can be adiabatically eliminated to give an effective Hamiltonian for the low-energy states. The interactions between particles in the low-energy subspace are described by an approximation to the two-body T-matrix.

We start from the many-body Hamiltonian for a system of structureless bosons with pairwise interactions [c.f. Eq.~(\ref{Hamiltonian})]
\begin{equation}
\hat{H} = \sum_{i} \hbar \omega_{i} \hat{a}^{\dagger}_{i}\hat{a}_{i} + \frac{1}{2} \sum_{ijkm}  \bra{i}{j} V \ket{k}{m}  \hat{a}^{\dagger}_{i}\hat{a}^{\dagger}_{j}\hat{a}_{k}\hat{a}_{m},
\end{equation}
where for convenience we have used the basis in which $\hat{H}_{sp}$ is diagonal,\footnote{We will subsequently make a unitary transformation to the basis orthogonal to the condensate which is used in the main text.} so $H^{sp}_{ij} = \hbar \omega_{i}\delta_{ij}$ and the index $i$ refers to the wave function $\zeta^{sp}_{i}({\bf r})$. The operators $\hat{a}^{\dagger}_{i}$ and $\hat{a}_{i}$ are respectively the creation and annihilation operators for the eigenstates of $\hat{H}^{sp}$ and have the usual Bose commutation relations
\begin{equation}
\begin{array}{ccc}
\left[ \hat{a_i}, \hat{a_j}^{\dagger} \right] = \delta_{ij},
&&\left[ \hat{a_i}, \hat{a_j} \right] = \left[ \hat{a_i}^{\dagger}, \hat{a_j}^{\dagger} \right] = 0.
\end{array}
\end{equation}
The equation of motion for any operator $\hat{O}$ is given by the Heisenberg formula
\begin{equation}
\imath \hbar \frac{d \hat{O}}{d t} = \left[\hat{O}, \hat{H} \right].
\label{Heisenbergmotion}
\end{equation}
Using this we obtain the equation of motion for the annihilation operator
\begin{equation}
\imath \hbar \frac{d \bar{a}_{i} }{dt} = \sum_{jkm}  \Vs{i}{j}{k}{m} \bar{a}^{\dagger}_{j}\bar{a}_{k}\bar{a}_{m}e^{\imath \omega_{ijkm} t},
\label{motion1abar}
\end{equation}
where $\omega_{ijkm} = \omega_{i} + \omega_{j} - \omega_{k} -\omega_{m}$ and we have eliminated the free evolution by defining $\bar{a}_{i} \equiv \hat{a}_{i} e^{\imath \omega_{i} t}$. Eq.~(\ref{motion1abar}) depends on products of three operators so we will also need the equation of motion for these, which is
\begin{eqnarray}
\imath \hbar \frac{d \, ( \bar{a}^{\dagger}_{j}\bar{a}_{k}\bar{a}_{m} ) }{dt} &=&  \sum_{pq}  \Vs{k}{m}{p}{q} \bar{a}^{\dagger}_{j}\bar{a}_{p}\bar{a}_{q}e^{\imath \omega_{kmpq} t} + \sum_{pqr} \Vs{p}{k}{q}{r} \bar{a}^{\dagger}_{j}\bar{a}^{\dagger}_{p}\bar{a}_{m}\bar{a}_{q}\bar{a}_{r}e^{\imath \omega_{pkqr} t}  \label{motion3abar} \\
&+& \sum_{pqr} \Vs{p}{m}{q}{r} \bar{a}^{\dagger}_{j}\bar{a}^{\dagger}_{p}\bar{a}_{k}\bar{a}_{q}\bar{a}_{r}e^{\imath \omega_{pmqr} t} - \sum_{pqr} \Vs{p}{q}{j}{r} \bar{a}^{\dagger}_{p}\bar{a}^{\dagger}_{q}\bar{a}_{r}\bar{a}_{k}\bar{a}_{m}e^{\imath \omega_{pqrj} t}. \nonumber 
\end{eqnarray}

We proceed by dividing the system into two subspaces, namely low (L) and high-energy states (H). The high-energy states evolve on a time scale which is short compared to the characteristic evolution at low energy. We can therefore adiabatically eliminate operators which act in the high-energy subspace by expressing them in terms of operators which act only on low-energy states. This results in the replacement of the bare interaction potential between particles in the low-energy subspace with an effective interaction which has the form of a two-body T-matrix.

In order to prove this result we need to integrate the equation of motion for the product of three operators using Eq.~(\ref{motion3abar}) and substitute the result into Eq.~(\ref{motion1abar}). Of course this is not possible in general because we do not have a closed set of equations. Eq.~(\ref{motion3abar}) depends on the equation of motion for the product of five operators which in turn depends on products of seven operators and so on ad infinitum. This infinite set of equations can be truncated to just Eqs.~(\ref{motion1abar}) and (\ref{motion3abar}), however, on the basis of two assumptions. Since we are interested in the properties of Bose gases at very low temperatures, we assume that all the particles are confined to the low-energy subspace. The high-energy states are only occupied in a virtual sense, as the intermediate states in collisions of low-energy particles. The dominant terms in any equation of motion therefore have low indices on all operators. We also assume that momentum conservation allows us to neglect all terms involving  matrix elements where only one index is high (in a trap such matrix elements will be non-zero but should still be negligible).

These assumptions, and the fact that we are only interested in the case that $i$ is a low index, mean that Eqs.~(\ref{motion1abar}) and (\ref{motion3abar}) can be reduced to
\begin{eqnarray}
\imath \hbar \frac{d \bar{a}_{i} }{dt} &=&  \sum^{L}_{jkm} \Vs{i}{j}{k}{m} \bar{a}^{\dagger}_{j}\bar{a}_{k}\bar{a}_{m}e^{\imath \omega_{ijkm} t}  +  \sum^{L}_{j}\sum^{H}_{km} \Vs{i}{j}{k}{m} \bar{a}^{\dagger}_{j}\bar{a}_{k}\bar{a}_{m}e^{\imath \omega_{ijkm} t}, \label{motion1abar2} \\
\imath \hbar \frac{d \, (\bar{a}^{\dagger}_{j}\bar{a}_{k}\bar{a}_{m} ) }{dt} &=&  \sum^{L}_{pq}  \Vs{k}{m}{p}{q} \bar{a}^{\dagger}_{j}\bar{a}_{p}\bar{a}_{q}e^{\imath \omega_{kmpq} t} + \sum^{H}_{pq}  \Vs{k}{m}{p}{q} \bar{a}^{\dagger}_{j}\bar{a}_{p}\bar{a}_{q}e^{\imath \omega_{kmpq} t}, \label{partmotion3a} \\
\mbox{($j$ Low, $k,m$ High)} && \nonumber
\end{eqnarray}
where $\sum^{L}_{pq}$ means that both the indices $p$ and $q$ are low while $\sum^{H}_{pq}$ means that both are high. In these equations, the first term on the right hand side gives the dominant contribution since it contains operators which act on populated states. The second term in Eq.~(\ref{partmotion3a}) contains operators of the same form as those on the left hand side, so this equation has a Lippmann-Schwinger form and its solution introduces a T-matrix.

We will solve Eq.~(\ref{partmotion3a}) using an approach based on Fourier transforms. We define the Fourier transform of $\bar{a}^{\dagger}_{j}\bar{a}_{k}\bar{a}_{m}$ by
\begin{eqnarray}
\hat{f}_{jkm}(\omega) &=& \frac{1}{2\pi} \int^{+\infty}_{-\infty} \! \! \! dt \; \bar{a}^{\dagger}_{j}\bar{a}_{k}\bar{a}_{m}(t) \, e^{\imath \omega t}, \\
\bar{a}^{\dagger}_{j}\bar{a}_{k}\bar{a}_{m}(t) &=& \int^{+\infty}_{-\infty} \! \! \! d\omega \; \hat{f}_{jkm}(\omega) \, e^{-\imath \omega t}.
\end{eqnarray}
Using these definitions in Eq.~(\ref{partmotion3a}) and initially considering only the first term on the right hand side leads to the result 
\begin{equation}
\bar{a}^{\dagger}_{j}\bar{a}_{k}\bar{a}_{m} = \sum^{L}_{pq} \Vs{k}{m}{p}{q} e^{\imath \omega_{kmpq} t} \int^{+\infty}_{-\infty} \! \! \! d \omega \; e^{-\imath \omega t} \frac{ \hat{f}_{jpq}(\omega)}{\hbar [\omega_{pqkm} + \omega]}.
\label{3a_sol1}
\end{equation}
We can choose the boundary between the high and low-energy states so that the evolution of a low-lying state has no significant frequency components in the high-energy subspace. We can therefore neglect the dependence on $\omega$, $\omega_{p}$ and $\omega_{q}$ in the energy denominator of Eq.~(\ref{3a_sol1}) to obtain
\begin{equation}
\bar{a}^{\dagger}_{j}\bar{a}_{k}\bar{a}_{m} = \sum^{L}_{pq} \Vs{k}{m}{p}{q} e^{\imath \omega_{kmpq} t} \frac{\bar{a}^{\dagger}_{j}\bar{a}_{p}\bar{a}_{q}}{-\hbar (\omega_{k} + \omega_{k})}.
\label{3a_sol2}
\end{equation}

We now postulate that the full solution to Eq.~(\ref{partmotion3a}) has this form but with the matrix element $\Vs{k}{m}{p}{q}$ replaced with $\bra{k}{m} T_{\mbox{{\scriptsize H}}} \ket{p}{q}$, where $T_{\mbox{{\scriptsize H}}}$ is an operator to be determined. We therefore have the ansatz
\begin{equation}
\bar{a}^{\dagger}_{j}\bar{a}_{k}\bar{a}_{m} = \sum^{L}_{pq}\bra{k}{m} T_{\mbox{{\scriptsize H}}} \ket{p}{q} e^{\imath \omega_{kmpq} t} \frac{\bar{a}^{\dagger}_{j}\bar{a}_{p}\bar{a}_{q}}{-\hbar (\omega_{k} + \omega_{k})}.
\label{3a_sol3}
\end{equation}
Repeating the procedure which led to Eq.~(\ref{3a_sol2}) we find that this is a solution of Eq.~(\ref{partmotion3a}) if $T_{\mbox{{\scriptsize H}}}$ satisfies the Lippmann-Schwinger equation
\begin{equation}
T_{\mbox{{\scriptsize H}}} = V +  \sum^{H}_{pq} V\ket{p}{q}\frac{1}{-\hbar (\omega_{p} + \omega_{q})}\bra{p}{q} T_{\mbox{{\scriptsize H}}}.
\label{TH_app}
\end{equation}
This is the usual expression for the two-body T-matrix except for the fact that the intermediate states $\ket{p}{q}$ in the collision are restricted to the high-energy subspace. For this reason we have denoted the T-matrix with the subscript H. We expect that $T_{\mbox{{\scriptsize H}}}$ will be approximately equal to $T_{\mbox{{\scriptsize 2b}}}$ if the density and temperature of the system are such that the states rapidly become perturbative (i.e. if the boundary between L and H can be taken at low energy). The relationship between $T_{\mbox{{\scriptsize H}}}$ and $T_{\mbox{{\scriptsize 2b}}}$ is discussed below.

If we now substitute Eq.~(\ref{3a_sol3}) into Eq.~(\ref{motion1abar}) we obtain the restricted T-matrix $T_{\mbox{{\scriptsize H}}}$ in the equation of motion for the annihilation operator $\bar{a}_{i}$
\begin{equation}
\imath \hbar \frac{d \bar{a}_{i} }{dt} = \sum^{L}_{jkm}  \bra{i}{j} T_{\mbox{{\scriptsize H}}} \ket{k}{m} \bar{a}^{\dagger}_{j}\bar{a}_{k}\bar{a}_{m}e^{\imath \omega_{ijkm} t}.
\end{equation}
This has the same form as Eq.~(\ref{motion1abar}) but with a summation which extends only over low-lying states, and a modified two-body interaction which is the restricted T-matrix. This equation of motion therefore corresponds to an effective Hamiltonian for the low-energy subspace which is
\begin{equation}
\hat{H}_{\mbox{{\scriptsize eff}}} = \sum^{L}_{i} \hbar \omega_{i} \hat{a}^{\dagger}_{i}\hat{a}_{i} + \frac{1}{2} \sum^{L}_{ijkm}  \bra{i}{j} T_{\mbox{{\scriptsize H}}} \ket{k}{m} \hat{a}^{\dagger}_{i}\hat{a}^{\dagger}_{j}\hat{a}_{k}\hat{a}_{m}.
\label{H_eff_app}
\end{equation}
Finally we can use a unitary transformation to the basis orthogonal to the condensate which is used in the main text. The effective Hamiltonian can therefore be written in the form of Eq.~(\ref{Hamiltonian}) but with the substitutions $\sum \rightarrow \sum^{L}$ and $\Vs{i}{j}{k}{m} \rightarrow \bra{i}{j} T_{\mbox{{\scriptsize H}}} \ket{k}{m}$. This is the result quoted in Sec.~\ref{C5}.

The restricted T-matrix $T_{\mbox{{\scriptsize H}}}$ is related to the two-body T-matrix $T_{\mbox{{\scriptsize 2b}}}$ of Eq.~(\ref{T2}) by
\begin{equation}
T_{\mbox{{\scriptsize 2b}}}(z) = T_{\mbox{{\scriptsize H}}} +  \sum^{L}_{pq} T_{\mbox{{\scriptsize H}}}\ket{p}{q}^{sp}\frac{1}{z-(\epsilon^{sp}_{p}+\epsilon^{sp}_{q})}\, ^{sp} \bra{p}{q} T_{\mbox{{\scriptsize 2b}}}(z).
\label{TH_T2_app}
\end{equation}
This result can be proved by substituting the definition of $T_{\mbox{{\scriptsize H}}}$ from Eq.~(\ref{TH_app}), which leads directly to Eq.~(\ref{T2}) for $T_{\mbox{{\scriptsize 2b}}}$ if we assume that $z$ can be neglected for high-energy states and that $\sum^{L} + \sum^{H} = \sum$. This last assumption means that terms involving one high and one low index are neglected, which is justified by momentum conservation as mentioned above.

We can solve Eq.~(\ref{TH_T2_app}) in the homogeneous limit if we assume that both $T_{\mbox{{\scriptsize H}}}$ and $T_{\mbox{{\scriptsize 2b}}}$ can be approximated by a contact potential. Writing $T_{\mbox{{\scriptsize H}}} = g \, \delta({\bf r})$ and $T_{\mbox{{\scriptsize 2b}}} = U_{0} \, \delta({\bf r})$ this equation becomes (for $i,j,k,m,z = 0$)
\begin{equation}
g = \frac{U_{0}}{1 - U_{0}\,\alpha_{K}},
\end{equation}
where $\alpha_{K}$ is defined by
\begin{equation}
\alpha_{K} = \frac{1}{(2\pi)^{3}} \int^{K}_{0} \!\! d^{3}{\bf k} \, \frac{m}{\hbar^{2} k} = \frac{m K}{2 \pi^{2} \hbar^{2}}\, ,
\label{alpha_app}
\end{equation}
and the integral has a high-energy cut-off at a wave vector $K$. If the introduction of $T_{\mbox{{\scriptsize H}}}$ is to be useful we need to have $g \approx U_{0}$ which means that $U_{0}\,\alpha_{K} \ll 1$. Eq.~(\ref{alpha_app}) therefore gives
\begin{equation}
Ka \ll 1.
\label{upper_cutoff}
\end{equation}

Thus the requirement that $T_{\mbox{{\scriptsize H}}} \approx T_{\mbox{{\scriptsize 2b}}}$ sets an upper limit to the boundary between high and low-energy states. A lower limit is provided by the requirement that the approximations we made in eliminating the high-energy states are valid. Specifically we require that the only significant population occurs in the low-energy subspace and that the evolution in this subspace occurs on a time scale which is long compared to that of the high-energy states. At zero temperature the second requirement is the critical factor, and the time scale for the low-energy evolution is determined by $n_{0}U_{0} = (\hbar k_{0})^{2}/2m$. The condition that the high-energy states evolve faster than this is therefore
\begin{equation}
K \gg k_{0}.
\label{lower_cutoff}
\end{equation}
$K$ can simultaneously satisfy the competing conditions set by Eqs.~(\ref{upper_cutoff}) and (\ref{lower_cutoff}) if $na^{3} \ll 1$, i.e. if the system is a dilute gas.

At finite temperature the lower limit to $K$ is set by the first of the two requirements given above, which can be written as
\begin{equation}
K \gg 1/\lambda_{dB},
\label{lower_cutoffT}
\end{equation}
where $\lambda_{dB}$ is the de~Broglie wave length. Thus $K$ must simultaneously satisfy Eqs.~(\ref{upper_cutoff}) and (\ref{lower_cutoffT}) which can be done if $a/\lambda_{dB} \ll 1$. This condition is satisfied for a dilute gas even at temperatures above the critical point.

\begin{thebibliography}{99}
\bibitem{Gardiner_orthogonal} The fact that excitations orthogonal to the condensate can be obtained by including the parameters $c_{j}$ and solving Eq.~(\ref{BdGposition}) rather than Eq.~(\ref{BdGposition2}) was brought to our attention by Prof. C. W. Gardiner.
\end{thebibliography}
\end{document}